%% file: main.tex
\documentclass[reprint, aps, prx, superscriptaddress]{revtex4-2}
\usepackage[utf8]{inputenc}
\usepackage{graphicx}
\usepackage{xcolor,colortbl}
\usepackage{amsmath}
\usepackage{amssymb}
\usepackage[colorlinks=false]{hyperref}
\usepackage{physics}
\usepackage{amsthm}
\usepackage[section]{placeins}

\newcommand{\vecx}{{\boldsymbol{x}}}
\newcommand{\veca}{{\boldsymbol{a}}}
\newcommand{\veck}{{\boldsymbol{k}}}

\begin{document}

\title{Machine learning discovery of new phases in programmable quantum simulator snapshots}
\author{Cole Miles}
\affiliation{Department of Physics, Cornell University, Ithaca, NY 14853, USA}
\author{Rhine Samajdar}
\affiliation{Department of Physics, Harvard University, Cambridge, MA 02138, USA}
\author{Sepehr Ebadi}
\affiliation{Department of Physics, Harvard University, Cambridge, MA 02138, USA}
\author{Tout T.~Wang}
\affiliation{Department of Physics, Harvard University, Cambridge, MA 02138, USA}
\author{Hannes Pichler}
\affiliation{Institute for Theoretical Physics, University of Innsbruck A-6020, Austria}
\affiliation{Institute for Quantum Optics and Quantum Information, Austrian Academy of Sciences, Innsbruck A-6020, Austria}
\author{Subir Sachdev}
\affiliation{Department of Physics, Harvard University, Cambridge, MA 02138, USA}
	\affiliation{School of Natural Sciences, Institute for Advanced Study, Princeton NJ 08540, USA}
\author{Mikhail D.~Lukin}
\affiliation{Department of Physics, Harvard University, Cambridge, MA 02138, USA}
\author{Markus Greiner}
\affiliation{Department of Physics, Harvard University, Cambridge, MA 02138, USA}
\author{Kilian Q. Weinberger}
\affiliation{Department of Computer Science, Cornell University, Ithaca, NY 14853, USA}
\author{Eun-Ah Kim}
\email{eun-ah.kim@cornell.edu}
\affiliation{Department of Physics, Cornell University, Ithaca, NY 14853, USA}
\date{\today}

\begin{abstract}
  Machine learning has recently emerged as a promising approach for studying complex phenomena characterized by rich datasets. 
   In particular, data-centric approaches lend to the possibility of automatically discovering structures in experimental datasets that manual inspection may miss. Here, we introduce an interpretable unsupervised-supervised hybrid machine learning approach, the hybrid-correlation convolutional neural network (Hybrid-CCNN), and apply it to experimental data generated using a programmable quantum simulator based on Rydberg atom arrays. Specifically,  we apply Hybrid-CCNN to analyze new quantum phases on square lattices with programmable interactions. 
   The initial unsupervised dimensionality reduction and clustering stage first reveals five distinct quantum phase regions. In a second supervised stage, we refine these phase boundaries and characterize each phase by training fully interpretable CCNNs and extracting the relevant correlations for each phase.
   The characteristic spatial weightings and snippets of correlations specifically recognized in each phase capture quantum fluctuations 
   in the striated phase and identify two previously undetected 
   phases, the rhombic and boundary-ordered phases. These observations demonstrate that a combination of programmable quantum simulators with 
   machine learning can be used as a powerful tool for detailed exploration of correlated quantum states of matter.
\end{abstract}

\maketitle

\section{Introduction}

Recent advances in realization of programmable quantum simulators (PQSs)
have opened up a new era in the exploration of strongly correlated quantum matter \cite{Schauss2012Nature, AntoineReview2020, Labuhn2016Nature, Bernien2017}, which calls for new approaches for analyzing large volumes of data generated by such quantum devices. Using optical techniques, it is possible to arrange a  large number of qubits in arbitrary lattice geometries \cite{Ebadi2021Nature} and to control the Hamiltonian evolution of the system \cite{AntoineReview2020} dynamically in real time. Remarkably, these simulators can probe states within an extremely large  Hilbert space. For example, 
in a $13$\,$\times$\,$13$ system, the quantum states live in a $2^{169}$-dimensional space while each measurement probabilistically projects to just a single dimension. Tomographically \cite{Vogel1989Phys.Rev.A, James2001Phys.Rev.A, Gross2010Phys.Rev.Lett.} inferring the entire many-body wavefunction from such measurements themselves is a formidable task. While certain types of many-body states can be easily identified by evaluating simple local observables, many exotic quantum phases that can be explored on programmable simulators cannot be characterized using conventional approaches.

In this work, we introduce  a hybrid unsupervised-supervised machine learning approach to analyze the  data generated using programmable quantum simulators. Specifically, we apply this method to a PQS based on Rydberg atoms  arrayed  on a square lattice  \cite{Ebadi2021Nature}. We show how this framework directly reveals the correct order parameters to diagnose multiple density-wave-ordered phases, thus providing a route to the construction of (\textit{a priori} unkown) order parameters for potentially more complicated symmetry-breaking phases \cite{scholl2021quantum}. Additionally, our analysis allows us to uncover several \textit{new} features including:
({\it i\/}) the pattern of quantum correlations in the ``striated phase''; 
({\it ii\/}) a previously undetected ``rhombic'' phase in consistency with the theoretical predictions of Ref.~\onlinecite{samajdar_complex_2020}; and ({\it iii\/}) a boundary-ordering quantum phase transition in which the edge of the system develops long-range order while the bulk remains trivial.

\begin{figure*}[ht!]
    \centering
    \includegraphics[width=2\columnwidth]{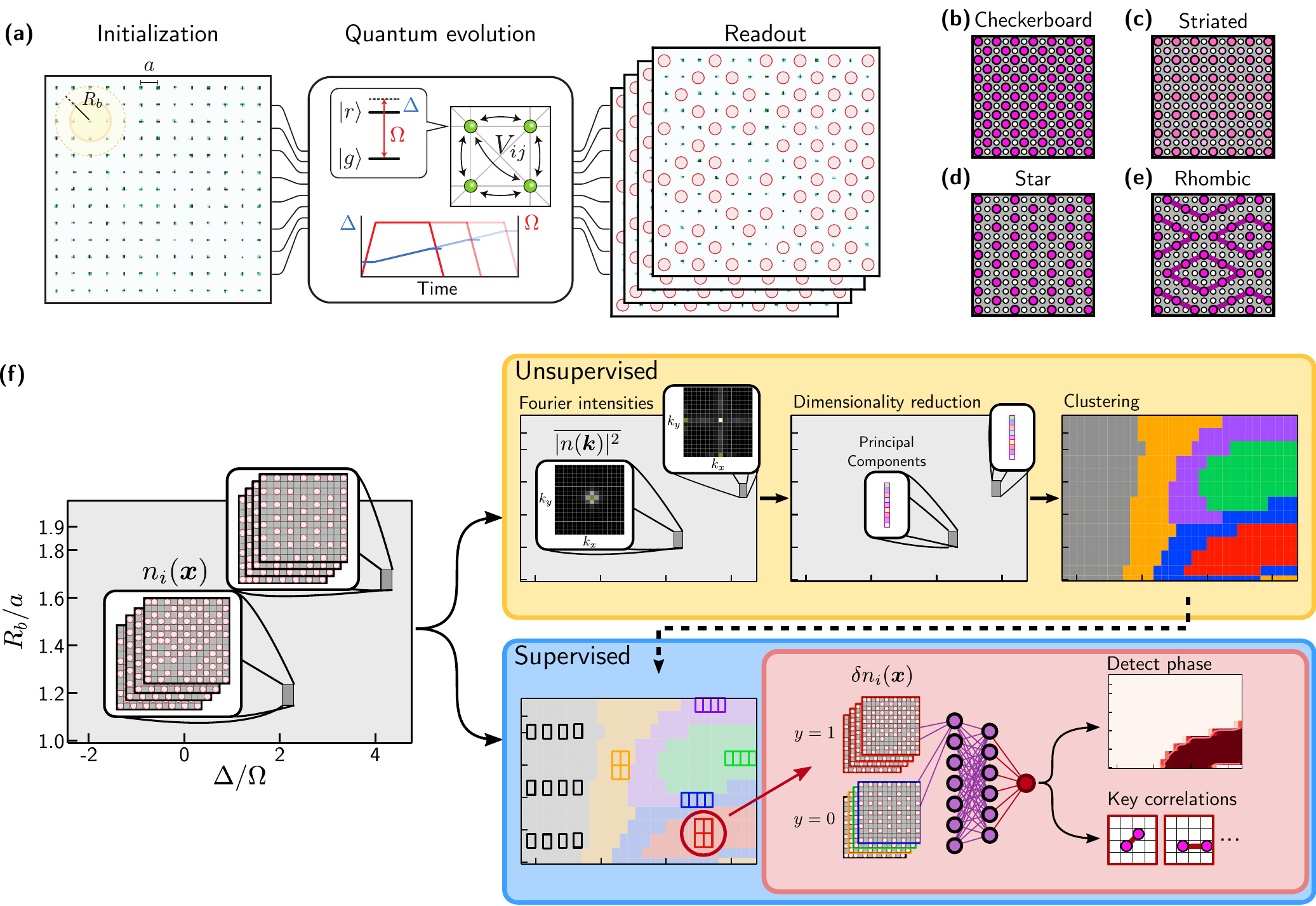}
    \caption{\textbf{(a)} Defect-free square lattices of neutral atoms undergo coherent quantum evolution for different values of blockade extent $R_b/a$ and linear detuning sweeps' endpoints $\Delta/\Omega$, followed by projective readout in which atoms excited to the Rydberg state are detected as loss (red circles).  \textbf{(b--e)} Idealized real-space patterns corresponding to phases predicted to be present at various regions of parameter space. Dark pink and white sites indicate $|r\rangle$ and $|g\rangle$ states, respectively, while the light pink sites in the striated phase are in a quantum superposition of $|r\rangle$ and $|g\rangle$. \textbf{(f)} A diagram outlining the Hybrid-CCNN approach. First, an unsupervised technique is used to generate a rough first-pass phase diagram. Here, we choose to measure average Fourier amplitudes $\overline{|n(\veck)|^2}$ at each $(\Delta, R_b)$, perform a dimensionality reduction using principal component analysis, and finally cluster using a Gaussian mixture model. The resulting phase diagram informs the starting ``seeds'' in the parameter space, from which snapshots are sampled in a second supervised stage. We then learn to distinguish these snapshots using interpretable classifiers, from which we can extract refined phase boundaries and key identifying features.}
    \label{fig:background}
\end{figure*}

Our programmable Rydberg quantum simulator [see Fig.~\ref{fig:background}(a)] consists of neutral atoms trapped in defect-free arrays of optical tweezers with 
programmable geometries 
\cite{Ebadi2021Nature}. Coherent laser excitations to atomic Rydberg states realize an Ising-like spin model \cite{Bernien2017,AntoineReview2020} described by the Hamiltonian
\begin{equation}
\frac{H}{\hbar} = \frac{\Omega}{2}\sum_i \left(|g^{}_i\rangle\langle r^{}_i| + |r^{}_i\rangle\langle g^{}_i|\right) - \Delta\sum_i n^{}_i + \sum_{i<j} V^{}_{ij}n^{}_i n^{}_j,
\label{ryd_ham}
\end{equation}
where atoms in the ground (Rydberg) state are denoted by $|g\rangle$ ($|r\rangle$), and  $n_i$\,$\equiv$\,$|r_i\rangle\langle r_i| $. The transverse field $\Omega$ corresponds to the Rabi frequency of the laser field, the longitudinal field $\Delta$ corresponds to the laser detuning, and $V_{ij} \equiv \mathcal{V}_0 / |\vecx_i - \vecx_j|^6$ is the long-range van der Waals interactions between Rydberg excitations at $\vecx_i$ and $\vecx_j$.

Density-matrix renormalization group (DMRG) calculations on the square lattice \cite{samajdar_complex_2020} have predicted a number of quantum phases of the Hamiltonian \eqref{ryd_ham} 
arising from the interplay between coherent laser driving and the long-range van der Waals interactions [see Fig.~\ref{fig:background}(b--e)]. These phases can be understood based on the Rydberg blockade phenomenon \cite{lukin2001}: the strong interactions $V_{ij}$ can prohibit (or ``blockade'') the simultaneous excitation of neighboring atoms to the Rydberg state. The spatial extent of this blockade (or equivalently, the interaction strength) is captured by the \textit{blockade radius}, defined as $R_b$\,$\equiv$\,$(\mathcal{V}_0/\Omega)^{1/6}$. The full phase diagram is thus parametrized by the ratio of the longitudinal to the transverse field, $\Delta/\Omega$, and $R_b / a$, where $a$ is the lattice spacing. For $\Delta/\Omega$\,$>$\,$0$, the system energetically favors maximizing the number of atoms in the Rydberg state. However, this is subject to the blockade constraint, so for $R_b/a$\,$\gtrsim$\,$1$, only one out of every pair of nearest neighbors can be excited; on a square lattice, this leads to the checkerboard phase with antiferromagnetic ordering of atoms in ground and Rydberg states. Higher values of $R_b/a$ result in various new density-wave-ordered phases. Some of these correspond to classical hard-sphere packing of Rydberg excitations \cite{samajdar_complex_2020}, while others are stabilized by quantum coherence between the ground and Rydberg states \cite{samajdar_complex_2020,Ebadi2021Nature}.

Recent experiments \cite{Ebadi2021Nature} have demonstrated three of these predicted states [Fig.~\ref{fig:background}(b--d)], namely, the checkerboard, striated, and star phases. In the  experiments, different values of $R_b/a$ are accessed by tuning the lattice spacing $a$ at fixed $\mathcal{V}_0$. By linearly ramping $\Delta/\Omega$ from negative to positive values, one can quasiadiabatically prepare different ordered states, which are probed by measuring order parameters derived from the Fourier transforms of the excitation densities.
However,
some of the more complex ordered phases are challenging to detect and analyze directly, especially in light of  
the finite system size and experimental imperfections. 
In particular, the finite system sizes introduce significant edge effects that impede the identification of phases based on manual inspection of the Fourier transforms.

To address the challenge associated with 
large volumes of data produced by PQS projective measurements, we 
introduce an unsupervised-supervised hybrid machine learning approach: Hybrid-CCNN. 
In Ref.~\onlinecite{Miles2021NatCommun}, some of us introduced the CCNN architecture, which modifies the standard convolutional network to allow direct interpretation of the key-feature correlations of a learned phase.
While this interpretability endowed the CCNN with the ability to reveal new theoretical insights, a direct supervised learning approach is ultimately limited by the subjective bias entering through training data selection.
Increasingly, the community has made efforts to develop ``phase recognition'' \cite{Carrasquilla2017NaturePhys} algorithms
to curtail such bias either by using fully unsupervised learning \cite{Wetzel2017Phys.Rev.Ea, hu_discovering_2017, Greplova2020NewJ.Phys., Liu2019Phys.Rev.B, Liu2018Phys.Rev.Lett., Liu2019Phys.Rev.B, Venderley2021ArXiv, kaming_unsupervised_2021, arnold_interpretable_2021, kottmann_unsupervised_2020, cole_quantitative_2021, huembeli_automated_2019} or by adopting an element of unsupervised learning
\cite{casert_interpretable_2019, huembeli_identifying_2018, rem_identifying_2019}. However, discovery of new phases or fluctuation effects from experimental data through such efforts are rare \cite{Venderley2021ArXiv} to date. Here, we introduce ``Hybrid-CCNN" that prefaces CCNN's supervised learning with an unsupervised discovery stage and apply the two-stage process shown in Fig.~\ref{fig:background}(f) on voluminous Rydberg PQS data to arrive at three discoveries.

An innovative feature of the Hybrid-CCNN is the use of different data preprocessing for the two stages to investigate different facets of the high-dimensional data: overall density modulations and snapshot-to-snapshot fluctuations. At each point in the $(\Delta,R_b)$ phase space, the available snapshot data consists of $250$ binary maps $n_i({\vecx };\{\Delta,R_b\}) \in \{0,1\}$, $i \in 1, ..., 250$ with
$\vecx \in [1, ..., 13]^2$. For the unsupervised learning stage, we focus on density modulations starting from the mean density-normalized snapshots $\tilde{n}_i(\vecx)$\,$=$\,$n_i(\vecx)$\,$-$\,$\bar{n}$, where $\bar{n}$ is the average density across all sites and snapshots. Then, Fourier transforming to $|\tilde{n}_i(\veck)|^2$, subtracting the background contribution $\sim$\,$\sum_{\veck} |\tilde{n}_i(\veck)|^2$,
and averaging over all snapshots leads us to a density-shift-invariant structure factor (refer to Appendix~A for details).
Beginning our analysis in Fourier space allows us to benchmark the Hybrid-CCNN findings against results obtained using a traditional measure of spatially modulated order for finite-size simulations (see Appendix B).

On the other hand, much information about the structure of many-body quantum fluctuations captured in each snapshot will be lost upon this averaging. Moreover, the Fourier basis is noisy in the presence of nonperiodic spatial variations and incommensurate domains.
Hence, in the final supervised phase, we choose to train our CCNNs to learn phase characteristics directly from the full density fluctuation maps $\delta n_i({\vecx})\equiv n_i({\vecx})-\bar{n}(\vecx)$ produced from each snapshot as independent training data. Here, $\bar{n}(\vecx)$\,$=$\,$\sum_{i}n_i(\vecx)/N$ is the snapshot-averaged Rydberg excitation density at each site, where $N$\,$=$\,$250$ is the number of snapshots at each $(\Delta, R_b)$. As detailed later, this normalization allows the CCNN to easily focus on fully connected components of low-order correlation functions and prevents the network from learning trivial overall excitation densities.

\section{Unsupervised Phase Discovery}

\begin{figure*}[htb]
    \centering
    \includegraphics[width=1.8\columnwidth]{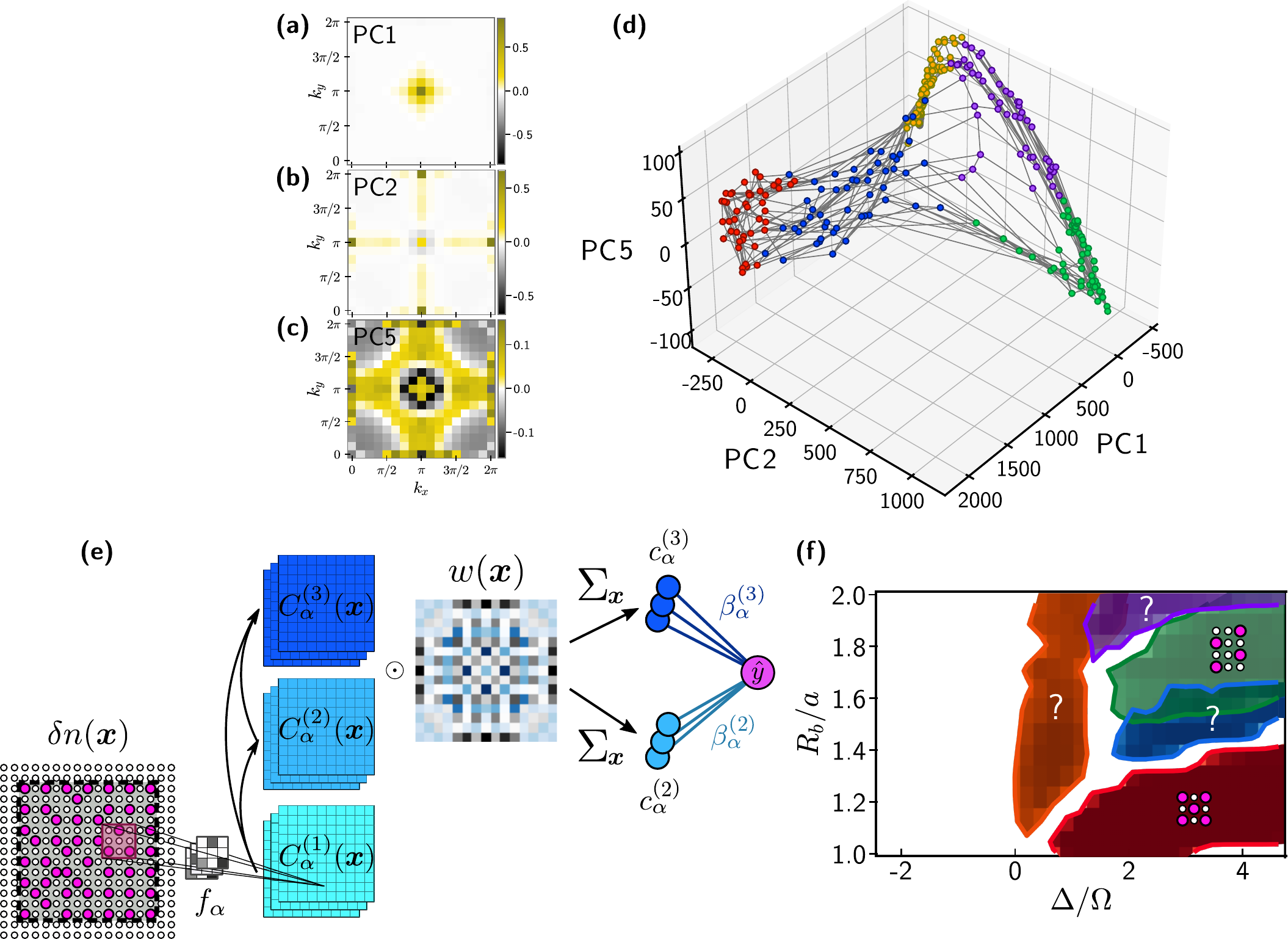}
    \caption{\textbf{(a--c)} Principal components $1, 2, 5$ in Fourier space. \textbf{(d)} Results of the GMM clustering performed in the reduced $10$-dimensional PCA space and projected for visualization into the space spanned by (a--c). \textbf{(e)} The CCNN architecture for the supervised learning stage, constructed here up to third order, $C_\alpha^{(3)}$, with three learned filters $f_\alpha$ and a learned spatial weighting $w(\vecx)$. First, a density-normalized snapshot $\delta n(\vecx)$ is convolved with the filters $f_\alpha$ to produce a convolutional map $C_\alpha^{(1)}(\vecx)$. The $\delta n(\vecx)$ are zero-padded to allow a convolution of the filters over the entire snapshot. Then, a series of polynomials are applied [Eqs.~\eqref{eq:ccnn-C2},\eqref{eq:ccnn-C3}] to produce maps $C_\alpha^{(m)}(\vecx)$, which measure $m$\textsuperscript{th}-order correlators near each $\vecx$. These maps are summed with a learned spatial weighting $w(\vecx)$ to produce features $c_\alpha^{(m)}$, which are used by a final logistic layer for classification. \textbf{(f)} Resulting phase diagram produced by supervised learning, obtained by cropping the classification confidence maps at level-set contours $\hat{y}=0.75$ and overlaying them.}
    \label{fig:techniques}
\end{figure*}

For the unsupervised learning stage, the aim is to form a collection of observables and use a clustering algorithm to map out different regions in the  $(\Delta,R_b)$ space where the observable set changes dramatically. 
The challenge is that clustering tends to be most robust in low-dimensional feature spaces.
On the other hand, even after averaging over snapshots, the full Fourier-space dimension is too large for standard clustering algorithms.
Specifically, working with a $16$\,$\times$\,$16$ grid in $\veck$-space \footnote{We choose to work with a larger $\veck$-space than the minimal $13 \times 13$ space so that we capture Fourier amplitudes at wavevectors $\pi$ and $\pi/2$.} implies a $256$-dimensional feature vector associated with each point in $(\Delta, R_b)$ space. To make this manageable for clustering, we further reduce the dimensionality of the feature space using principal component analysis (PCA)\cite{Bishop2006}.

PCA identifies vectors called principal components (PCs) in this $256$-dimensional feature space along which the data varies the most dramatically across the full phase diagram. Specifically, each of the principal components is a linear combination of multiple points in $\veck$-space that vary together across the phase space, visualized in Fig.~\ref{fig:techniques}(a--c). PC1 and PC2 show considerable overlap with the theoretical order parameters for checkerboard and star phases, respectively. While PC3 and PC4 do not immediately offer interpretation as an order parameter in Fourier space, possibly capturing peak broadening, PC5 resembles the Fourier-space order parameter associated with the rhombic phase predicted in DMRG simulations on a cylinder~\cite{samajdar_complex_2020}. This is tantalizing since the previous manual analysis of the data only detected three phases: checkerboard, star, and striated \cite{Ebadi2021Nature}. 

The final step of the unsupervised learning stage is to cluster the phase-space points $(\Delta, R_b)$ in the reduced feature space spanned by the first few dominant PCA components. As our clustering algorithm, we use a Gaussian Mixture Model \cite{Bishop2006} (GMM) for its robustness and invariance to the scale of each feature. The two choices we must make are the number of principal components to keep for the clustering, and the number of clusters to fit. We choose the first by increasing the number of retained principal components one-by-one until the clusters stabilize, finding $10$ to be sufficient in the process. We then determine the optimal number of clusters to be six as the Bayesian information criterion \cite{Bishop2006} plateaus past this number (see Appendix~A).

The clustering result shown in Fig.~\ref{fig:background}(f) partially resembles the manually obtained phase diagram based on evaluation of three target order parameters for the checkerboard, striated, and star phases \cite{Ebadi2021Nature} (see Appendix~C), except that the unsupervised learning indicates that there are five phases distinct from the disordered phase [shown in grey in Fig.~\ref{fig:background}(f)].
A visualization of the clusters projected to the subspace spanned by PC1, PC2, and PC5 shows that the clusters are tightly defined [Fig.~\ref{fig:techniques}(d)]. Although the clusters reside in submanifolds unrestricted to a single principal component axis, the red, green, and purple clusters are located far along the PC1, PC2, and PC5 axes, respectively. Hence, the unsupervised learning stage of Hybrid-CCNN identifies the checkerboard phase (red region and PC1) and the star phase (green region and PC2) with no prior knowledge. Moreover, it \textit{discovers} evidence of a new phase associated with a large PC5 component (purple). At this stage, the identities of the orange and blue phases also remain unclear from the unsupervised learning alone. 

\section{Supervised Phase Characterization}
\subsection{Architecture and training}
To better characterize each of the phases from the full dataset of snapshot-to-snapshot fluctuations, $\delta n_i(\vecx; \Delta, R_b)$, we now turn to a supervised learning stage. 
The unsupervised learning results indicate five phases distinct from the disordered phase and inform the starting choice of the training points. However, since the CCNN has access to the full information about quantum fluctuations, we anticipate changes to the precise phase boundaries. Hence, we optimize the choice of training points by iteratively exploring to minimize training loss, overlap between different phases, and understandibility \footnote{This is somewhat similar in concept to the ``learning by confusion'' scheme \cite{vanNieuwenburg2017NaturePhys, Liu2018Phys.Rev.Lett.}, though---for this complex multiphase system---we carry out these adjustments manually.}. 
To learn the distinct identity of each phase, we train multiple neural networks, with each given the task of identifying snapshots of a single phase against the rest through a binary classification \footnote{This is distinct from the common choice of using one neural network with multineuron output for multiphase detection\cite{Venderley2018Phys.Rev.Lett., Rem2019Nat.Phys., Bohrdt2019Nat.Phys.}.}. For the neural network architecture, we adapt the CCNN introduced in Ref.~\onlinecite{Miles2021NatCommun} to learn the spatial structure of correlations specific to each phase [see Fig.~\ref{fig:techniques}(e)].
CCNNs recognize $m$-site correlations \cite{Miles2021NatCommun} from site-normalized density fluctuation snapshots $\delta n_i(\vecx)$ through $m$\textsuperscript{th} order polynomials of convolutional maps produced with learned filters $f_\alpha(\veca)$, $C^{(m)}_\alpha(\vecx)$, defined by
    $ C^{(m)}_\alpha(\vecx) = \sum_{\veca_1 \ne \dots \ne \veca_m} \prod_{j=1}^m f_{\alpha}(\veca_j) \delta n(\vecx + \veca_j)$.
The use of local features $C_\alpha^{(m)}(\vecx)$ allows the CCNN to discover higher-order and spatially inhomogeneous correlations that are often overlooked, but which can be crucial for identification of a given phase \cite{Miles2021NatCommun}. 
For the present dataset, we find truncating to $m=3$
to be sufficient. Specifically, we have
\begin{widetext}
\begin{align}
  C^{(2)}_\alpha(\vecx) &=  \sum_{\veca_1 \ne \veca_2} f_\alpha(\veca_1)f_\alpha(\veca_2)\delta n(\vecx + \veca_1) \delta n(\vecx + \veca_2), \label{eq:ccnn-C2} \\
 C^{(3)}_\alpha(\vecx) &=  \sum_{\veca_1 \ne \veca_2 \ne \veca_3}f_\alpha(\veca_1)f_\alpha(\veca_2)f_\alpha(\veca_3)\delta n(\vecx + \veca_1) \delta n(\vecx + \veca_2) \delta n(\vecx + \veca_3).
    \label{eq:ccnn-C3}
\end{align}
\end{widetext}
For interpretability, we fix the filters to be positive definite. 
In addition, we learn spatial structures beyond the length scale of the filters using a follow-up spatial weighting $w(\vecx)$ \footnote{We restrict $w(\vecx)$ to be symmetric under reflections and rotations of the spatial coordinates for simplicity of parametrization.} applied to the correlation maps $C_\alpha^{(m)}(\vecx)$, resulting in scalar features 
\begin{equation}
 c_\alpha^{(m)} \equiv \sum_{\vecx}w(\vecx)C^{(m)}_\alpha(\vecx). 
    \label{eq:ccnn-w}
\end{equation}
These $c_\alpha^{(m)}$ features enter a final logistic layer coupled by weights $\beta_\alpha^{(m)}$ with bias $\epsilon$ to produce the output
\begin{equation}
\hat{y} = \left[1+\exp(-\sum_{\alpha,m}\beta_\alpha^{(m)} c_\alpha^{(m)} + \epsilon)\right]^{-1}.
    \label{eq:y}
\end{equation}
The sign and magnitude of the learned weights $\beta_\alpha^{(m)}$ can then be investigated to reveal the ferromagnetic $(+)$ or antiferromagnetic $(-)$ nature and significance of the identified correlations. This whole process is pictorially summarized in Fig.~\ref{fig:techniques}(e).

For training, the snapshots from the target training points are labeled with $y$\,$=$\,$1$ and those from the remaining training points are labeled with $y$\,$=$\,$0$.
During training, 
the filters $f_\alpha(\veca)$, logistic weights $\beta_\alpha^{(n)}$, bias $\epsilon$, and spatial weighting $w(\vecx)$ are all simultaneously learned by stochastic gradient descent to minimize the cross-entropy loss, which drives the predicted $\hat{y}$s towards their correct labels $y$ (Appendix~A). 
In order to arrive at a minimally complex set of features describing each phase, we perform ablation studies \cite{Meyes2019ArXiv} (described in Appendix~A) to
progressively determine the minimal number of components necessary for the CCNN's successful training. The resulting nuggets of each phase allows us to connect the full snapshot dataset to theoretical insight. 

\subsection{Interpreting the learned phases}

\begin{figure*}[t!]
    \centering
    \includegraphics[width=1.5\columnwidth]{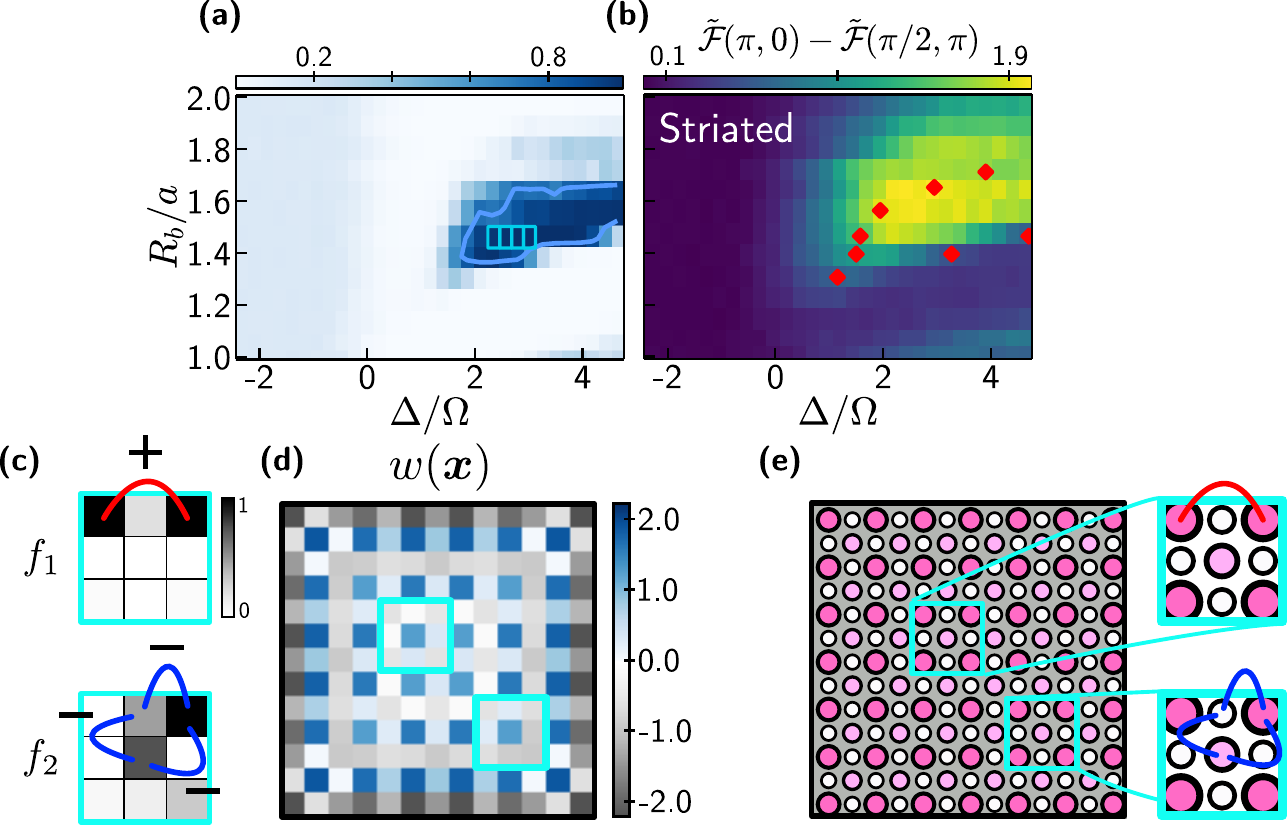}
    \caption{\textbf{(a)} CCNN-learned region of support for the striated phase, with highlighted boxes indicating the training points. \textbf{(b)} Previously used approximate order parameter detecting the striated phase. Red markers indicate phase boundaries obtained from DMRG simulations on a $9\times 9$ array \cite{Ebadi2021Nature}. \textbf{(c)} The filters learned by a third-order nonuniform CCNN to identify the striated phase in (a) and the signs on the $\beta_\alpha^{(2)}$ coefficients connecting the corresponding $c_\alpha^{(2)}$ to the output. For ease of display, the filter weights are normalized such that the maximum is $1$ within each filter. \textbf{(d)} The spatial weighting $w(\vecx)$ learned by a third-order CCNN identifying the striated phase. A single-pixel outer layer, corresponding to where the filter lands on the zero-padded region, is omitted for clarity. \textbf{(e)} A diagram showing example patches of the idealized striated phase whose correlations are measured by the CCNN of (c,d).
    }
    \label{fig:striated_ccnn}
\end{figure*}

We first focus on the red and green phase-space regions in Fig.~\ref{fig:techniques}(f). We find that for these phases, a fixed uniform spatial weighting ($w(\vecx)$\,$=$\,$1$) and truncation to second-order correlations are sufficient for characterization.
In Appendix~C, we show by the convolution theorem that a linear combination of uniform $c_\alpha^{(2)}$ features amounts to a sum of the structure factor $|\delta n(\veck)|^2$ weighted by the Fourier transform of the filters $f_\alpha(\veca)$. Inspecting the learned effective weightings in Fourier space (shown in Appendix~C), we remarkably find that these CCNNs identify the correct order parameters traditionally used to characterize the respective density-wave orderings.
Indeed, comparing these two regions to the earlier result \cite{Ebadi2021Nature} based on manual evaluation of order parameters in Fourier space, the red and green regions of Fig.~\ref{fig:techniques}(f) clearly map to checkerboard [Fig.~\ref{fig:background}(b)] and star [Fig.~\ref{fig:background}(d)] phases, respectively. 
For these phases, the key advantages of CCNN-based phase recognition are an unbiased discovery of the simplest order parameter suitable for the complexity of fluctuations present in the data (see Appendix~B), as well as noticeably sharper phase boundaries. We emphasize the nontriviality of the former observation as the Hybrid-CCNN is able to reveal the correct order parameters \textit{without} any prior input about the physics or structure of the density-wave-ordered ground states. This highlights the utility of our method for applications to potentially more complicated symmetry-breaking phases, for which the correct order parameters may not be immediately obvious.

\subsubsection{Fluctuations in the striated phase}
Next, we turn to the blue region in Fig.~\ref{fig:techniques}(f). From the CCNN trained to identify this phase, we produce the confidence map shown in Fig.\ref{fig:striated_ccnn}(a) by ``classifying'' each point in parameter space by averaging the correlator features $c_\alpha^{(m)}$ across all available snapshots and then predicting using Eq.~\ref{eq:y}. We find that resulting map overlaps significantly with the region previously demarcated as the striated phase in Ref.~\onlinecite{Ebadi2021Nature} using an approximate Fourier order parameter [Fig.~\ref{fig:striated_ccnn}(b)]. However, the CCNN confidence map produces a remarkably sharper region of support for the phase.
In the parameter regime of interest, the striated phase is unique in that it is an intrinsically quantum phase stabilized by the quantum fluctuations driven by the transverse field $\sim\Omega$; in fact, the Rydberg interaction term alone energetically favors star order, so the striated phase does not exist in the classical limit of $\Omega/\Delta\rightarrow 0$ \cite{samajdar_complex_2020}.
Motivated by this, previous work \cite{Ebadi2021Nature} characterized this phase using a mean-field ansatz consisting of a product of single-particle states in quantum superpositions of $\ket{g}$ and $\ket{r}$.
However, such an ansatz describes a product state.
In contrast, the two and three-site connected correlations that the CCNN learned to recognize using two filters $f_1$ and $f_2$
[Fig.~\ref{fig:striated_ccnn}(c)], combined with the long-ranged structure learned by the weight map $w(\vecx)$ [Fig.~\ref{fig:striated_ccnn}(d)], offer a first glimpse at potentially nontrivial
quantum many-body correlations in this region.

\begin{figure*}[t]
    \centering
    \includegraphics[width=1.9\columnwidth]{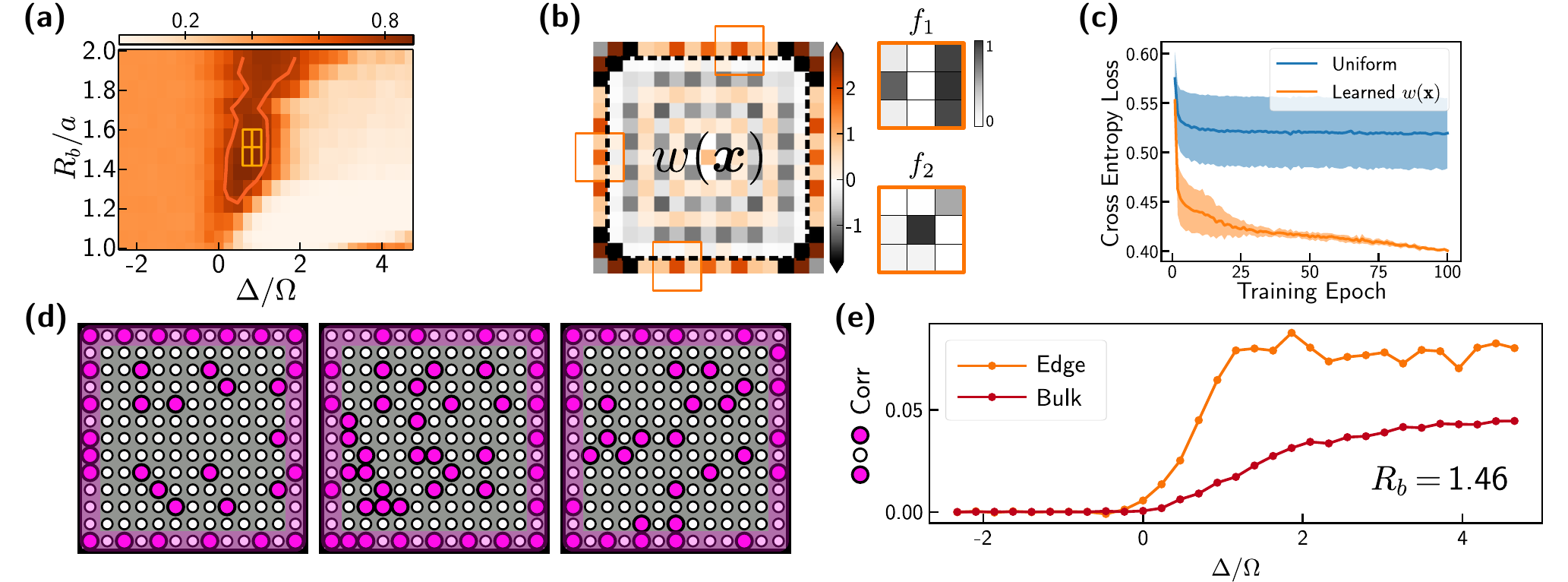}
    \caption{\textbf{(a)} CCNN-learned region of support for the edge phase, with highlighted boxes indicating the training points. \textbf{(b)} The learned spatial weighting and filters for the model trained to identify the edge-ordered phase, with the spatial extent of the snapshot indicated by the dashed line. The outermost pixels correspond to where the filter is centered on zero-padding but ``clip" the edge pixels of the snapshot. For display purposes, the filter weights are normalized such that the maximum is $1$ within each filter. \textbf{(c)} Measured performance discrepancy between second-order models with a fixed uniform $w(\vecx)$\,$=$\,$1$ and a freely learned spatial weighting. The central lines and bands show the mean and standard deviation across five randomly initialized models of each type, respectively. \textbf{(d)} Experimental snapshots from the training set, showing $\bullet\circ\bullet$ motifs primarily only along the single-site border, with the interior highly disordered. \textbf{(e)} Measurement of the third-nearest-neighbor $\langle \delta n_{i,j} \delta n_{i+2, j}\rangle$ connected correlation function within the edge and bulk (all sites but the outermost two-site strips) along a cut at $R_b$\,$=$\,$1.46$, averaged across translations and other symmetries.}
    \label{fig:edge_ccnn}
\end{figure*}

\begin{figure*}[t]
    \centering
    \includegraphics[width=1.9\columnwidth]{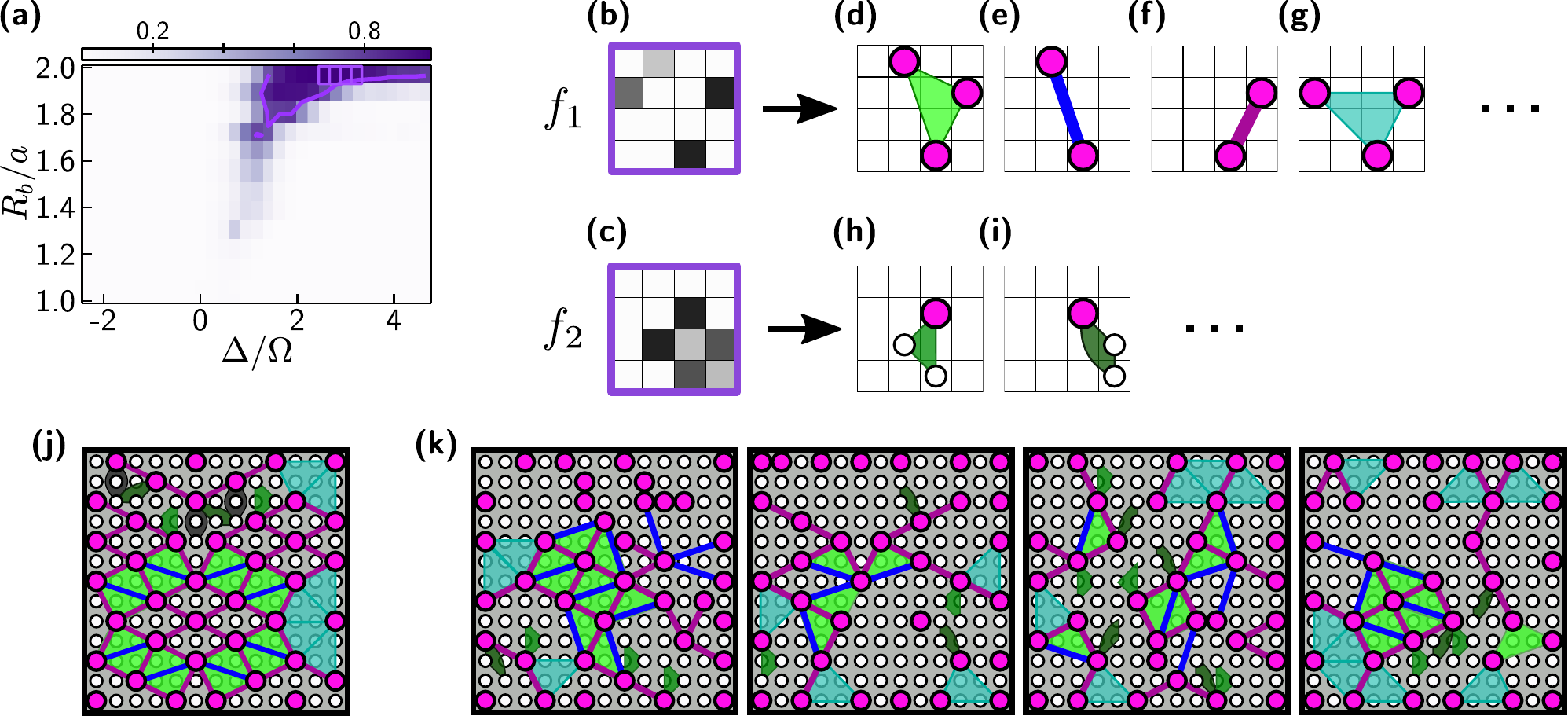}
    \caption{\textbf{(a)} The region of support for the rhombic phase as learned by the full CCNN model of Fig.~1(n), with $5\times 5$ filters and highlighted boxes indicating the training points. \textbf{(b,c)} The two learned $4$\,$\times$\,$4$ convolutional filters for a simplified model ($w(\vecx) = 1, \beta_\alpha^{(n)} \ge 0$) trained to identify the rhombic phase. \textbf{(d--i)} High-weight two- and three-point connected correlators measured by $c_\alpha^{(2)}, c_\alpha^{(3)}$ resulting from the filters in (a,b). We find $\beta_2^{(2)}$ to be nearly zero, so we omit two-point correlators stemming from the filter $f_2$. Our CCNN is symmetrized, (see Appendix~B) and so measures all correlators symmetry-equivalent under rotations and flips to those shown. \textbf{(j)} Identification of these two- and three-point motifs in the idealized rhombic ordering with boundary defects (light blue). \textbf{(k)} Identification of local occurrences of these motifs in experimental snapshots sampled from the training set.}
    \label{fig:rhombic_ccnn}
\end{figure*}

First, the weight map $w(\vecx)$ learned to identify a specific sublattice in the bulk, and correspondingly activates the filters only when they are centered on this sublattice. As a result, the CCNN measures correlations within repeating $3\times3$ blocks that span the system [see Fig~\ref{fig:striated_ccnn}(e)]. The learned positive sign of the logistic weight $\beta_1^{(2)}$ associated with the filter $f_1$ implies that the corner sites will tend to be jointly excited, which matches the expected pattern of excitations in the striated phase.
Moreover, the weight map identifies correlations within this effectively larger unit cell via the filter $f_2$. Specifically, the learned negative logistic weight $\beta_2^{(2)}$, combined with the pattern in filter $f_2$, indicates that the CCNN identified pairwise anticorrelations between three sites in the unit cell as a relevant feature of the striated phase.
Note that such pairwise anticorrelation of three sites is not possible for any perfectly polarized product state, in which the atom at each site is exclusively in either $\vert g \rangle$ or $\vert r \rangle$. Rather, Fig.~\ref{fig:striated_ccnn}(c) captures the tomographic information that on each site, the atom is in the state $\vert g \rangle$ ($\vert r \rangle$) with a small admixture of $\vert r \rangle$ ($\vert g \rangle$).

The persistence of nonzero second-order connected correlations (which should ideally vanish deep in an ordered phase) calls for a description beyond a simple  product-state ansatz. Quantum correlations and entanglement can arise from two primary sources. Firstly, they might be present in the ground state itself, particularly in the vicinity of a second-order quantum phase transition \cite{metlitski2009entanglement}. Second, they might be generated in the dynamical state preparation process due to the quantum Kibble-Zurek mechanism \cite{Zurek2005PRL}, where nonadiabatic processes can coherently generate superpositions including excited states that generically result in entanglement.
Our Hybrid-CCNN approach cannot discern between these scenarios as it is agnostic to the actual origin of the correlations. Nonetheless, to gain further insight into the potential entanglement structure, we inspect the bipartite entanglement entropy and correlations within a $9\times 9$ system using the density-matrix renormalization group (DMRG) in Appendix~D. We find that the calculated von Neumann entanglement entropy $\mathcal{S}$ peaks sharply along transition lines, before plateauing to a small but nonzero value within the phase. This is accompanied by anticorrelations between the excited sublattices as found by the CCNN, though the state preparation process appears to significantly extend the support of these correlations as compared to the DMRG ground state (see Appendix D).

\subsubsection{The boundary-ordered phase}

Having analyzed the properties of the striated phase, next, we turn to the first of the two mysterious phases, depicted in dark orange in Fig.~\ref{fig:edge_ccnn}(a). The first clue regarding the identity of this phase comes from the learned weight map $w(\vecx)$, which focuses strongly on the edges of the snapshots. As shown in 
Fig.~\ref{fig:edge_ccnn}(b), the CCNN
learned to measure the differences in correlations between the bulk and the boundary by having large $w(\vecx)$\,$>$\,$0$ along the edge and predominantly $w(\vecx)$\,$<$\,$0$ in the interior. The learned filters focus the CCNN's attention on specific short-range two-point correlations that differ significantly between the edge and bulk.
Figure~\ref{fig:edge_ccnn}(c) demonstrates the dramatic performance gain enabled by the learned edge-centered weight map.
Inspection of the experimental snapshots in this orange phase [Fig.~\ref{fig:edge_ccnn}(d)] indeed confirms a regular occurrence of local $\mathbb{Z}_2$ patterns of $(\bullet\circ\bullet)$ along the edges of the snapshots. In contrast, the bulk of the snapshots appear disordered, further evidenced by explicit measurements of correlation functions along the edge and in the bulk in Fig.~\ref{fig:edge_ccnn}(e).

While the local $\mathbb{Z}_2$ pattern is commensurate with the neighboring checkerboard and striated phases, the reduced energetic cost of having Rydberg excitations along the boundary (relative to the bulk), due to the presence of fewer neighbors, actually allows the edge to order \textit{before} the bulk. Hence, we identify this mystery phase as a boundary-ordered phase characterized by the edge ordering in the absence of long-range order in the bulk. The subsequent onset of bulk order, in the presence of preexisting edge order, defines an ``extraordinary'' boundary universality class~\cite{metlitski2020boundary}. We highlight that the existence of this boundary-ordered phase is a new \textit{discovery} of the present work since this phase cannot be obtained on geometries with fully periodic or cylindrical boundary conditions, as was used for earlier DMRG calculations \cite{samajdar_complex_2020}. Critical to the identification of this phase  is the real-space nature of the CCNN analysis as the edge ordering introduces a large number of artifacts into $\langle |n(\veck)|^2 \rangle$, which can challenge traditional Fourier-based analysis.
Interestingly, a complementary work \cite{Kalinowski2021} independently detected this edge ordering in quantum Monte Carlo simulations of the system with open boundary conditions, and confirmed the first-order nature of several transitions.

\subsubsection{The rhombic phase}
Finally, we examine the other mystery phase identified by the Hybrid-CCNN: the purple swath in Fig.~\ref{fig:rhombic_ccnn}(a). To identify the defining characteristic of this phase, we restrict
the CCNN to learn positive correlation functions by enforcing $\beta_\alpha^{(n)} \ge 0$ during training, increase the filter size to $4\times 4$, and fix uniform $w(\vecx)$\,$=$\,$1$ (see Appendices C and D). 
The CCNN learns the two filters $f_1$ and $f_2$ shown in Fig.~\ref{fig:rhombic_ccnn}(b,c) and uses a combination of second- and third-order correlations $c^{(2)}_\alpha, c^{(3)}_\alpha$ to recognize this phase as shown in Fig.~\ref{fig:rhombic_ccnn}(d--i).  
Remarkably, these learned motifs strongly point towards the rhombic phase from among the candidate ordering patterns in Fig.~\ref{fig:background}(b--e).

The rhombic phase is an intricately patterned density-wave-ordered phase characterized by Fourier peaks at $\pm(\pi,\pi/4)$, $\pm(2\pi/5,\pi)$ (and their $\mathcal{C}_4$-rotated copies) [see Fig.~\ref{fig:background}(e)], which was originally predicted by Ref.~\onlinecite{samajdar_complex_2020}. However, given its large unit cell comprising 40 sites, the robustness of this phase in the actual experimental system of \citet{Ebadi2021Nature} is \textit{a priori} unclear due to both the long-ranged tails of the van der Waals interactions and the incompatibility of the ideal ordering pattern with the dimensions of the lattices used. Our results illustrate that, interestingly, we can still find characteristic remnants of this phase.
In particular, the three-point motif of Fig.~\ref{fig:rhombic_ccnn}(d) provides a unique signature of the rhombic phase as a fragment of a full rhombic crystal while Fig.~\ref{fig:rhombic_ccnn}(g) occurs as an edge defect when the rhombic pattern is embedded in the finite incommensurate system
as shown in Fig.~\ref{fig:rhombic_ccnn}(j).
Additionally, the shorter-range three-point motifs of Fig.~\ref{fig:rhombic_ccnn}(h) and (i) occur most frequently in the rhombic phase (see Appendix~E). The virtue of these motifs is that they signify the tendency of fluctuations towards rhombic ordering even when extended ordered portions cannot form inside a finite system (due to the large and incommensurate $6\times 5$ unit cell). Indeed, these motifs are ubiquitous in the experimental snapshots sampled from this phase region as we showcase in Fig.~\ref{fig:rhombic_ccnn}(k). Hence, we can unambiguously identify the second mystery phase to be the finite-size manifestation of the rhombic phase.

\section{Discussion}
In summary, we developed a supervised-unsupervised hybrid machine learning approach, the Hybrid-CCNN, to reveal collective quantum phenomena in voluminous quantum snapshot datasets and applied our approach to square-lattice Rydberg PQS data. The initial unsupervised stage used Fourier intensities $\overline{\langle|n(\veck)|^2\rangle}$ and clustering in a low-dimensional feature space obtained using PCA to reveal a rough initial phase diagram. This first pass reveals the number of phases to expect and informs the initial location of training points for the supervised stage. The phase diagram is then refined in the second supervised stage by training nonuniform CCNNs to define sharper phase boundaries and uncover the identities of each phase. The identities revealed in this way not only confirmed the previously known phases \cite{Ebadi2021Nature, samajdar_complex_2020} but also resulted in new insight into potential quantum entanglement structures in the striated phase and the discovery of two previously undetected phases: the edge-ordered phase and the rhombic phase. To the authors' knowledge, this is the first time new insights and discoveries were gained from quantum simulator data using machine learning tools trained \textit{entirely} on experimental data.

With the rapid progress towards
probing more exotic quantum many-body phenomena using PQS \cite{Samajdar.2021, Verresen.2021, semeghini_probing_2021},
the need for new data-centric approaches to extracting insight from large volumes of quantum snapshot data will only grow.
Here, we demonstrated that the Hybrid-CCNN can not only meet this need but also enable the discovery and identification of new quantum states. 
The Hybrid-CCNN's ability to extract spatial structures of a quantum state at multiple length 
scales is particularly valuable given the limited spatial extent and incommensurate domains of phases produced by finite systems under quasiadiabatic state preparation \cite{Zurek2005PRL}.
In particular, this ability enabled the discovery of the edge-ordered phase and rhombic phases in our present work. Given the challenges to establishing and detecting new ordered phases due to the inevitable breakdown of adiabaticity when sweeping across a quantum phase transition \cite{Zurek2005PRL}, it will be interesting to compare the motifs learned by Hybrid-CCNN with rigorous finite-size scaling studies of corresponding quantum Monte Carlo simulations. Also tantalizing is the Hybrid-CCNN's ability to detect nonclassical correlations (that defy description in terms of product states) in the snapshot data of pure states; for instance, this hinted at nontrivial quantum entanglement among diagonal bonds in the striated phase. Comparing CCNN-detected entanglement against traditional measures is an interesting direction for future exploration. 
We anticipate such probes of entanglement to critically contribute to the identification and understanding of new exotic states that are beginning to become experimentally accessible \cite{semeghini_probing_2021}.

\section{Acknowledgements}

C.M.\ acknowledges funding support by the U.S. Department of Energy, Office of Science, Office of Advanced Scientific Computing Research, Department of Energy Computational Science Graduate Fellowship under Award Number DE-SC002034. K.W.\, M.G.\ and E-A.K.\ acknowledge support by the National Science Foundation through grant No.~OAC-1934714. This research was supported in part by a New Frontier
Grant from Cornell University’s College of Arts and Sciences.
R.S.\ and S.S.\ acknowledge support by the U.S. Department of Energy, Grant DE-SC0019030. S.E., T.T.W., H.P., and M.D.L.\ acknowledge financial support from the Center for Ultracold Atoms, the National Science Foundation, the U.S. Department of Energy (DE-SC0021013 \& LBNL QSA Center), the Army Research Office, ARO MURI, an ESQ discovery grant, and the DARPA ONISQ program. The authors acknowledge helpful discussions with Soonwon Choi, Marcin Kalinowski, and Roger Melko. We would also like to thank H. Levine, A. Keesling, G. Semeghini, A. Omran, D. Bluvstein for the use of experimental data presented in this work.  The DMRG calculations in this paper were performed using the ITensor package \cite{itensor} and were run on the FASRC Odyssey cluster supported by the FAS Division of Science Research Computing Group at Harvard University.

\input{supp}

\bibliography{RydbergCCNN}

\end{document}

%% file: supp.tex

\appendix
\renewcommand{\theequation}{A\arabic{equation}}
\renewcommand{\thefigure}{A\arabic{figure}}

\newcommand{\vecb}{{\boldsymbol{b}}}
\newcommand{\vecc}{{\boldsymbol{c}}}
\newcommand{\veci}{{\boldsymbol{i}}}
\newcommand{\vecj}{{\boldsymbol{j}}}

\newpage

\section{Unsupervised learning details}
To perform our initial rough-estimate unsupervised phase discovery, we first collect features at each experimental parameter point $(\Delta, R_b)$. For easy interpretability, we choose to work with simple features that represent all second-order fluctuations, but are blind to the overall Rydberg excitation density present in the snapshots. We find that blinding our entire machine learning pipeline to the overall density results in phase boundaries that are more closely aligned to meaningful changes in spatial orderings rather than simple increases in Rydberg excitation densities. More complex phase transitions would require higher-order correlations or nonlinear unsupervised learning techniques, but we find the following approach to be sufficient for our data. To allow for direct comparison to previous experimental work \cite{Ebadi2021Nature}, we would like to work with the average squared Fourier amplitudes of the snapshots at each $(\Delta, R_b)$. In order to make the process blind to overall density changes, we must take two steps. 

First, in this section, we work with snapshots which are overall density-normalized as $\delta n_i(\vecx) \equiv n_i(\vecx) - \langle n \rangle$, where the expectation value is computed over all sites of all snapshots available at the same $(\Delta, R_b)$. Note that this is different than the per-site density normalization used in the supervised follow-up. In the supervised phase, we found that per-site density normalization was necessary to build interpretable order parameters for phases with subtle correlation structures such as the striated and rhombic phases, which could be masked by average density modulations induced by the edge ordering. Per-site density normalization subtracts out this average density modulation, allowing connected correlators to be easily measured. However, while difficult to interpret, these very same average density modulations appear to be key to the success of our unsupervised phase, which is restricted to measuring raw Fourier amplitudes.

We then measure the average squared Fourier amplitudes of these normalized snapshots:
\begin{equation}
    \delta\hat{n}(\veck; \Delta, R_b) = \frac{1}{N(\Delta, R_b)} \sum_{n_i \in (\Delta, R_b)} \bigg| \sum_{\vecx} e^{-i\veck\cdot\vecx} \delta n_i(\vecx)\bigg|^2,
    \label{eq_supp_ft1}
\end{equation}
where $N(\Delta, R_b)$ is the number of available snapshots at the parameter value $(\Delta, R_b)$.

\begin{figure*}[]
    \centering
    \includegraphics[width=1.95\columnwidth]{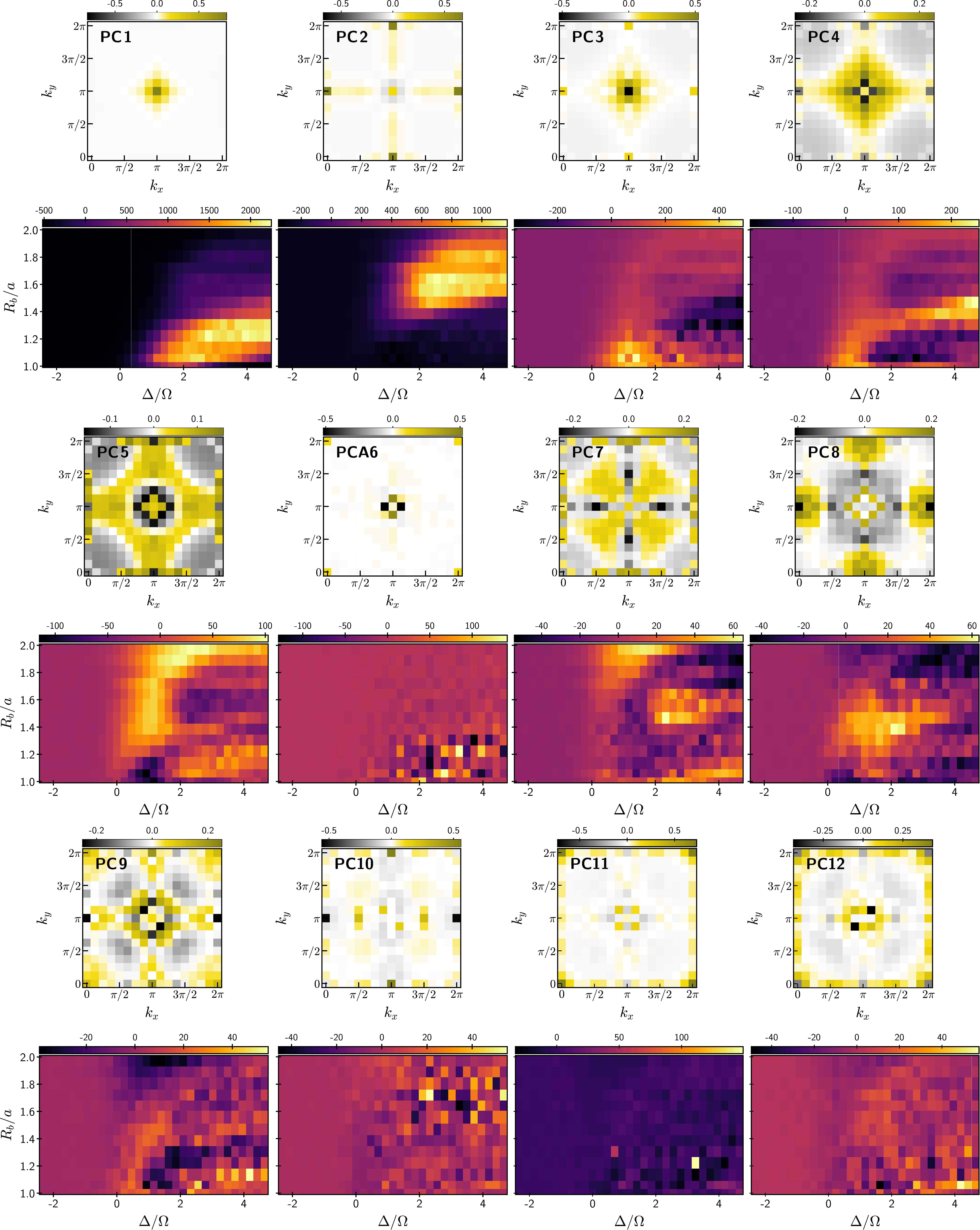}
    \caption{The top twelve principal components, shown as weightings in $\veck$-space, and the average projection onto each component across the experimental parameter space.}
    \label{fig:supp_unsupervised}
\end{figure*}

Our first density normalization does not make $\delta \hat{n}(\veck)$ invariant under an overall shift in the density of the $n(\vecx)$s, even away from $\veck$\,$=$\,$\mathbf{0}$, due to the input data being sampled from a bosonic system with the values of $n(\vecx)$ restricted to $0$ or $1$. In this case, the marginal distribution of each individual site's density is a Bernoulli distribution whose variance is linked to its mean as $\langle \delta n(\vecx)^2 \rangle = \langle n(\vecx) \rangle (1 - \langle n(\vecx) \rangle)$. In this way, information about the density can ``bleed'' through to the learning process at all $\veck$-points.
We can see this explicitly by expanding out the square in Eq.~\eqref{eq_supp_ft1} and rewriting it as
\begin{multline}
    \delta\hat{n}(\veck; \Delta, R_b) = \frac{1}{N(\Delta, R_b)} \sum_{n_i} \Big[
        \sum_{\vecx} \delta n_i(\vecx)^2 \\ + \sum_{\vecx\ne\vecx'} e^{-i\veck \cdot (\vecx - \vecx')} \delta n(\vecx) \delta n(\vecx')
    \Big].
\end{multline}
To make our Fourier-space features invariant under density shifts we need to remove the first term. Using Plancherel's theorem, we can achieve this this by normalizing once more in $\veck$-space as
\begin{equation}
    \hat{p}(\veck; \Delta, R_b) = \delta\hat{n}(\veck; \Delta, R_b) - \frac{1}{L^2} \sum_{\veck'} \delta\hat{n}(\veck'; \Delta, R_b),
\end{equation}
with $L$ being the number of $\veck$-points sampled along each direction of the discrete Fourier transform. These resulting features $\hat{p}(\veck; \Delta, R_b)$ are then provided to the principal component analysis, the results of which are summarized in Fig.~\ref{fig:supp_unsupervised}. Figure~\ref{fig:supp_unsupervised} shows the resulting top 12 principal components, which together capture $>99.9\%$ of the variance of the dataset. Notably, since all of the input features lie within the $\sum_{\veck}\hat{p}(\veck) = 0$ subspace, so do the resulting principal component vectors.

Below each PCA vector in Fig.~\ref{fig:supp_unsupervised}, we show its average inner product with each $\hat{p}(\veck; \Delta, R_b)$ across parameter space. These maps reveal which areas of parameter space have average Fourier-space intensity patterns that match the principal component vectors well. Importantly, many of these maps vary smoothly in parameter space and as such, correspond to meaningful changes in Fourier structure. Meanwhile, at first glance PCA6, and partially PCA10, seem visually noisy in parameter space. However, this is due to these PCA components breaking a rotational symmetry in Fourier space. In phase regions where the relevant Fourier peaks are present, these PCA components are either strongly postive or strongly negative depending on which way the symmetry is broken, while in other regions they remain close to zero. These components could be improved by further post-processing; however, for simplicity, we do not do so here.

To robustly perform the GMM clustering, we initialize the clusters using the k-Means algorithm \cite{Bishop2006}, repeat the initialization and clustering with different random seeds until 500 sequential clusterings do not improve the final log-likelihood, and keep the best-found clustering. Interestingly, random initializations (rather than k-Means) often produce clusterings with higher final log-likelihoods but poor structures in parameter space, with one cluster often a seemingly-random collection of points across parameter space. This ambiguity points to the persisting need to have a physicist ``in the loop'' verifying machine learning (ML) results at each stage, which is made easier when using interpretable techniques.

\begin{figure*}[t]
    \centering
    \includegraphics[width=1.7\columnwidth]{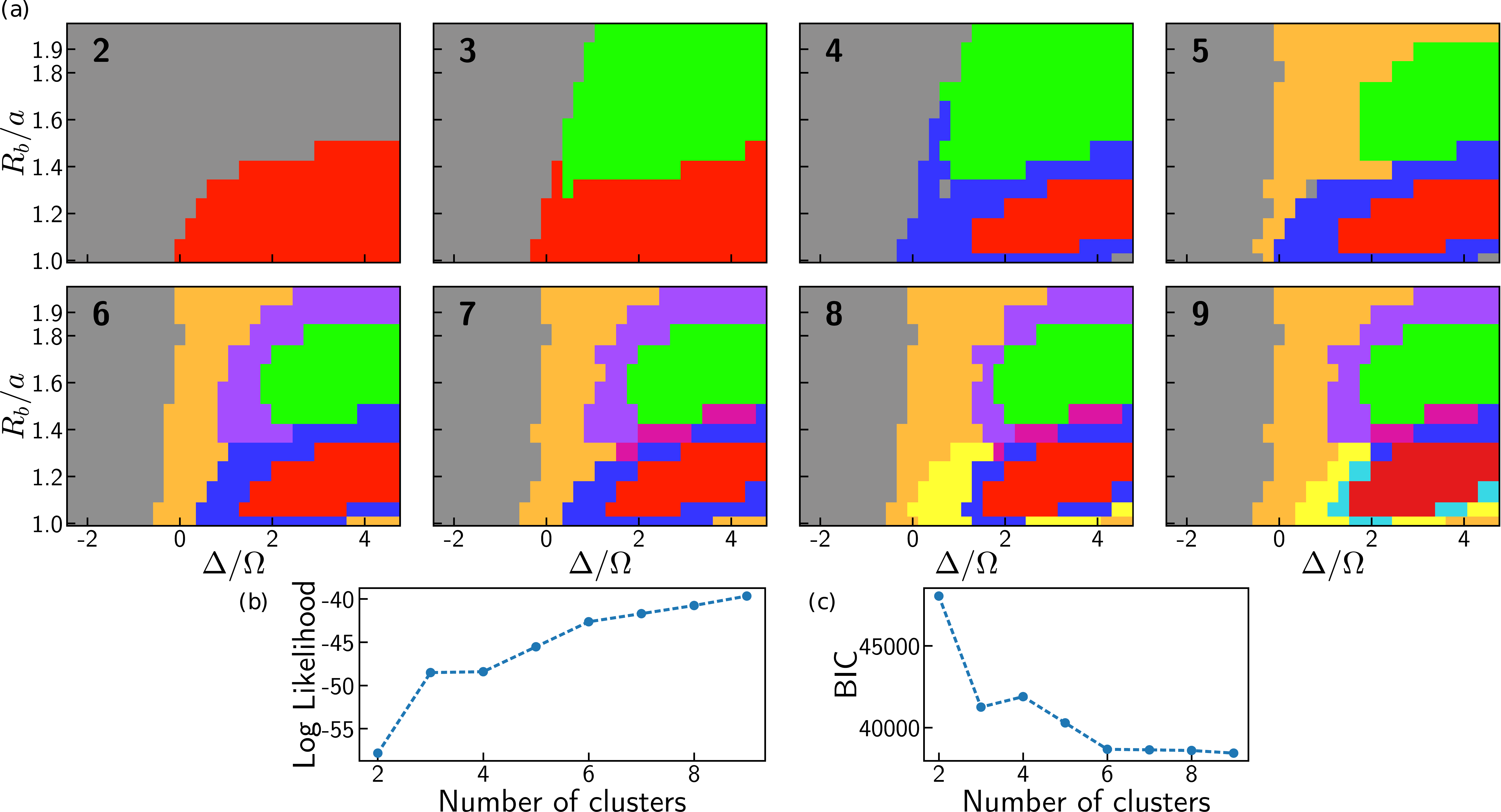}
    \caption{(a) Clusters obtained when performing a Gaussian mixture model clustering with varying number of clusters (from 2--9), and the top 14 PCA components retained. Colors of each cluster are manually chosen for visual continuity. (b,c) The log-likelihood and Bayesian information criterion (BIC) for the best mixture model at each number of clusters.}
    \label{fig:supp_unsupervised_numclusts}
\end{figure*}

To determine the appropriate number of clusters and PCA components to retain, we perform the entire clustering process while varying the number of clusters and the number of retained PCA components. In Fig.~\ref{fig:supp_unsupervised_numclusts}(a), we find that starting from two clusters, each additional cluster meaningfully captures the next-strongest phase in the dataset, with topology similar to the clustering of the main text with $N_{\mathrm{clust}} = 6$. As expected, each additional cluster improves the final log-likelihood [Fig.~\ref{fig:supp_unsupervised_numclusts}(b)]; however, a distinct kink and reduction in slope is observed after $N_{\mathrm{clust}} = 6$. This is reflected additionally by the Bayesian information criterion  (BIC) \cite{Bishop2006}, a standard heuristic metric for determining the correct number of clusters, plateauing past this point. As we should pick the simplest model which minimizes the BIC, this points to $N_{\mathrm{clust}} = 6$ as being optimal. Clusterings with $N_{\mathrm{clust}} > 7$ are increasingly noisy and can be seen to simply be chunking off transition regions between phases, rather than dramatically changing the phase diagram's topology.

\begin{figure*}[h!]
    \centering
    \includegraphics[width=1.42\columnwidth]{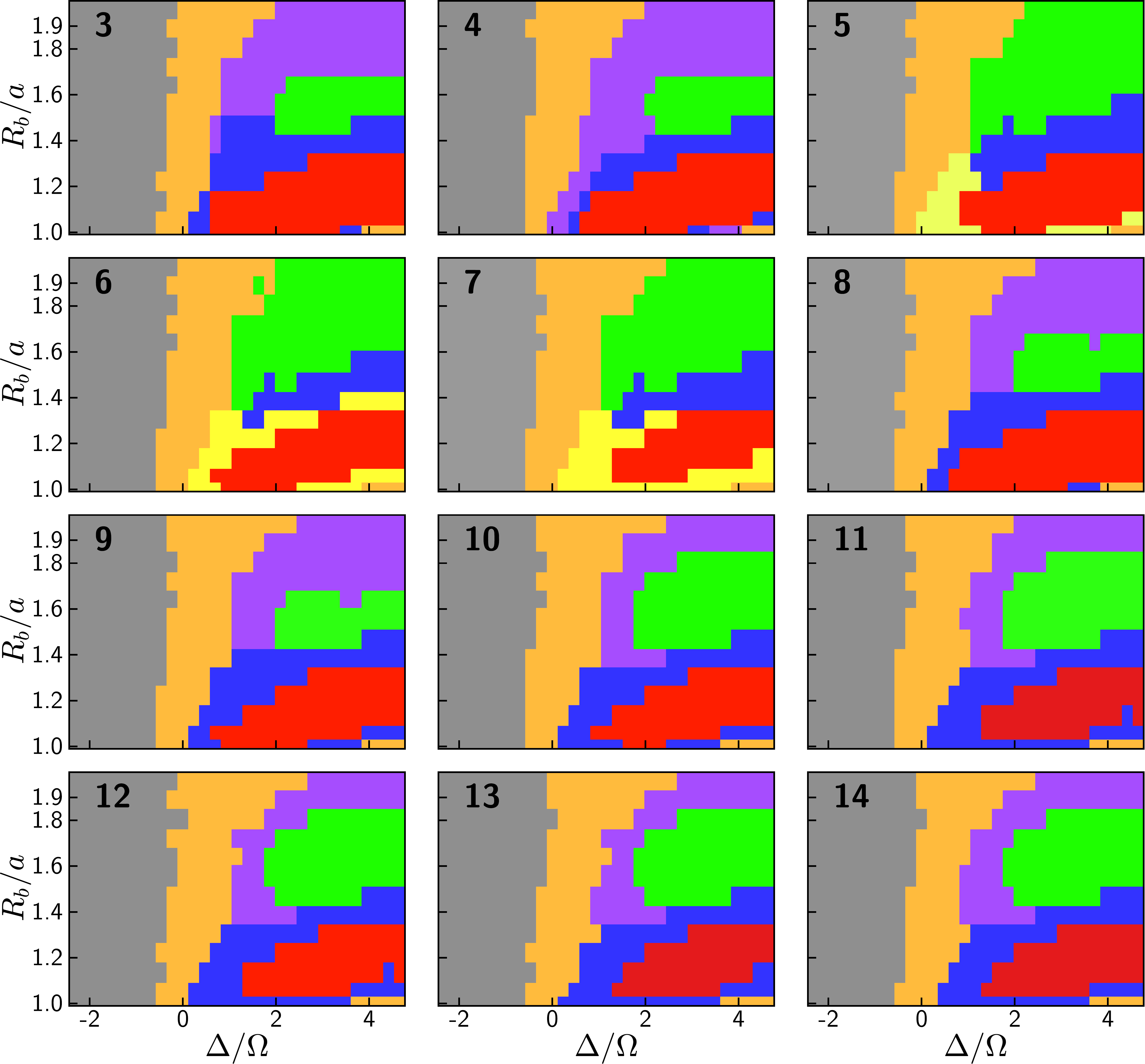}
    \caption{Clusters obtained when performing a Gaussian mixture model clustering with six clusters, and varying the number of PCA components retained (from 3--14). Colors of each cluster are manually chosen for visual continuity.}
    \label{fig:supp_unsupervised_numpca}
\end{figure*}

In Fig.~\ref{fig:supp_unsupervised_numpca}, we show clustering results with a fixed $N_{\mathrm{clust}} = 6$ but varying number of PCA components kept. Across all settings, we predominately see the topology presented in the main text (similar to $N_{\mathrm{PCA}} = 9$), with exceptions being $N_{\mathrm{PCA}} = 5,6,7$ which substitute the rhombic phase with a cluster around the broadened-checkerboard region characterized by PCA3. The relative stability and understandability of these clusterings give us confidence that they indicate real regions of varying orderings suitable for a follow-up analysis. As truncating the number of PCA components kept is primarily a cost-saving and robustness-improving technique, we increase the number of principal components kept until the clustering is seen to stabilize, which can be seen in Fig.~\ref{fig:supp_unsupervised_numpca} to be roughly 10.

\section{Supervised learning details} \label{sec:training_details}

\subsection{CCNN Training} \label{sec:ccnn_training}

We implement our Convolutional Correlator Neural Networks using the Pytorch \cite{paszke_pytorch_2019} library, and our code is made available at \url{github.com/KimGroup/QGasML}. Each CCNN is trained to identify a single phase: all snapshots sampled within that phase are labeled $0$, while snapshots sampled from all other phases are labeled $1$. The list of parameter-space points sampled to form each phase's training set is given in Table \ref{tab:supp_training_pts}. From each point, we randomly take 90\% of the snapshots as the training dataset shown to the network, and take the remaining 10\% as a validation dataset which is not shown during training, and is only used to verify that the network has not overfit. To handle the uneven distribution of snapshots available from each phase, when training a CCNN to identify a phase $\mathcal{P}$ each presented snapshot has a $50\%$ probability to be sampled from phase $\mathcal{P}$, and a $10\%$ probability to be sampled from each of the $5$ other phases. This ensures that there is an equal representation of ``within-phase-$\mathcal{P}$'' snapshots and ``out-of-phase-$\mathcal{P}$'' snapshots, as well as equal representation among all classes within the ``out-of-phase'' distribution.

The training points in Table \ref{tab:supp_training_pts} are obtained by starting with points suggested by the unsupervised clustering, and modifying iteratively based on the results of training until all phases overlap minimally and have clear distinctions. For the striated phase, whose support was very narrow in the unsupervised learning results, we explored a wider range in phase space to identify the training points yielding physically meaningful features. 

\begin{table}[]
    \centering
    \begin{tabular}{|c|c|c|}
    \hline
    \textbf{Phase} & $\Delta/\Omega$ & $R_b/a$ \\
    \hline
    Checkerboard & $3.02, 3.26$ & $1.13, 1.23$\\
    \hline
    Striated     & $2.33, 2.56, 2.79, 3.02$ & $1.46$ \\
    \hline
    Star         & $3.95, 4.19, 4.42, 4.65$ & $1.71$ \\
    \hline
    Rhombic      & $2.32, 2.56, 2.79, 3.02$ & $1.97$ \\
    \hline
    Edge         & $0.69, 0.93$ & $1.46, 1.56$ \\
    \hline
    Disordered   & $-2.09, -1.62, -1.16, -0.4$ & $1.13, 1.46, 1.81$\\
    \hline
    \end{tabular}
    \caption{Parameter-space points used to train models identifying each phase. Training sets are formed by
             all points in the Cartesian product of the $\Delta/\Omega$ and $R_b/a$ columns.}
    \label{tab:supp_training_pts}
\end{table}

During training, the free parameters of our CCNNs, $\{f_\alpha(\veca)$, $\beta_\alpha^{(m)}, \epsilon, w(\vecx)\}$, are all simultaneously learned so as to minimize the training loss averaged across the dataset, chosen to be the standard cross entropy loss used for classification tasks with an additional L1 regularization on the filter weights:
\begin{equation}
    \mathcal{L} = \frac{1}{N}\sum_{i}\Big(-y_i\log\hat{y_i} - (1-y_i)\log(1-\hat{y_i})\Big) + \gamma \sum_{\alpha, \veca} |f_\alpha(\veca)| . \label{eq:supp_loss}
\end{equation}
where $i$ runs over all snapshots in the training dataset, and $N$ is the total number of snapshots. We optimize this loss using the ADAM \cite{kingma_adam_2017} optimization algorithm, with a minibatch size of $128$ snapshots, an initial learning rate of $0.01$, and a cosine-annealed learning rate schedule as implemented by Pytorch's \texttt{CosineAnnealingLR} \cite{paszke_pytorch_2019}.

As in Ref.~\onlinecite{Miles2021NatCommun}, we also place a BatchNorm \cite{ioffe_batch_2015} layer (without the optional affine transformation) after the correlator features $c_\alpha^{(m)}$, which introduces no extra free parameters but aids in rapid and stable convergence of the training. Similar to Ref.~\onlinecite{Miles2021NatCommun}, we additionally spatially symmetrize $C_\alpha^{(m)}(\vecx)$ and $w(\vecx)$ to improve generalization and interpretability. Specifically, on each forward pass, we symmetrize the maps by summing over all symmetry transformations as
\begin{equation}
    C_\alpha^{(m)}(\vecx) \leftarrow \sum_{g \in D_4} g\, C_\alpha^{(m)}(\vecx),
\end{equation}
with the sum running over all group elements $g\in D_4$ acting on the convolutional maps, and the same transformation additionally applied to $w(\vecx)$. This symmetrization procedure aids in generalization, as the model's predictions are made invariant under symmetry transformations of the input snapshots, and reduces the effective total number of parameters to learn. In particular, this allows the model to avoid having to learn symmetry-equivalent versions of the filters $f_\alpha(\veca)$, and makes the spatial weighting $w(\vecx)$ easier to visually interpret. To make interpretation simpler, we also restrict $f_\alpha(\veca)$ to take only positive values by applying an absolute value function on every forward pass.

\begin{table*}[t!]
    \centering
    \begin{tabular}{|r|c|c|c|c|c|}
    \hline
                                      & Checkerboard & Star & Striated & Rhombic & Edge \\
    \hline
    2nd Order, $w(\vecx) = 1$, $4\times4$ Filters, $\gamma=0.0$   & 99.81(1) & 84.9(3) & 75.7(3) & 89.3(3) & 78.0(3) \\
    \hline
    3rd Order, $w(\vecx) = 1$, $4\times 4$ Filters, $\gamma=0.1$    & 98.5(2)   & 73(1) & 86.1(7) & 87(1) & 71.4(9) \\
    \hline
    3rd Order, $\beta_\alpha^{(m)} \ge 0$, $w(\vecx)=1$, $4\times 4$ Filters, $\gamma=0.1$     & 98.7(2) & 78(1) & 83.3(5) & \textit{86.3(4)} & 71.6(8)   \\
    \hline
    2nd Order, Learned $w(\vecx)$, $3\times 3$ Filters, $\gamma=0.1$    & 98.4(3) & 83.6(9) & 87.3(5) & 88.5(4) & \textit{82.9(3)}   \\
    \hline
    3rd Order, Learned $w(\vecx)$, $3 \times 3$ Filters, $\gamma=0.1$ & 100. & 85.8(5)  & 91.9(2)  & 91.7(4)  & 88.0(2)  \\
    \hline
    \end{tabular}
    \caption{Final validation accuracies for models identifying snapshots of each phase using various architectures. Measurements are made with 10-fold cross-validation \cite{Bishop2006}, with 5 random seeds run at each train/validation split. Reported errors for each model type are the standard error across all runs. All models are trained with $3$ convolutional filters. Figure~2(f) of the main text is produced by the final, most expressive, row of models. Table entries in italics are simpler models studied in Figs.~\ref{fig:edge_ccnn},\ref{fig:rhombic_ccnn} of the main text.
    Models sampled from the ``2nd Order, $w(\vecx) = 1$'' row produce the order parameters shown in Fig.~\ref{fig:supp_second_order_maps} of Appendix~\ref{sec:second_fourier}.
    }
    \label{tab:supp_perf}
\end{table*}

To ensure that all pixels of the learned filters hit all pixels of the snapshot in the convolution, we zero-pad each $\delta n_i(\vecx)$ with a sufficient number of zeros. In particular, if the convolutional filters $f_\alpha(\veca)$ are of spatial extent $F \times F$, we pad the input snapshots with $F-1$ zeros on all edges.

After training is completed, a CCNN has learned a collection of convolutional filters $f_\alpha(\veca)$, as well as a set of coefficients $\beta_\alpha^{(m)}$ connecting the $m$\textsuperscript{th} order feature $c_\alpha^{(m)}$ derived from filter $f_\alpha$ to the output, and an overall bias $\epsilon$. The output of the model is then
\begin{equation}
    \hat{y} = \left[1 + \exp\left(-\sum_{\alpha, m} \beta_\alpha^{(m)} c_\alpha^{(m)} + \epsilon\right)\right]^{-1},
\end{equation}
where $c_\alpha^{(m)}$ are constructed from the learned filters $f_\alpha$ and the input by Eqs.~\eqref{eq:ccnn-C2} and \eqref{eq:ccnn-C3} of the main text.

\subsection{Ablation testing} \label{sec:ccnn_ablation}

When building a specific CCNN, there is a lot of flexibility in the architectural choices. These architectural hyperparameters include the order $m$ to truncate at, whether to include a learnable spatial weighting $w(\vecx)$, and how many filters (and of what size) to use. For the current work, we take the approach of building the simplest architecture (second-order, uniform $w(\vecx)$) first, and adding on architectural complexity piece-by-piece. If a large gain in accuracy is achieved by a single architectural addition, then we keep that piece and attribute a quality of the phase to requiring the additional expressibility. This approach is commonly referred to as \textit{ablation testing} in the ML literature \cite{Meyes2019ArXiv}, though for large neural networks, ML practitioners commonly \textit{remove} modules piece-by-piece rather than adding them as we do for our shallow networks.

To improve interpretability, an additional hyperparameter that we have at our disposal is the coefficient $\gamma$ on the filter L1 loss in Eq.~\ref{eq:supp_loss}. Intuitively, larger $\gamma$ results in simpler filters with more pixels deactivated but decreased classification performance. For all models except for the uniform second-order models (which we can easily interpret in Fourier space, see Appendix~\ref{sec:second_fourier}), we increase $\gamma$ until the filters identifying all phases are sparse enough to easily interpret while performance is maintained sufficiently high.

Along these lines, in Table~\ref{tab:supp_perf}, we summarize validation accuracy measurements of several variations of our CCNNs trained to identify each phase.
From these measurements, in conjunction with observing the quality of the resulting phase diagram, we can determine what is required to form a good order parameter for each of the identified phases. For example, the checkerboard, star, and rhombic phases show roughly uniform or even decreasing (due to overfitting) validation performance as the CCNN's truncation order is increased, indicating that second-order features are sufficient to distinguish these from other phases. However, we find the second-order features identifying the rhombic phase to be somewhat uninteresting as they primarily just measure the tendency for longer-range density correlations; see Appendix \ref{sec:second_fourier}. The star phase shows some improvement when incorporating spatial inhomogeneity, but we find that the phase diagram changes little while the second order model is simplest to interpret (see Appendix \ref{sec:second_fourier}).

The most striking changes are observed for the two remaining phases, both of which benefit heavily from learning a nonuniform spatial weighting $w(\vecx)$. For the striated phase, we observe a dramatic $11.6\%$ jump in classification accuracy between uniform and nonuniform second-order models. This difference reflects that due to finite-size effects, many striated-like correlations persist throughout the system within the star phase. Appendix~\ref{sec:second_fourier} presents an alternate interpretation in Fourier space, where the difficulty originates from the relevant Fourier peaks being overly diffuse.

However, many of these striated-like correlations in the star phase occur on the even sublattice, while the striated ordering entirely occupies the odd sublattice. If the network has the ability to focus on a specific sublattice matching the ideal striated ordering (making all other sites contribute with a negative weight to the output), it can better distinguish the striated from the star phase. Alternatively, we find that increasing the order of the model while keeping $w(\vecx) = 1$ also results in a good classifier for the striated phase but one which is more difficult to understand.

In contrast, for the edge-ordered phase, we find that simply increasing the order of the model while keeping uniform $w(\vecx)=1$ makes negative changes to generalization performance. Meanwhile, keeping the model at second order, we see a $4.9\%$ jump in accuracy when allowing for a spatially varying $w(\vecx)$. While this may not seem dramatic, we find that all models with uniform $w(\vecx)$ produce order parameters which persist deep into the disordered region. This indicates that successfully identifying this phase \textit{requires} measuring spatially-inhomogeneous correlation functions, which reflects that this phase is itself \textit{defined} by this inhomogeneity.

\section{Extracting Fourier-space order parameters from uniform second-order CCNNs} \label{sec:second_fourier}

Second-order CCNN features (under a uniform spatial weighting $w(\vecx) = 1$) have an alternate interpretation as performing weighted sums of the Fourier spectrum of the input. Specifically, each feature $c_\alpha^{(2)}$ can be written as
\begin{widetext}
\begin{align}
        c_\alpha^{(2)} = \sum_{\vecx} C_\alpha^{(2)}(\vecx) &\equiv \sum_{\vecx} \left[\left(\sum_{\veca} f(\veca) \delta n(\vecx+\veca)\right)^2 - \sum_{\veca} f(\veca)^2 \delta n(\vecx+\veca)^2\right] \label{eq:supp_c2}\\
                          &= \frac{1}{N^2}\left[
                          \sum_{\veck} \left(\left|\hat{f}(\veck)\right|^2 - \frac{1}{N^2} \sum_{\veck'}\left|\hat{f}(\veck')^2\right|\right)\left|\delta \hat{n}(\veck)\right|^2
                          \right], \label{eq:supp_c2_fourier}
\end{align}
\end{widetext}
where Eq.~\eqref{eq:supp_c2} is just the expanded definition of \(C^{(2)}(\vecx)\), Eq.~\eqref{eq:supp_c2_fourier} is the Fourier-equivalent form, and discrete Fourier transforms are defined by
\begin{equation}
    \hat{f}(\veck) \equiv \sum_{\veca=\boldsymbol{0}}^{(L_f, L_f)} e^{-i\veck\cdot\veca} f(\veca), \hspace{0.25cm} \delta \hat{n}(\veck) \equiv \sum_{\vecx=\boldsymbol{0}}^{(L_n, L_n)} e^{-i\veck\cdot\vecx} \delta n(\vecx) \label{eq_ft},
\end{equation}
where $L_f, L_n$ are the lengths of each dimension of the convolutional filter $f$ and the input snapshot \(\delta n\), respectively. The wavevectors in the discrete Fourier transform are defined as $k_i \in \{0, 2\pi/N,4\pi/N, \dots, (N-1)2\pi/N\}$ with $N\ge L$ defining the resolution of the Fourier transform. Choosing $N > L$ encodes no new information in the result, but produces higher-resolution Fourier spectra for plotting.

Note that for all terms in Eq.~\eqref{eq:supp_c2} to be well-defined, we take an ``infinite zero-padding'' convention, where the domain of $\delta n$ is expanded to all $\vecx$ by defining \(\delta n(\vecx) = 0\) $\forall$ \(\vecx \notin [0,L_n]\times[0,L_n]\). Under this convention, we take the $\vecx$ sum to be over all space. In practice, our CCNNs equivalently pad $\delta n$ with enough zeros to recover all nonzero $C^{(2)}_\alpha(\vecx)$. Equation~\eqref{eq:supp_c2_fourier} can be obtained from Eq.~\eqref{eq:supp_c2} by straightforward application of the convolution theorem and Plancherel's theorem.

A simple interpretation of this result is that a uniform-weighted $c_\alpha^{(2)}$ measures a weighted sum of $\delta n(\veck)$, with the weights given by the Fourier transform of $f_\alpha$, normalized to be zero-mean in $\veck$-space. Since second-order CCNNs linearly combine multiple filters $f_\alpha$ with coefficients $\beta_\alpha$ to produce the input to the final logistic function, we can understand the full action of the network by the weighted sum of the input in $\veck$-space with effective weights $\tilde{f}(\veck) = \sum_\alpha \beta_\alpha^{(2)} \tilde{f}_\alpha(\veck)$, with normalized Fourier intensities of each filter defined as $\tilde{f}_\alpha(\veck) = |\hat{f}_\alpha(\veck)|^2 - \frac{1}{N^2}\sum_{\veck'} |\hat{f}_\alpha(\veck')|^2\textbf{}$. Due to the symmetrization procedure outlined in Appendix \ref{sec:training_details}, this map must be then symmetrized over all symmetries of $D_4$. This ``order parameter map'' defines our effective second-order CCNN order parameter for a given phase as
\begin{equation}
    O = \sigma \left(\sum_{\veck} \tilde{f}(\veck)^{\mathrm{sym}} |\delta \hat{n}(\veck)|^2 - \epsilon\right),
\end{equation}
with $\sigma(x) = (1 + \exp(-x))^{-1}$ the logistic sigmoid function, and $\epsilon$ the learned bias.

\begin{figure*}[t!]
    \centering
    \includegraphics[width=1.95\columnwidth]{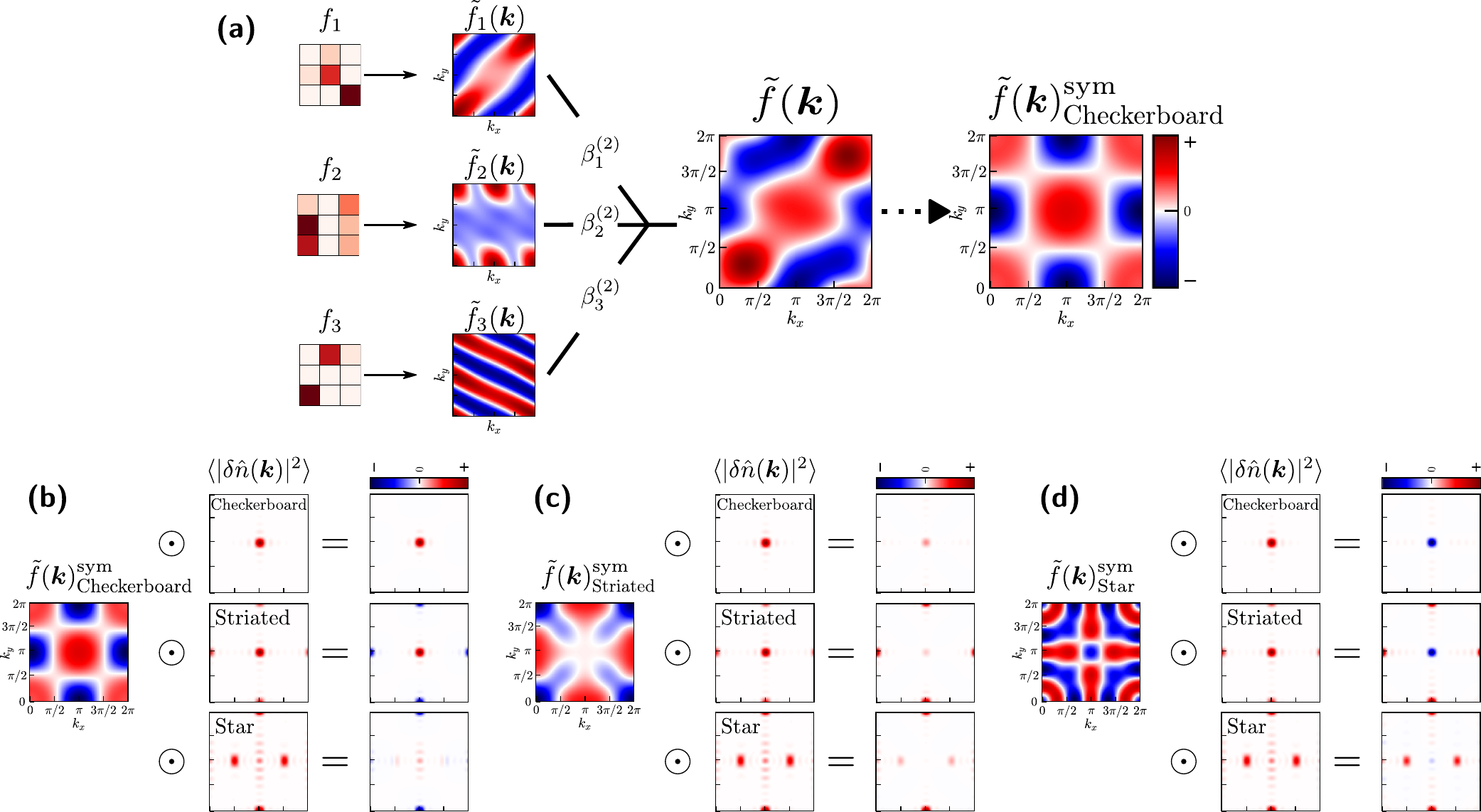}
    \caption{(a) The process by which the action of a uniform-$w(\vecx)$ second-order CCNN can be interpreted in Fourier space. Each filter is discrete Fourier transformed and normalized to form $\tilde{f}_\alpha(\veck)$, linearly combined pointwise in $\veck$-space with learned coefficients $\beta_\alpha^{(2)}$, then symmetrized to form the final map. (b--d) The order parameter maps $\tilde{f}(\veck)^{\mathrm{sym}}$ learned to identify the checkboard, striated, and star phases. We apply each order parameter map as weightings to the idealized Fourier intensities for each of the checkerboard, striated, and star phases. If learning is successful, applying this weighting and then summing in $\veck$-space should produce large positive numbers for the target phase, and small or negative numbers for every other phase.}
    \label{fig:second_order_fourier}
\end{figure*}

In Fig.~\ref{fig:second_order_fourier}(a), we provide a visual diagram demonstrating this process to produce an interpretable $\tilde{f}(\veck)^{\mathrm{sym}}$. In Fig.~\ref{fig:second_order_fourier}(b--d), we show exemplary resulting Fourier weighting maps learned for each of the checkerboard, striated, and star phases, and demonstrate the intuition to interpret these maps. Applying the weighting $\tilde{f}(\veck)^{\mathrm{sym}}$ to the average Fourier intensity of the target phase should produce a large positive number, while applying it to the intensities of all other phases should produce smaller or negative numbers (after subtracting the learned $\epsilon$, only the target phase should remain positive).

We find that the checkerboard, striated, and star Fourier weightings look strikingly similar to smeared-out versions of the hand-crafted order parameters discussed in Refs.~\onlinecite{Ebadi2021Nature,samajdar_complex_2020}, which can be interpreted in our framework as $\tilde{f}(\veck)$ that are nonzero only at finitely many $\veck$-points.
In particular, our learned checkerboard order parameter is strongly positive at $(\pi, \pi)$, the striated at $(\pm \pi, 0), (0, \pm \pi)$, and the star order parameter spreads positive weight towards $(\pi/2, 0), (0, \pi/2)$ while putting negative weight at $(\pi, \pi)$. This observation is reassuring as it demonstrates that these models identify each phase by an ordering similar to the ideal density waves.

\begin{figure*}[]
    \centering
    \includegraphics[width=1.3\columnwidth]{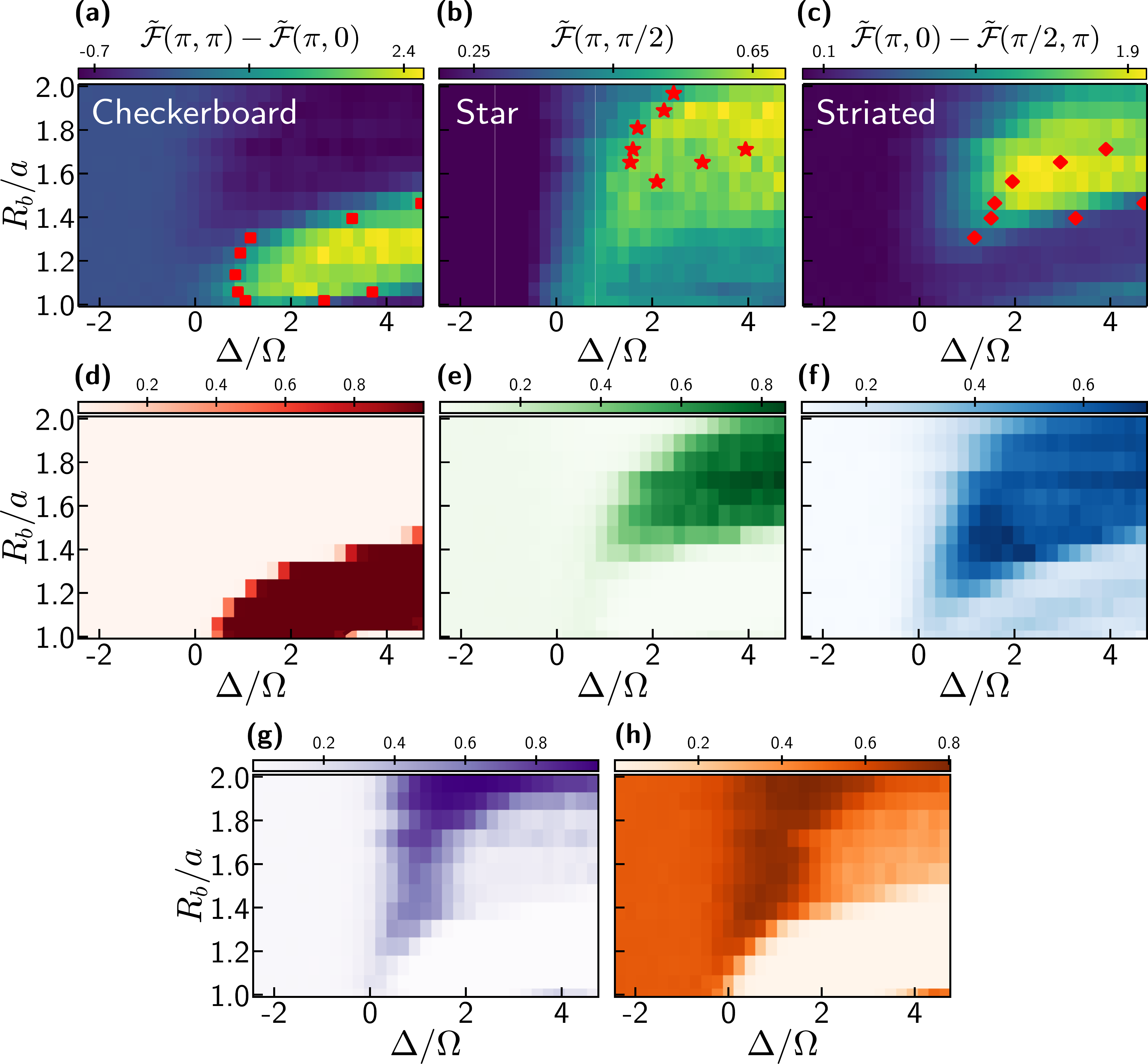}
    \caption{(a--c) Measurements of Fourier-space order parameters manually crafted \cite{Ebadi2021Nature} to capture the checkerboard, star, and striated phases. Red symbols mark phase boundaries from DMRG simulations of a $9\times 9$ open boundary system \cite{Ebadi2021Nature}. (d--h) CCNN order parameter maps of uniform second-order CCNNs trained to identify each of the phases examined in the main text. At each point in $(\Delta, R_b)$ space, $c_\alpha^{(2)}$ is averaged across all available snapshots (mimicking ``measuring the correlation functions'' from the snapshots) and then fed into the final logistic layer. Only the checkerboard (red) and star (green) phases are well resolved by this minimal model.}
    \label{fig:supp_second_order_maps}
\end{figure*}

In Fig.~\ref{fig:supp_second_order_maps}(a--c), we show measurements of Fourier-space order parameters manually crafted to capture each of the checkerboard, star, and striated phases. We can see that due to the blurring of the relevant Fourier peaks resulting from the finite-size system, the order parameters for the star and striated phases [Fig.~\ref{fig:supp_second_order_maps}(b,c)] heavily overlap. For comparison, in Fig.~\ref{fig:supp_second_order_maps}(d--h), we show confidence maps from uniform second-order CCNNs with $4\times 4$ convolutional filters trained to recognize each phase examined in the main text. We can observe that only the checkerboard (red) and star (green) phases are well-resolved by these reduced models, in accordance with the intuition that these phases are classical crystals easily identified in Fourier space. Meanwhile, just as with manually crafted Fourier-space order parameters, the uniform second-order CCNN's predictions for the striated phase significantly overlap with those of the star phase. These combined observations point to a fundamental limit to phase resolution using Fourier-space order parameters in finite, open-boundary, noisy systems. The rhombic (purple) phase is somewhat well-resolved, but it too suffers from overly broad peaks in Fourier space. Meanwhile, we see that uniform second-order models fail dramatically for the edge-ordered (orange) phase, which by its nature fundamentally requires the ability to learn spatially inhomogeneous functions, as measured in Appendix~\ref{sec:training_details}.

\begin{figure*}[]
    \centering
    \includegraphics[width=1.8\columnwidth]{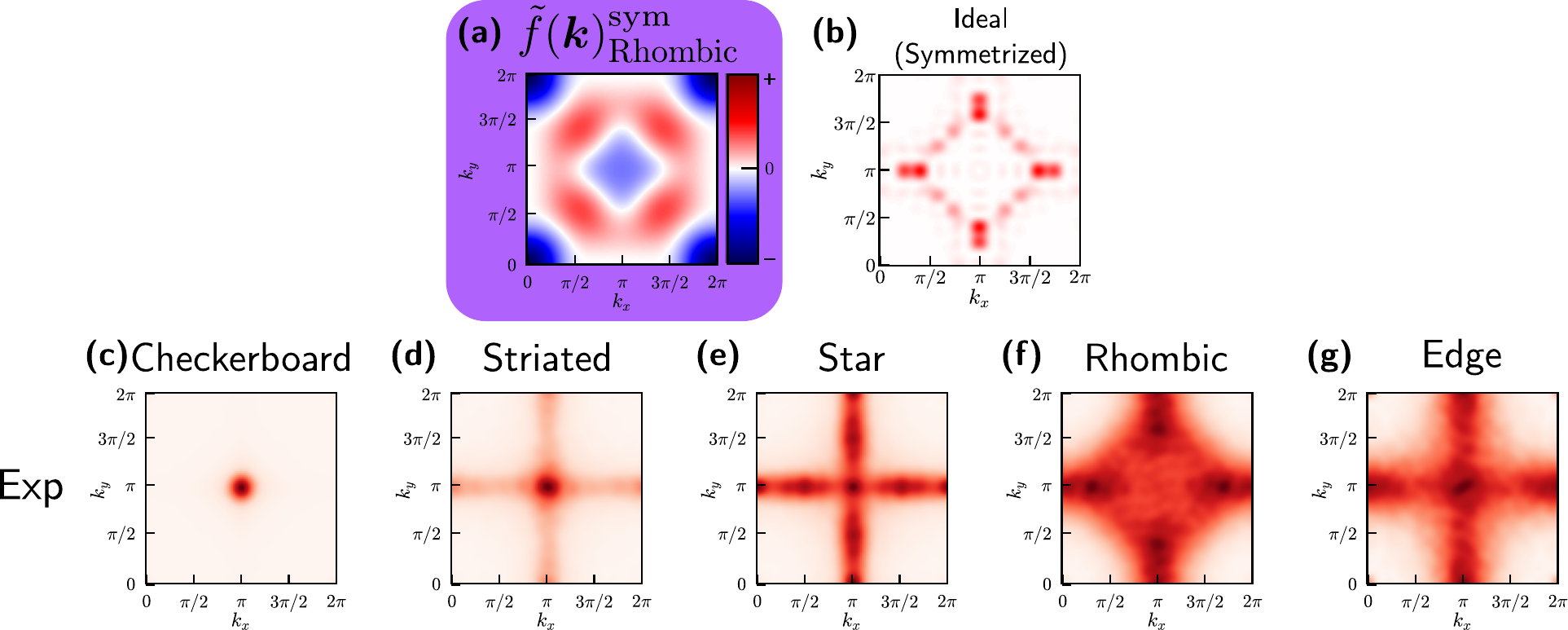}
    \caption{(a) Fourier intensity weighting map derived from a second-order model trained to identify the rhombic phase. (b) Symmetrized Fourier intensities resulting from the density-normalized ideal long-range rhombic ordering. (c--g) Fourier intensities of per-site density-normalized experimental data $\delta \hat{n}(\veck)$ sampled deep in each of the identified phases.}
    \label{fig:supp_second_rhombic}
\end{figure*}

To uncover what is being learned by the rhombic second-order CCNN, we repeat the above analysis to produce the Fourier-space order parameter shown in Fig.~\ref{fig:supp_second_rhombic}(a). We see that the learned order parameter attempts to identify Fourier intensity along the diagonals connecting the $(\pm 2\pi/5, 0)$ and $(0, \pm\pi/2)$ peaks (and symmetry equivalents), as the intensity uniquely appears here only in this phase. Due to broadening resulting from experimental noise and the finite-size system, the CCNN does not attempt to directly measure the rhombic $(\pm 2\pi/5, 0)$ peaks as these blur too strongly into the star phase's $(\pm \pi/2, 0)$ peaks. Nevertheless, these peaks can be visually resolved when averaging over a large number of snapshots as in Fig.~\ref{fig:supp_second_rhombic}(c).

Manual inspection of the learned filters and $\beta_\alpha^{(2)}$ coefficients reveals that the rhombic CCNN obtains this order parameter by placing negative $\beta_\alpha^{(2)}$ on filters which contain short-range patterns. This makes intuitive sense, as density fluctuations between nearby sites are anticorrelated at high $R_b$. However, this does not give us clear insight into the actual Rydberg crystal being realized, other than that it is constructed from longer-range displacements. This motivated our choice in the main text of investigating third-order models which enforce $\beta_\alpha^{(m)} > 0$ to ensure that the rhombic CCNN must rely on positive correlations to identify the phase. Due to three-point correlators having a nontrivial sign structure (see Appendix~\ref{sec:third_order_signs}), the CCNN still measures some short-range patterns, but also learns several key longer-range patterns uniquely characterizing the phase.

\begin{figure*}[]
    \centering
    \includegraphics[width=1.95\columnwidth]{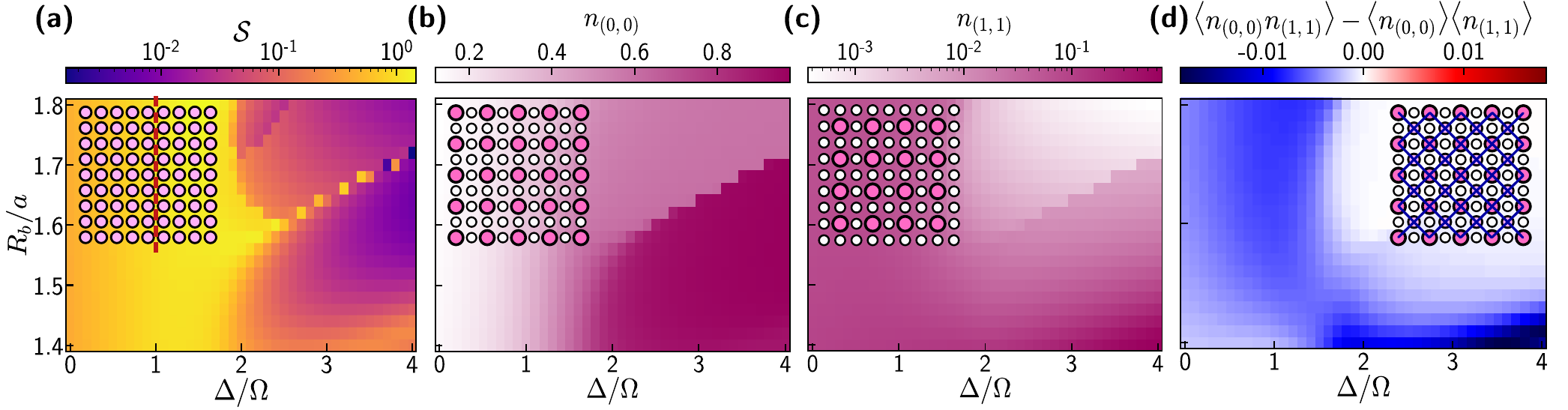}
    \caption{Density-matrix renormalization group results for a $9\times 9$ system. (a) Bipartite entanglement entropy $\mathcal{S}$ between two halves of the system. (b, c) Average density on $(0, 0)$ and $(1, 1)$ sublattices, respectively. (d) Average connected density-density correlator between next-nearest-neighboring sites on the so-called $(0, 0)$ and $(1,1)$ sublattices, marked in pink and purple in the inset, respectively.}
    \label{fig:supp_DMRG}
\end{figure*}

\section{Entanglement generated by Rydberg interactions} \label{sec:striated_entangle}

As often exploited by Rydberg-atom quantum simulators, the Rydberg blockade effect can generate entanglement between interacting atoms \cite{Saffman2010Rev.Mod.Phys., Weimer2010NaturePhys, Wilk2010Phys.Rev.Lett.}. Here, we perform simple calculations to examine the interplay between entanglement and correlations in the vicinity of the transition to the striated phase and draw connections to the correlations in the experimental data uncovered by our CCNN analysis. To this end, we perform density-matrix renormalization group (DMRG) calculations using a ``snakelike'' matrix product state ansatz for a $9\times 9$ system with open boundaries. The Rydberg Hamiltonian realized by the experiment examined in this work is given by Eq.~\eqref{ryd_ham}, and for notational simplicity, we will refer to the first, second, and third terms therein as $\hat{\Omega}, \hat{\Delta},$ and $\hat{V}$, respectively. Entanglement can be generated in the ground state of $\hat{H} (\equiv H/\hbar)$ due to energetic competition between these terms. For any two sites, the interaction term $\hat{V}$ prefers small overlap with basis states containing $\ket{rr}$, while $\hat{\Delta}$ desires large overlap with all of $\ket{gr}, \ket{rg}$, and $\ket{rr}$. Crucially, $\hat{\Omega}$ favors weight to be present with opposite phase between basis states with a single site flipped as $\ket{g}\leftrightarrow\ket{r}$. As a result, for a two-site system, the ground state as $R_b/x_{12} \rightarrow \infty$ places weight across all of the $\ket{gg}, \ket{rg}, \ket{gr}$ basis states, but not $\ket{rr}$, resulting in an entangled, anticorrelated state.

To examine this behavior more closely, we now turn to the results of the DMRG computations. In Fig.~\ref{fig:supp_DMRG}, we show the bipartite entanglement entropy between two halves of the system. Within the disordered phase ($\Delta \lesssim 1$), the entanglement entropy of the ground state increases monotonically as one approaches the quantum critical points, and, at large $R_b$, the density on any site is anticorrelated with that of its next-nearest neighbors in the corners [Fig.~\ref{fig:supp_DMRG}(d)]. As we transition deep into the classically ordered phases, both the entanglement and the connected correlations vanish due to the density on each site approaching either $0$ or $1$ [Fig.~\ref{fig:supp_DMRG}(b,c)]. However, in a narrow region at $R_b/a \approx 1.4$ where quantum fluctuations stabilize a significant density on the $(1,1)$ sublattice [see Fig.~\ref{fig:supp_DMRG}(d)], both entanglement and diagonal anticorrelations survive. We emphasize that this entanglement is dependent on the state on each sublattice remaining in a quantum superposition of $\ket{g}$ and $\ket{r}$, as is uniquely characteristic of the striated phase.

\begin{figure*}
    \centering
    \includegraphics[width=1.8\columnwidth]{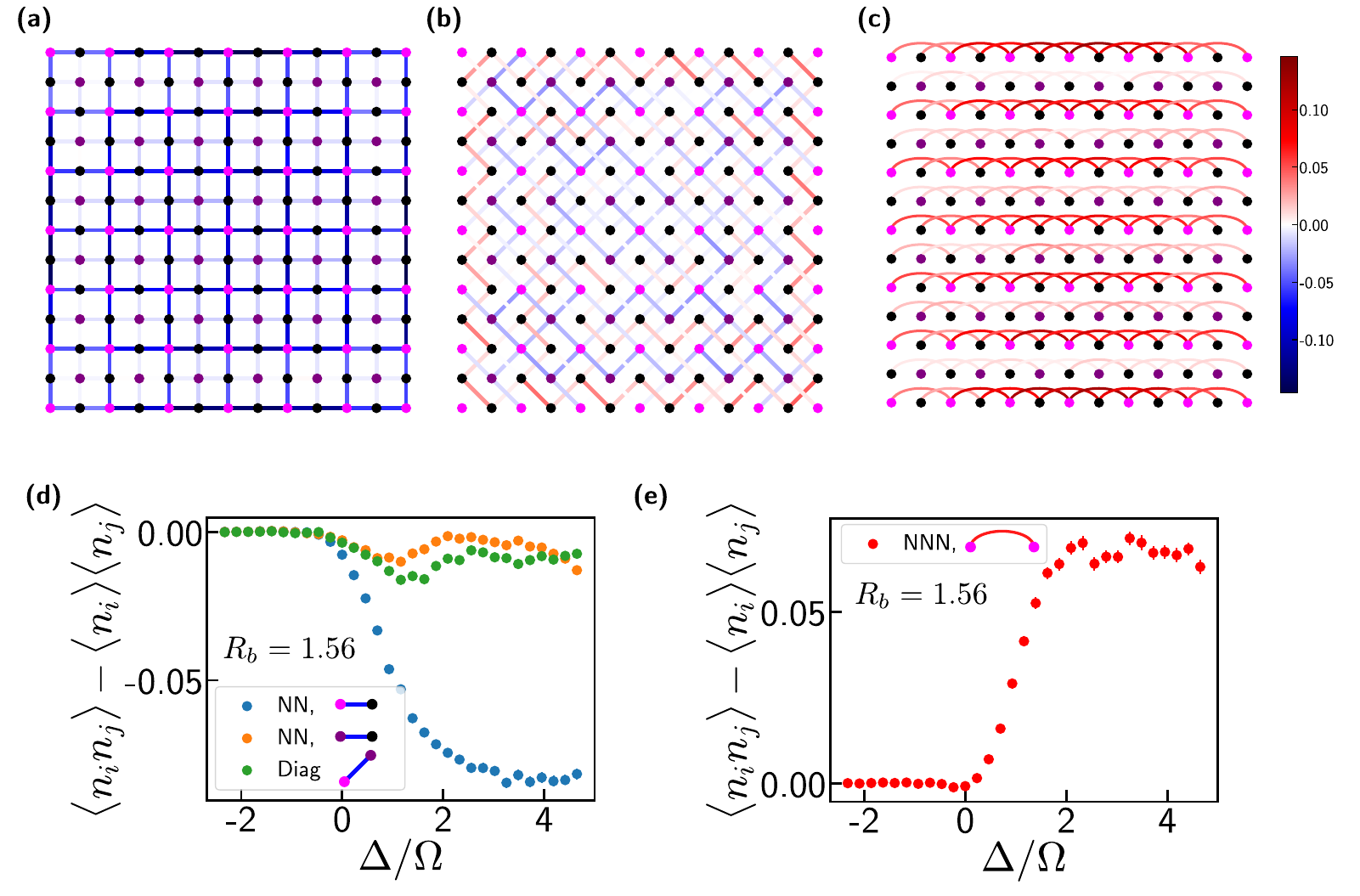}
    \caption{(a, b, c) Measurements of connected correlations from the striated experimental training dataset (see Table \ref{tab:supp_training_pts}), performed on each individual nearest-neighbor (NN), next-nearest-neighbor (NNN), or next-to-next-nearest-neighbor (NNNN) bond, respectively. The colors of each site are for visual aid, showing the sites expected to be in the mostly excited (pink), mostly ground (purple), and entirely ground (black) states in the ideal striated limit. (d, e) Averaged correlations across all bonds of different symmetry classes, tracked as a function of $\Delta/\Omega$ for a cut at $R_b = 1.56$.}
    \label{fig:striated_entangle}
\end{figure*}

Indeed, as shown in Fig.~\ref{fig:striated_entangle}, in the experiment, the connected part of many short-range correlations remains finite and negative upon transitioning into the striated phase. In particular, all nearest-neighbor correlators remain anticorrelated, along with next-nearest-neighbor correlators between the two excited sublattices. However, there are many confounding effects which make it difficult to pinpoint with certainty the origin of these correlations. First, due to decoherence and experimental noise, the experiment, in principle, produces a mixed rather than a pure state. Given a mixed state $\rho$, nonzero connected correlations can emerge either by entanglement, or by ``classical'' correlations between the pure states comprising $\rho$ \cite{Vedral1997Phys.Rev.Lett.}. Efficient means for unambiguously revealing entanglement in experimental settings without full tomography is an active field of research \cite{Elben2020Phys.Rev.Lett., Huang2020Nat.Phys., Ketterer2020Quantum, Knips2020npjQuantumInf, Neven2021ArXiv, PhysRevLett.125.200502, Yu2021Phys.Rev.Lett.}.

Moreover, quasiadiabatic sweeps across phase boundaries produce final states which are not perfect ground states but are dependent on the original state starting from which the phase boundary was crossed. As the system is theoretically transitioning from a region of high entanglement, where nearby neighbors have anticorrelated densities, it is likely that some of the magnitude of correlations captured by the CCNN is not directly representative of the true ground state, instead having been ``frozen in'' from before the transition \cite{Zurek2005PRL}. Nevertheless, the nature (and, in particular, the signs) of these correlations still reveals qualitative structures of each identified phase.

\section{Sign structures of third-order correlators} \label{sec:third_order_signs}

\begin{figure*}[t]
    \centering
    \includegraphics[width=1.9\columnwidth]{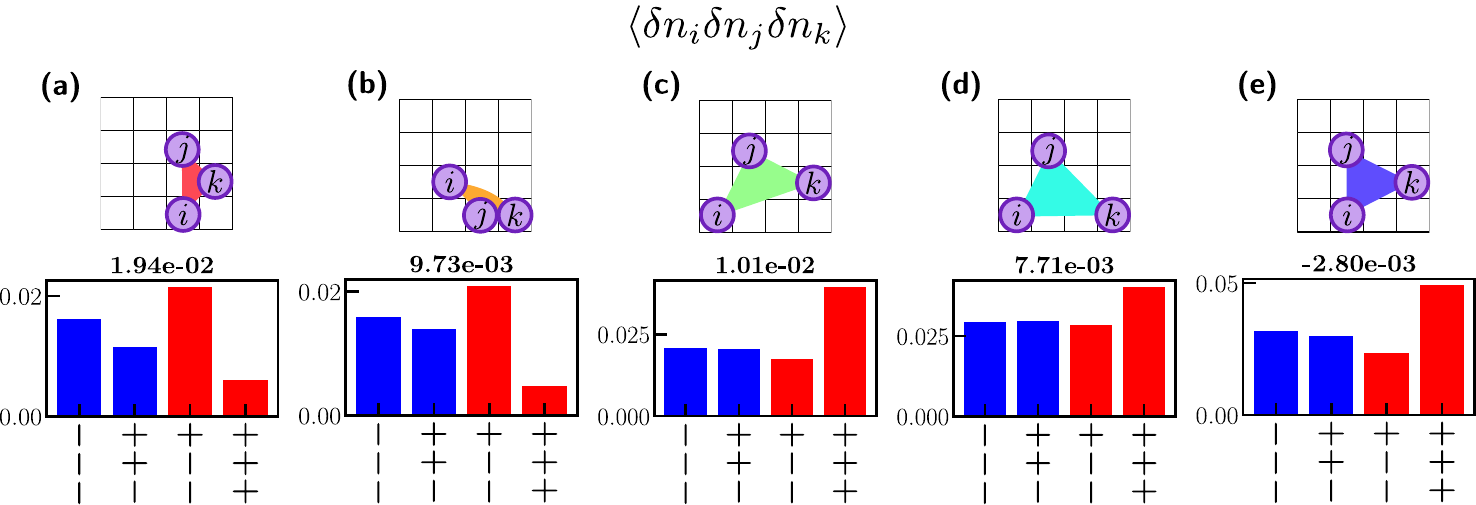}
    \caption{(a--e) The three-point connected correlators identified from the third-order CCNN trained to identify the rhombic phase (a--d) in the main text, along with another notable correlations characterizing the star and rhombic phases (e). For each, we show the contributions to the correlator from three-site density fluctuation patterns of different sign configurations, measured from the snapshots in the rhombic training dataset (see Table \ref{tab:supp_training_pts}). Contributions are summed across all 8 reflection/rotation-transformed correlators, and averaged across all translations which leave all three sites within the snapshot. Red bars show magnitudes of positive contributions, while blue bars correspond to negative contributions. For simplicity, the bar graph shows the $({+}{+}{-})$ and $({+}{-}{-})$ contributions averaged over all three possible sign combinations. Above each bar graph, we show the total value of the correlator, obtained by summing all contributions with the appropriate sign and multiplying the averaged mixed-sign contributions by $3$.}
    \label{fig:three_point_signs}
\end{figure*}

By inspecting the learned $\beta_\alpha^{(3)}$ and the patterns in the learned associated filters $f_\alpha$, we can determine which three-point correlations are being measured for a target phase, and whether they should be positive or negative within the phase. However, the sign on a three-point contribution can be somewhat confusing to interpret, as there are multiple ways to obtain positive/negative three-point correlations. The CCNN itself does not inherently point out \textit{how} to interpret these correlators---simply that they are strongly positive/negative within the phase. Manual follow-up and investigation is always necessary to understand the underlying physics. This section attempts to clarify the subtleties of these measured three-point correlators, and presents explicit measurements thereof from the data to confirm that the CCNN's identification was meaningful.

Within this section, for notational brevity, spatial indices are written as (vector) subscripts. Given density-normalized snapshots $\delta n_{\veci} \equiv n_{\veci} - \langle n_{\veci} \rangle$, uniform third-order CCNN features measure weighted sums of connected three-point correlation functions, averaged over all spatial translations:
\begin{equation}
    \langle c_\alpha^{(3)} \rangle = \sum_{\vecx} \sum_{\veci\vecj\veck = (0, 0)}^{(L_f-1, L_f-1)} f^{}_{\veci} f^{}_{\vecj} f^{}_{\veck} \langle \delta n^{}_{\vecx+\veci} \delta n^{}_{\vecx+\vecj} \delta n^{}_{\vecx+\veck} \rangle,
\end{equation}
where the inner sum runs over all configurations of displacements $(\veci, \vecj, \veck)$ within the spatial extent of the filter $f$ of length $L_f$.
Due to the symmetrization process outlined in Appendix \ref{sec:ccnn_training}, this must also be averaged over all rotations and flips of the three-point pattern.

\begin{figure}
    \centering
    \includegraphics[width=\columnwidth]{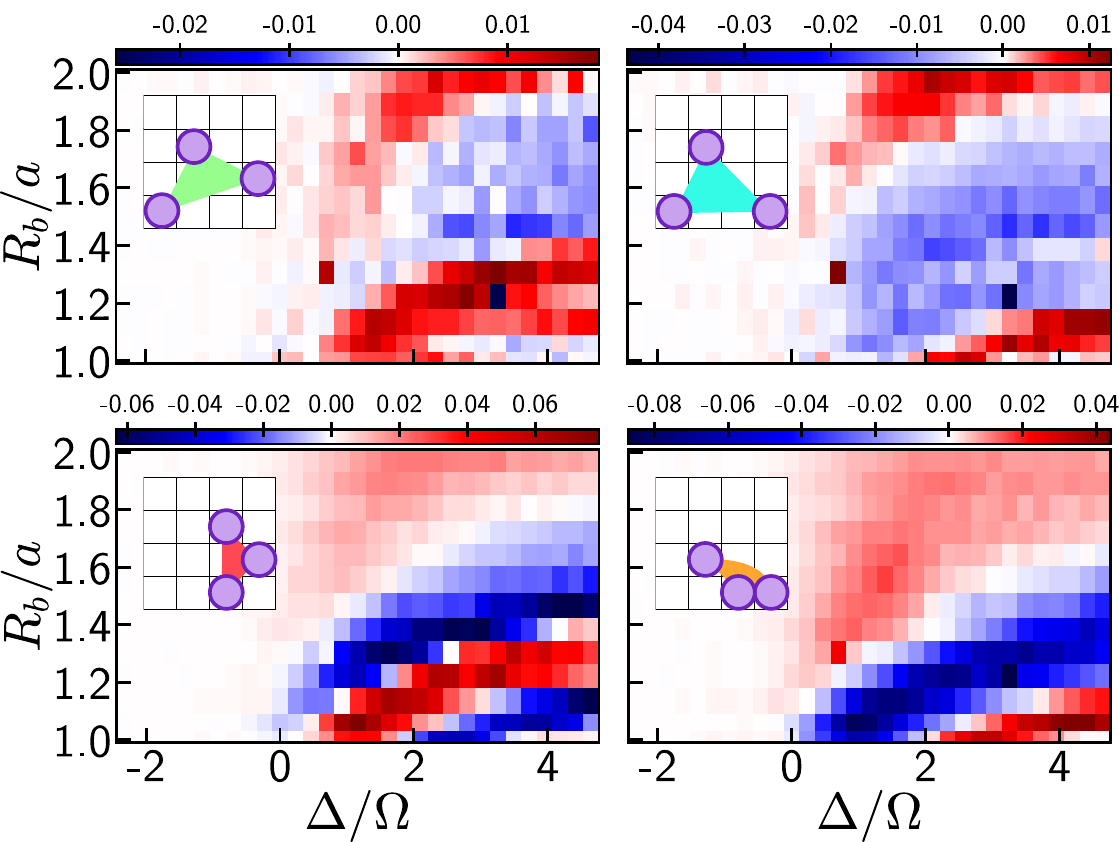}
    \caption{Connected three-point correlators that were discovered by our CCNNs to capture the rhombic phase, used in tandem with other two- and three-point correlators not shown. All correlators are summed across all 8 reflection/rotation-transformed versions and averaged over all translations which leave all three sites within the snapshot. At low $R_b$, these correlators become positive again due to different sign contributions [$({+}{-}{-})$ for the top two and $({+}{+}{+})$ for the bottom two].}
    \label{fig:three_point_maps}
\end{figure}

In terms of individual density configurations, each of these correlators acquire positive contributions either when all of $(\delta n_\veci, \delta n_\vecj, \delta n_\veck)$ are positive $({+}{+}{+})$, or when only one is $({+}{-}{-}), ({-}{+}{-}), ({-}{-}{+})$. The CCNN does not directly inform us as to how the correlator became positive---manual follow-up is necessary to uncover this information. For example, in Fig.~\ref{fig:three_point_signs}, we show the statistics of these different sign contributions for the dominant three-point correlators learned by the CCNN within the rhombic phase. For conciseness, the heights of the bars corresponding to mixed-sign contributions are averaged over all configurations which produce the same sign. In Fig.~\ref{fig:three_point_maps}, we show the value of a handful of these key correlators across the $(\Delta/\Omega, R_b/a)$ parameter space in the data.

From this, we can see that the short-range patterns learned by the CCNN are actually producing positive signals due to a large number of $({+}{-}{-})$ density configurations, as shown in Fig.~\ref{fig:three_point_signs}(a,b). This points to the longer-range packing of Rydberg excitations within this phase---if one the sites in the indicated triples is excited, it is more likely that the other two are in the ground state. Meanwhile, the patterns of Fig.~\ref{fig:three_point_signs}(c,d) are positive dominantly due to $({+}{+}{+})$ configurations, indicating that these motifs signal actual common configurations of joint Rydberg excitations. Figure~\ref{fig:three_point_signs} shows that the star-like configurations also have a large $({+}{+}{+})$ signal in this phase, as expected from the idealized pattern, but the other sign contributions cause this correlator to turn negative and not be picked up by the CCNN. Together, we can infer from these correlations that we are probing a rhombic-like phase which is failing to develop long-range order due to the incommensurate geometry of the system.

In Fig.~\ref{fig:three_point_maps}, we show the extent in parameter space of the key identified connected three-point correlators, and observe that the region colored by the CCNN as the rhombic phase does indeed correspond to the region where the long-range three-point motifs uniquely produce a positive signal. Further theoretical and numerical analysis is needed to confirm that this phase exists in the thermodynamic limit and that the rough region of parameter space identified by our CCNN corresponds to the true region hosting this phase, as well as to determine if these higher-order signals remain good indicators of the phase in the thermodynamic limit.

%% file: main.bbl
\begin{thebibliography}{59}%
\makeatletter
\providecommand \@ifxundefined [1]{%
 \@ifx{#1\undefined}
}%
\providecommand \@ifnum [1]{%
 \ifnum #1\expandafter \@firstoftwo
 \else \expandafter \@secondoftwo
 \fi
}%
\providecommand \@ifx [1]{%
 \ifx #1\expandafter \@firstoftwo
 \else \expandafter \@secondoftwo
 \fi
}%
\providecommand \natexlab [1]{#1}%
\providecommand \enquote  [1]{``#1''}%
\providecommand \bibnamefont  [1]{#1}%
\providecommand \bibfnamefont [1]{#1}%
\providecommand \citenamefont [1]{#1}%
\providecommand \href@noop [0]{\@secondoftwo}%
\providecommand \href [0]{\begingroup \@sanitize@url \@href}%
\providecommand \@href[1]{\@@startlink{#1}\@@href}%
\providecommand \@@href[1]{\endgroup#1\@@endlink}%
\providecommand \@sanitize@url [0]{\catcode `\\12\catcode `\$12\catcode
  `\&12\catcode `\#12\catcode `\^12\catcode `\_12\catcode `\%12\relax}%
\providecommand \@@startlink[1]{}%
\providecommand \@@endlink[0]{}%
\providecommand \url  [0]{\begingroup\@sanitize@url \@url }%
\providecommand \@url [1]{\endgroup\@href {#1}{\urlprefix }}%
\providecommand \urlprefix  [0]{URL }%
\providecommand \Eprint [0]{\href }%
\providecommand \doibase [0]{https://doi.org/}%
\providecommand \selectlanguage [0]{\@gobble}%
\providecommand \bibinfo  [0]{\@secondoftwo}%
\providecommand \bibfield  [0]{\@secondoftwo}%
\providecommand \translation [1]{[#1]}%
\providecommand \BibitemOpen [0]{}%
\providecommand \bibitemStop [0]{}%
\providecommand \bibitemNoStop [0]{.\EOS\space}%
\providecommand \EOS [0]{\spacefactor3000\relax}%
\providecommand \BibitemShut  [1]{\csname bibitem#1\endcsname}%
\let\auto@bib@innerbib\@empty
\bibitem [{\citenamefont {Schau{\ss}}\ \emph {et~al.}(2012)\citenamefont
  {Schau{\ss}}, \citenamefont {Cheneau}, \citenamefont {Endres}, \citenamefont
  {Fukuhara}, \citenamefont {Hild}, \citenamefont {Omran}, \citenamefont
  {Pohl}, \citenamefont {Gross}, \citenamefont {Kuhr},\ and\ \citenamefont
  {Bloch}}]{Schauss2012Nature}%
  \BibitemOpen
  \bibfield  {author} {\bibinfo {author} {\bibfnamefont {P.}~\bibnamefont
  {Schau{\ss}}}, \bibinfo {author} {\bibfnamefont {M.}~\bibnamefont {Cheneau}},
  \bibinfo {author} {\bibfnamefont {M.}~\bibnamefont {Endres}}, \bibinfo
  {author} {\bibfnamefont {T.}~\bibnamefont {Fukuhara}}, \bibinfo {author}
  {\bibfnamefont {S.}~\bibnamefont {Hild}}, \bibinfo {author} {\bibfnamefont
  {A.}~\bibnamefont {Omran}}, \bibinfo {author} {\bibfnamefont
  {T.}~\bibnamefont {Pohl}}, \bibinfo {author} {\bibfnamefont {C.}~\bibnamefont
  {Gross}}, \bibinfo {author} {\bibfnamefont {S.}~\bibnamefont {Kuhr}},\ and\
  \bibinfo {author} {\bibfnamefont {I.}~\bibnamefont {Bloch}},\ }\bibfield
  {title} {\bibinfo {title} {Observation of spatially ordered structures in a
  two-dimensional {{Rydberg}} gas},\ }\href
  {https://doi.org/10.1038/nature11596} {\bibfield  {journal} {\bibinfo
  {journal} {Nature}\ }\textbf {\bibinfo {volume} {491}},\ \bibinfo {pages}
  {87} (\bibinfo {year} {2012})}\BibitemShut {NoStop}%
\bibitem [{\citenamefont {Browaeys}\ and\ \citenamefont
  {Lahaye}(2020)}]{AntoineReview2020}%
  \BibitemOpen
  \bibfield  {author} {\bibinfo {author} {\bibfnamefont {A.}~\bibnamefont
  {Browaeys}}\ and\ \bibinfo {author} {\bibfnamefont {T.}~\bibnamefont
  {Lahaye}},\ }\bibfield  {title} {\bibinfo {title} {Many-body physics with
  individually controlled {R}ydberg atoms},\ }\href@noop {} {\bibfield
  {journal} {\bibinfo  {journal} {Nature Physics}\ }\textbf {\bibinfo {volume}
  {16}},\ \bibinfo {pages} {132} (\bibinfo {year} {2020})}\BibitemShut
  {NoStop}%
\bibitem [{\citenamefont {Labuhn}\ \emph {et~al.}(2016)\citenamefont {Labuhn},
  \citenamefont {Barredo}, \citenamefont {Ravets}, \citenamefont {{de
  L{\'e}s{\'e}leuc}}, \citenamefont {Macr{\`i}}, \citenamefont {Lahaye},\ and\
  \citenamefont {Browaeys}}]{Labuhn2016Nature}%
  \BibitemOpen
  \bibfield  {author} {\bibinfo {author} {\bibfnamefont {H.}~\bibnamefont
  {Labuhn}}, \bibinfo {author} {\bibfnamefont {D.}~\bibnamefont {Barredo}},
  \bibinfo {author} {\bibfnamefont {S.}~\bibnamefont {Ravets}}, \bibinfo
  {author} {\bibfnamefont {S.}~\bibnamefont {{de L{\'e}s{\'e}leuc}}}, \bibinfo
  {author} {\bibfnamefont {T.}~\bibnamefont {Macr{\`i}}}, \bibinfo {author}
  {\bibfnamefont {T.}~\bibnamefont {Lahaye}},\ and\ \bibinfo {author}
  {\bibfnamefont {A.}~\bibnamefont {Browaeys}},\ }\bibfield  {title} {\bibinfo
  {title} {Tunable two-dimensional arrays of single {{Rydberg}} atoms for
  realizing quantum {{Ising}} models},\ }\href
  {https://doi.org/10.1038/nature18274} {\bibfield  {journal} {\bibinfo
  {journal} {Nature}\ }\textbf {\bibinfo {volume} {534}},\ \bibinfo {pages}
  {667} (\bibinfo {year} {2016})}\BibitemShut {NoStop}%
\bibitem [{\citenamefont {Bernien}\ \emph {et~al.}(2017)\citenamefont
  {Bernien}, \citenamefont {Schwartz}, \citenamefont {Keesling}, \citenamefont
  {Levine}, \citenamefont {Omran}, \citenamefont {Pichler}, \citenamefont
  {Choi}, \citenamefont {Zibrov}, \citenamefont {Endres}, \citenamefont
  {Greiner}, \citenamefont {Vuleti\'c},\ and\ \citenamefont
  {Lukin}}]{Bernien2017}%
  \BibitemOpen
  \bibfield  {author} {\bibinfo {author} {\bibfnamefont {H.}~\bibnamefont
  {Bernien}}, \bibinfo {author} {\bibfnamefont {S.}~\bibnamefont {Schwartz}},
  \bibinfo {author} {\bibfnamefont {A.}~\bibnamefont {Keesling}}, \bibinfo
  {author} {\bibfnamefont {H.}~\bibnamefont {Levine}}, \bibinfo {author}
  {\bibfnamefont {A.}~\bibnamefont {Omran}}, \bibinfo {author} {\bibfnamefont
  {H.}~\bibnamefont {Pichler}}, \bibinfo {author} {\bibfnamefont
  {S.}~\bibnamefont {Choi}}, \bibinfo {author} {\bibfnamefont {A.~S.}\
  \bibnamefont {Zibrov}}, \bibinfo {author} {\bibfnamefont {M.}~\bibnamefont
  {Endres}}, \bibinfo {author} {\bibfnamefont {M.}~\bibnamefont {Greiner}},
  \bibinfo {author} {\bibfnamefont {V.}~\bibnamefont {Vuleti\'c}},\ and\
  \bibinfo {author} {\bibfnamefont {M.~D.}\ \bibnamefont {Lukin}},\ }\bibfield
  {title} {\bibinfo {title} {Probing many-body dynamics on a 51-atom quantum
  simulator},\ }\href@noop {} {\bibfield  {journal} {\bibinfo  {journal}
  {Nature}\ }\textbf {\bibinfo {volume} {551}},\ \bibinfo {pages} {579}
  (\bibinfo {year} {2017})}\BibitemShut {NoStop}%
\bibitem [{\citenamefont {Ebadi}\ \emph {et~al.}(2021)\citenamefont {Ebadi},
  \citenamefont {Wang}, \citenamefont {Levine}, \citenamefont {Keesling},
  \citenamefont {Semeghini}, \citenamefont {Omran}, \citenamefont {Bluvstein},
  \citenamefont {Samajdar}, \citenamefont {Pichler}, \citenamefont {Ho},
  \citenamefont {Choi}, \citenamefont {Sachdev}, \citenamefont {Greiner},
  \citenamefont {Vuleti{\'c}},\ and\ \citenamefont {Lukin}}]{Ebadi2021Nature}%
  \BibitemOpen
  \bibfield  {author} {\bibinfo {author} {\bibfnamefont {S.}~\bibnamefont
  {Ebadi}}, \bibinfo {author} {\bibfnamefont {T.~T.}\ \bibnamefont {Wang}},
  \bibinfo {author} {\bibfnamefont {H.}~\bibnamefont {Levine}}, \bibinfo
  {author} {\bibfnamefont {A.}~\bibnamefont {Keesling}}, \bibinfo {author}
  {\bibfnamefont {G.}~\bibnamefont {Semeghini}}, \bibinfo {author}
  {\bibfnamefont {A.}~\bibnamefont {Omran}}, \bibinfo {author} {\bibfnamefont
  {D.}~\bibnamefont {Bluvstein}}, \bibinfo {author} {\bibfnamefont
  {R.}~\bibnamefont {Samajdar}}, \bibinfo {author} {\bibfnamefont
  {H.}~\bibnamefont {Pichler}}, \bibinfo {author} {\bibfnamefont {W.~W.}\
  \bibnamefont {Ho}}, \bibinfo {author} {\bibfnamefont {S.}~\bibnamefont
  {Choi}}, \bibinfo {author} {\bibfnamefont {S.}~\bibnamefont {Sachdev}},
  \bibinfo {author} {\bibfnamefont {M.}~\bibnamefont {Greiner}}, \bibinfo
  {author} {\bibfnamefont {V.}~\bibnamefont {Vuleti{\'c}}},\ and\ \bibinfo
  {author} {\bibfnamefont {M.~D.}\ \bibnamefont {Lukin}},\ }\bibfield  {title}
  {\bibinfo {title} {Quantum phases of matter on a 256-atom programmable
  quantum simulator},\ }\href {https://doi.org/10.1038/s41586-021-03582-4}
  {\bibfield  {journal} {\bibinfo  {journal} {Nature}\ }\textbf {\bibinfo
  {volume} {595}},\ \bibinfo {pages} {227} (\bibinfo {year}
  {2021})}\BibitemShut {NoStop}%
\bibitem [{\citenamefont {Vogel}\ and\ \citenamefont
  {Risken}(1989)}]{Vogel1989Phys.Rev.A}%
  \BibitemOpen
  \bibfield  {author} {\bibinfo {author} {\bibfnamefont {K.}~\bibnamefont
  {Vogel}}\ and\ \bibinfo {author} {\bibfnamefont {H.}~\bibnamefont {Risken}},\
  }\bibfield  {title} {\bibinfo {title} {Determination of quasiprobability
  distributions in terms of probability distributions for the rotated
  quadrature phase},\ }\href {https://doi.org/10.1103/PhysRevA.40.2847}
  {\bibfield  {journal} {\bibinfo  {journal} {Physical Review A}\ }\textbf
  {\bibinfo {volume} {40}},\ \bibinfo {pages} {2847} (\bibinfo {year}
  {1989})}\BibitemShut {NoStop}%
\bibitem [{\citenamefont {James}\ \emph {et~al.}(2001)\citenamefont {James},
  \citenamefont {Kwiat}, \citenamefont {Munro},\ and\ \citenamefont
  {White}}]{James2001Phys.Rev.A}%
  \BibitemOpen
  \bibfield  {author} {\bibinfo {author} {\bibfnamefont {D.~F.~V.}\
  \bibnamefont {James}}, \bibinfo {author} {\bibfnamefont {P.~G.}\ \bibnamefont
  {Kwiat}}, \bibinfo {author} {\bibfnamefont {W.~J.}\ \bibnamefont {Munro}},\
  and\ \bibinfo {author} {\bibfnamefont {A.~G.}\ \bibnamefont {White}},\
  }\bibfield  {title} {\bibinfo {title} {Measurement of qubits},\ }\href
  {https://doi.org/10.1103/PhysRevA.64.052312} {\bibfield  {journal} {\bibinfo
  {journal} {Physical Review A}\ }\textbf {\bibinfo {volume} {64}},\ \bibinfo
  {pages} {052312} (\bibinfo {year} {2001})}\BibitemShut {NoStop}%
\bibitem [{\citenamefont {Gross}\ \emph {et~al.}(2010)\citenamefont {Gross},
  \citenamefont {Liu}, \citenamefont {Flammia}, \citenamefont {Becker},\ and\
  \citenamefont {Eisert}}]{Gross2010Phys.Rev.Lett.}%
  \BibitemOpen
  \bibfield  {author} {\bibinfo {author} {\bibfnamefont {D.}~\bibnamefont
  {Gross}}, \bibinfo {author} {\bibfnamefont {Y.-K.}\ \bibnamefont {Liu}},
  \bibinfo {author} {\bibfnamefont {S.~T.}\ \bibnamefont {Flammia}}, \bibinfo
  {author} {\bibfnamefont {S.}~\bibnamefont {Becker}},\ and\ \bibinfo {author}
  {\bibfnamefont {J.}~\bibnamefont {Eisert}},\ }\bibfield  {title} {\bibinfo
  {title} {Quantum {{State Tomography}} via {{Compressed Sensing}}},\ }\href
  {https://doi.org/10.1103/PhysRevLett.105.150401} {\bibfield  {journal}
  {\bibinfo  {journal} {Physical Review Letters}\ }\textbf {\bibinfo {volume}
  {105}},\ \bibinfo {pages} {150401} (\bibinfo {year} {2010})}\BibitemShut
  {NoStop}%
\bibitem [{\citenamefont {Scholl}\ \emph {et~al.}(2021)\citenamefont {Scholl},
  \citenamefont {Schuler}, \citenamefont {Williams}, \citenamefont
  {Eberharter}, \citenamefont {Barredo}, \citenamefont {Schymik}, \citenamefont
  {Lienhard}, \citenamefont {Henry}, \citenamefont {Lang}, \citenamefont
  {Lahaye}, \citenamefont {L\"auchli},\ and\ \citenamefont
  {Browaeys}}]{scholl2021quantum}%
  \BibitemOpen
  \bibfield  {author} {\bibinfo {author} {\bibfnamefont {P.}~\bibnamefont
  {Scholl}}, \bibinfo {author} {\bibfnamefont {M.}~\bibnamefont {Schuler}},
  \bibinfo {author} {\bibfnamefont {H.~J.}\ \bibnamefont {Williams}}, \bibinfo
  {author} {\bibfnamefont {A.~A.}\ \bibnamefont {Eberharter}}, \bibinfo
  {author} {\bibfnamefont {D.}~\bibnamefont {Barredo}}, \bibinfo {author}
  {\bibfnamefont {K.-N.}\ \bibnamefont {Schymik}}, \bibinfo {author}
  {\bibfnamefont {V.}~\bibnamefont {Lienhard}}, \bibinfo {author}
  {\bibfnamefont {L.-P.}\ \bibnamefont {Henry}}, \bibinfo {author}
  {\bibfnamefont {T.~C.}\ \bibnamefont {Lang}}, \bibinfo {author}
  {\bibfnamefont {T.}~\bibnamefont {Lahaye}}, \bibinfo {author} {\bibfnamefont
  {A.~M.}\ \bibnamefont {L\"auchli}},\ and\ \bibinfo {author} {\bibfnamefont
  {A.}~\bibnamefont {Browaeys}},\ }\bibfield  {title} {\bibinfo {title}
  {{Quantum simulation of 2D antiferromagnets with hundreds of Rydberg
  atoms}},\ }\href {https://doi.org/10.1038/s41586-021-03585-1} {\bibfield
  {journal} {\bibinfo  {journal} {Nature}\ }\textbf {\bibinfo {volume} {595}},\
  \bibinfo {pages} {233} (\bibinfo {year} {2021})}\BibitemShut {NoStop}%
\bibitem [{\citenamefont {Samajdar}\ \emph {et~al.}(2020)\citenamefont
  {Samajdar}, \citenamefont {Ho}, \citenamefont {Pichler}, \citenamefont
  {Lukin},\ and\ \citenamefont {Sachdev}}]{samajdar_complex_2020}%
  \BibitemOpen
  \bibfield  {author} {\bibinfo {author} {\bibfnamefont {R.}~\bibnamefont
  {Samajdar}}, \bibinfo {author} {\bibfnamefont {W.~W.}\ \bibnamefont {Ho}},
  \bibinfo {author} {\bibfnamefont {H.}~\bibnamefont {Pichler}}, \bibinfo
  {author} {\bibfnamefont {M.~D.}\ \bibnamefont {Lukin}},\ and\ \bibinfo
  {author} {\bibfnamefont {S.}~\bibnamefont {Sachdev}},\ }\bibfield  {title}
  {\bibinfo {title} {Complex {{Density Wave Orders}} and {{Quantum Phase
  Transitions}} in a {{Model}} of {{Square}}-{{Lattice Rydberg Atom Arrays}}},\
  }\href {https://doi.org/10.1103/PhysRevLett.124.103601} {\bibfield  {journal}
  {\bibinfo  {journal} {Physical Review Letters}\ }\textbf {\bibinfo {volume}
  {124}},\ \bibinfo {pages} {103601} (\bibinfo {year} {2020})}\BibitemShut
  {NoStop}%
\bibitem [{\citenamefont {Lukin}\ \emph {et~al.}(2001)\citenamefont {Lukin},
  \citenamefont {Fleischhauer}, \citenamefont {Cote}, \citenamefont {Duan},
  \citenamefont {Jaksch}, \citenamefont {Cirac},\ and\ \citenamefont
  {Zoller}}]{lukin2001}%
  \BibitemOpen
  \bibfield  {author} {\bibinfo {author} {\bibfnamefont {M.~D.}\ \bibnamefont
  {Lukin}}, \bibinfo {author} {\bibfnamefont {M.}~\bibnamefont {Fleischhauer}},
  \bibinfo {author} {\bibfnamefont {R.}~\bibnamefont {Cote}}, \bibinfo {author}
  {\bibfnamefont {L.~M.}\ \bibnamefont {Duan}}, \bibinfo {author}
  {\bibfnamefont {D.}~\bibnamefont {Jaksch}}, \bibinfo {author} {\bibfnamefont
  {J.~I.}\ \bibnamefont {Cirac}},\ and\ \bibinfo {author} {\bibfnamefont
  {P.}~\bibnamefont {Zoller}},\ }\bibfield  {title} {\bibinfo {title} {{Dipole
  Blockade and Quantum Information Processing in Mesoscopic Atomic
  Ensembles}},\ }\href {https://doi.org/10.1103/PhysRevLett.87.037901}
  {\bibfield  {journal} {\bibinfo  {journal} {Physical Review Letters}\
  }\textbf {\bibinfo {volume} {87}},\ \bibinfo {pages} {037901} (\bibinfo
  {year} {2001})}\BibitemShut {NoStop}%
\bibitem [{\citenamefont {Miles}\ \emph {et~al.}(2021)\citenamefont {Miles},
  \citenamefont {Bohrdt}, \citenamefont {Wu}, \citenamefont {Chiu},
  \citenamefont {Xu}, \citenamefont {Ji}, \citenamefont {Greiner},
  \citenamefont {Weinberger}, \citenamefont {Demler},\ and\ \citenamefont
  {Kim}}]{Miles2021NatCommun}%
  \BibitemOpen
  \bibfield  {author} {\bibinfo {author} {\bibfnamefont {C.}~\bibnamefont
  {Miles}}, \bibinfo {author} {\bibfnamefont {A.}~\bibnamefont {Bohrdt}},
  \bibinfo {author} {\bibfnamefont {R.}~\bibnamefont {Wu}}, \bibinfo {author}
  {\bibfnamefont {C.}~\bibnamefont {Chiu}}, \bibinfo {author} {\bibfnamefont
  {M.}~\bibnamefont {Xu}}, \bibinfo {author} {\bibfnamefont {G.}~\bibnamefont
  {Ji}}, \bibinfo {author} {\bibfnamefont {M.}~\bibnamefont {Greiner}},
  \bibinfo {author} {\bibfnamefont {K.~Q.}\ \bibnamefont {Weinberger}},
  \bibinfo {author} {\bibfnamefont {E.}~\bibnamefont {Demler}},\ and\ \bibinfo
  {author} {\bibfnamefont {E.-A.}\ \bibnamefont {Kim}},\ }\bibfield  {title}
  {\bibinfo {title} {Correlator convolutional neural networks as an
  interpretable architecture for image-like quantum matter data},\ }\href
  {https://doi.org/10.1038/s41467-021-23952-w} {\bibfield  {journal} {\bibinfo
  {journal} {Nature Communications}\ }\textbf {\bibinfo {volume} {12}},\
  \bibinfo {pages} {3905} (\bibinfo {year} {2021})}\BibitemShut {NoStop}%
\bibitem [{\citenamefont {Carrasquilla}\ and\ \citenamefont
  {Melko}(2017)}]{Carrasquilla2017NaturePhys}%
  \BibitemOpen
  \bibfield  {author} {\bibinfo {author} {\bibfnamefont {J.}~\bibnamefont
  {Carrasquilla}}\ and\ \bibinfo {author} {\bibfnamefont {R.~G.}\ \bibnamefont
  {Melko}},\ }\bibfield  {title} {\bibinfo {title} {Machine learning phases of
  matter},\ }\href {https://doi.org/10.1038/nphys4035} {\bibfield  {journal}
  {\bibinfo  {journal} {Nature Physics}\ }\textbf {\bibinfo {volume} {13}},\
  \bibinfo {pages} {431} (\bibinfo {year} {2017})}\BibitemShut {NoStop}%
\bibitem [{\citenamefont {Wetzel}(2017)}]{Wetzel2017Phys.Rev.Ea}%
  \BibitemOpen
  \bibfield  {author} {\bibinfo {author} {\bibfnamefont {S.~J.}\ \bibnamefont
  {Wetzel}},\ }\bibfield  {title} {\bibinfo {title} {Unsupervised learning of
  phase transitions: {{From}} principal component analysis to variational
  autoencoders},\ }\href {https://doi.org/10.1103/PhysRevE.96.022140}
  {\bibfield  {journal} {\bibinfo  {journal} {Physical Review E}\ }\textbf
  {\bibinfo {volume} {96}},\ \bibinfo {pages} {022140} (\bibinfo {year}
  {2017})}\BibitemShut {NoStop}%
\bibitem [{\citenamefont {Hu}\ \emph {et~al.}(2017)\citenamefont {Hu},
  \citenamefont {Singh},\ and\ \citenamefont
  {Scalettar}}]{hu_discovering_2017}%
  \BibitemOpen
  \bibfield  {author} {\bibinfo {author} {\bibfnamefont {W.}~\bibnamefont
  {Hu}}, \bibinfo {author} {\bibfnamefont {R.~R.~P.}\ \bibnamefont {Singh}},\
  and\ \bibinfo {author} {\bibfnamefont {R.~T.}\ \bibnamefont {Scalettar}},\
  }\bibfield  {title} {\bibinfo {title} {Discovering phases, phase transitions,
  and crossovers through unsupervised machine learning: {{A}} critical
  examination},\ }\href {https://doi.org/10.1103/PhysRevE.95.062122} {\bibfield
   {journal} {\bibinfo  {journal} {Physical Review E}\ }\textbf {\bibinfo
  {volume} {95}},\ \bibinfo {pages} {062122} (\bibinfo {year}
  {2017})}\BibitemShut {NoStop}%
\bibitem [{\citenamefont {Greplova}\ \emph {et~al.}(2020)\citenamefont
  {Greplova}, \citenamefont {Valenti}, \citenamefont {Boschung}, \citenamefont
  {Sch{\"a}fer}, \citenamefont {L{\"o}rch},\ and\ \citenamefont
  {Huber}}]{Greplova2020NewJ.Phys.}%
  \BibitemOpen
  \bibfield  {author} {\bibinfo {author} {\bibfnamefont {E.}~\bibnamefont
  {Greplova}}, \bibinfo {author} {\bibfnamefont {A.}~\bibnamefont {Valenti}},
  \bibinfo {author} {\bibfnamefont {G.}~\bibnamefont {Boschung}}, \bibinfo
  {author} {\bibfnamefont {F.}~\bibnamefont {Sch{\"a}fer}}, \bibinfo {author}
  {\bibfnamefont {N.}~\bibnamefont {L{\"o}rch}},\ and\ \bibinfo {author}
  {\bibfnamefont {S.~D.}\ \bibnamefont {Huber}},\ }\bibfield  {title} {\bibinfo
  {title} {Unsupervised identification of topological phase transitions using
  predictive models},\ }\href {https://doi.org/10.1088/1367-2630/ab7771}
  {\bibfield  {journal} {\bibinfo  {journal} {New Journal of Physics}\ }\textbf
  {\bibinfo {volume} {22}},\ \bibinfo {pages} {045003} (\bibinfo {year}
  {2020})}\BibitemShut {NoStop}%
\bibitem [{\citenamefont {Liu}\ \emph {et~al.}(2019)\citenamefont {Liu},
  \citenamefont {Greitemann},\ and\ \citenamefont
  {Pollet}}]{Liu2019Phys.Rev.B}%
  \BibitemOpen
  \bibfield  {author} {\bibinfo {author} {\bibfnamefont {K.}~\bibnamefont
  {Liu}}, \bibinfo {author} {\bibfnamefont {J.}~\bibnamefont {Greitemann}},\
  and\ \bibinfo {author} {\bibfnamefont {L.}~\bibnamefont {Pollet}},\
  }\bibfield  {title} {\bibinfo {title} {Learning multiple order parameters
  with interpretable machines},\ }\href
  {https://doi.org/10.1103/PhysRevB.99.104410} {\bibfield  {journal} {\bibinfo
  {journal} {Physical Review B}\ }\textbf {\bibinfo {volume} {99}},\ \bibinfo
  {pages} {104410} (\bibinfo {year} {2019})}\BibitemShut {NoStop}%
\bibitem [{\citenamefont {Liu}\ and\ \citenamefont {{van
  Nieuwenburg}}(2018)}]{Liu2018Phys.Rev.Lett.}%
  \BibitemOpen
  \bibfield  {author} {\bibinfo {author} {\bibfnamefont {Y.-H.}\ \bibnamefont
  {Liu}}\ and\ \bibinfo {author} {\bibfnamefont {E.~P.~L.}\ \bibnamefont {{van
  Nieuwenburg}}},\ }\bibfield  {title} {\bibinfo {title} {Discriminative
  {{Cooperative Networks}} for {{Detecting Phase Transitions}}},\ }\href
  {https://doi.org/10.1103/PhysRevLett.120.176401} {\bibfield  {journal}
  {\bibinfo  {journal} {Physical Review Letters}\ }\textbf {\bibinfo {volume}
  {120}},\ \bibinfo {pages} {176401} (\bibinfo {year} {2018})}\BibitemShut
  {NoStop}%
\bibitem [{\citenamefont {Venderley}\ \emph {et~al.}(2021)\citenamefont
  {Venderley}, \citenamefont {Matty}, \citenamefont {Mallayya}, \citenamefont
  {Krogstad}, \citenamefont {Ruff}, \citenamefont {Pleiss}, \citenamefont
  {Kishore}, \citenamefont {Mandrus}, \citenamefont {Phelan}, \citenamefont
  {Poudel}, \citenamefont {Wilson}, \citenamefont {Weinberger}, \citenamefont
  {Upreti}, \citenamefont {Norman}, \citenamefont {Rosenkranz}, \citenamefont
  {Osborn},\ and\ \citenamefont {Kim}}]{Venderley2021ArXiv}%
  \BibitemOpen
  \bibfield  {author} {\bibinfo {author} {\bibfnamefont {J.}~\bibnamefont
  {Venderley}}, \bibinfo {author} {\bibfnamefont {M.}~\bibnamefont {Matty}},
  \bibinfo {author} {\bibfnamefont {K.}~\bibnamefont {Mallayya}}, \bibinfo
  {author} {\bibfnamefont {M.}~\bibnamefont {Krogstad}}, \bibinfo {author}
  {\bibfnamefont {J.}~\bibnamefont {Ruff}}, \bibinfo {author} {\bibfnamefont
  {G.}~\bibnamefont {Pleiss}}, \bibinfo {author} {\bibfnamefont
  {V.}~\bibnamefont {Kishore}}, \bibinfo {author} {\bibfnamefont
  {D.}~\bibnamefont {Mandrus}}, \bibinfo {author} {\bibfnamefont
  {D.}~\bibnamefont {Phelan}}, \bibinfo {author} {\bibfnamefont
  {L.}~\bibnamefont {Poudel}}, \bibinfo {author} {\bibfnamefont {A.~G.}\
  \bibnamefont {Wilson}}, \bibinfo {author} {\bibfnamefont {K.}~\bibnamefont
  {Weinberger}}, \bibinfo {author} {\bibfnamefont {P.}~\bibnamefont {Upreti}},
  \bibinfo {author} {\bibfnamefont {M.~R.}\ \bibnamefont {Norman}}, \bibinfo
  {author} {\bibfnamefont {S.}~\bibnamefont {Rosenkranz}}, \bibinfo {author}
  {\bibfnamefont {R.}~\bibnamefont {Osborn}},\ and\ \bibinfo {author}
  {\bibfnamefont {E.-A.}\ \bibnamefont {Kim}},\ }\bibfield  {title} {\bibinfo
  {title} {Harnessing {{Interpretable}} and {{Unsupervised Machine Learning}}
  to {{Address Big Data}} from {{Modern X}}-ray {{Diffraction}}},\ }\href@noop
  {} {\bibfield  {journal} {\bibinfo  {journal} {arXiv:2008.03275 [cond-mat]}\
  } (\bibinfo {year} {2021})}\BibitemShut {NoStop}%
\bibitem [{\citenamefont {K{\"a}ming}\ \emph {et~al.}(2021)\citenamefont
  {K{\"a}ming}, \citenamefont {Dawid}, \citenamefont {Kottmann}, \citenamefont
  {Lewenstein}, \citenamefont {Sengstock}, \citenamefont {Dauphin},\ and\
  \citenamefont {Weitenberg}}]{kaming_unsupervised_2021}%
  \BibitemOpen
  \bibfield  {author} {\bibinfo {author} {\bibfnamefont {N.}~\bibnamefont
  {K{\"a}ming}}, \bibinfo {author} {\bibfnamefont {A.}~\bibnamefont {Dawid}},
  \bibinfo {author} {\bibfnamefont {K.}~\bibnamefont {Kottmann}}, \bibinfo
  {author} {\bibfnamefont {M.}~\bibnamefont {Lewenstein}}, \bibinfo {author}
  {\bibfnamefont {K.}~\bibnamefont {Sengstock}}, \bibinfo {author}
  {\bibfnamefont {A.}~\bibnamefont {Dauphin}},\ and\ \bibinfo {author}
  {\bibfnamefont {C.}~\bibnamefont {Weitenberg}},\ }\bibfield  {title}
  {\bibinfo {title} {Unsupervised machine learning of topological phase
  transitions from experimental data},\ }\href
  {https://doi.org/10.1088/2632-2153/abffe7} {\bibfield  {journal} {\bibinfo
  {journal} {Machine Learning: Science and Technology}\ }\textbf {\bibinfo
  {volume} {2}},\ \bibinfo {pages} {035037} (\bibinfo {year}
  {2021})}\BibitemShut {NoStop}%
\bibitem [{\citenamefont {Arnold}\ \emph {et~al.}(2021)\citenamefont {Arnold},
  \citenamefont {Sch\"afer}, \citenamefont {\ifmmode~\check{Z}\else
  \v{Z}\fi{}onda},\ and\ \citenamefont {Lode}}]{arnold_interpretable_2021}%
  \BibitemOpen
  \bibfield  {author} {\bibinfo {author} {\bibfnamefont {J.}~\bibnamefont
  {Arnold}}, \bibinfo {author} {\bibfnamefont {F.}~\bibnamefont {Sch\"afer}},
  \bibinfo {author} {\bibfnamefont {M.}~\bibnamefont {\ifmmode~\check{Z}\else
  \v{Z}\fi{}onda}},\ and\ \bibinfo {author} {\bibfnamefont {A.~U.~J.}\
  \bibnamefont {Lode}},\ }\bibfield  {title} {\bibinfo {title} {Interpretable
  and unsupervised phase classification},\ }\href
  {https://doi.org/10.1103/PhysRevResearch.3.033052} {\bibfield  {journal}
  {\bibinfo  {journal} {Physical Review Research}\ }\textbf {\bibinfo {volume}
  {3}},\ \bibinfo {pages} {033052} (\bibinfo {year} {2021})}\BibitemShut
  {NoStop}%
\bibitem [{\citenamefont {Kottmann}\ \emph {et~al.}(2020)\citenamefont
  {Kottmann}, \citenamefont {Huembeli}, \citenamefont {Lewenstein},\ and\
  \citenamefont {Ac\'in}}]{kottmann_unsupervised_2020}%
  \BibitemOpen
  \bibfield  {author} {\bibinfo {author} {\bibfnamefont {K.}~\bibnamefont
  {Kottmann}}, \bibinfo {author} {\bibfnamefont {P.}~\bibnamefont {Huembeli}},
  \bibinfo {author} {\bibfnamefont {M.}~\bibnamefont {Lewenstein}},\ and\
  \bibinfo {author} {\bibfnamefont {A.}~\bibnamefont {Ac\'in}},\ }\bibfield
  {title} {\bibinfo {title} {Unsupervised {Phase} {Discovery} with {Deep}
  {Anomaly} {Detection}},\ }\href
  {https://doi.org/10.1103/PhysRevLett.125.170603} {\bibfield  {journal}
  {\bibinfo  {journal} {Physical Review Letters}\ }\textbf {\bibinfo {volume}
  {125}},\ \bibinfo {pages} {170603} (\bibinfo {year} {2020})}\BibitemShut
  {NoStop}%
\bibitem [{\citenamefont {Cole}\ \emph {et~al.}(2021)\citenamefont {Cole},
  \citenamefont {Loges},\ and\ \citenamefont {Shiu}}]{cole_quantitative_2021}%
  \BibitemOpen
  \bibfield  {author} {\bibinfo {author} {\bibfnamefont {A.}~\bibnamefont
  {Cole}}, \bibinfo {author} {\bibfnamefont {G.~J.}\ \bibnamefont {Loges}},\
  and\ \bibinfo {author} {\bibfnamefont {G.}~\bibnamefont {Shiu}},\ }\bibfield
  {title} {\bibinfo {title} {Quantitative and interpretable order parameters
  for phase transitions from persistent homology},\ }\href
  {https://doi.org/10.1103/PhysRevB.104.104426} {\bibfield  {journal} {\bibinfo
   {journal} {Physical Review B}\ }\textbf {\bibinfo {volume} {104}},\ \bibinfo
  {pages} {104426} (\bibinfo {year} {2021})}\BibitemShut {NoStop}%
\bibitem [{\citenamefont {Huembeli}\ \emph {et~al.}(2019)\citenamefont
  {Huembeli}, \citenamefont {Dauphin}, \citenamefont {Wittek},\ and\
  \citenamefont {Gogolin}}]{huembeli_automated_2019}%
  \BibitemOpen
  \bibfield  {author} {\bibinfo {author} {\bibfnamefont {P.}~\bibnamefont
  {Huembeli}}, \bibinfo {author} {\bibfnamefont {A.}~\bibnamefont {Dauphin}},
  \bibinfo {author} {\bibfnamefont {P.}~\bibnamefont {Wittek}},\ and\ \bibinfo
  {author} {\bibfnamefont {C.}~\bibnamefont {Gogolin}},\ }\bibfield  {title}
  {\bibinfo {title} {Automated discovery of characteristic features of phase
  transitions in many-body localization},\ }\href
  {https://doi.org/10.1103/PhysRevB.99.104106} {\bibfield  {journal} {\bibinfo
  {journal} {Physical Review B}\ }\textbf {\bibinfo {volume} {99}},\ \bibinfo
  {pages} {104106} (\bibinfo {year} {2019})}\BibitemShut {NoStop}%
\bibitem [{\citenamefont {Casert}\ \emph {et~al.}(2019)\citenamefont {Casert},
  \citenamefont {Vieijra}, \citenamefont {Nys},\ and\ \citenamefont
  {Ryckebusch}}]{casert_interpretable_2019}%
  \BibitemOpen
  \bibfield  {author} {\bibinfo {author} {\bibfnamefont {C.}~\bibnamefont
  {Casert}}, \bibinfo {author} {\bibfnamefont {T.}~\bibnamefont {Vieijra}},
  \bibinfo {author} {\bibfnamefont {J.}~\bibnamefont {Nys}},\ and\ \bibinfo
  {author} {\bibfnamefont {J.}~\bibnamefont {Ryckebusch}},\ }\bibfield  {title}
  {\bibinfo {title} {Interpretable machine learning for inferring the phase
  boundaries in a nonequilibrium system},\ }\href
  {https://doi.org/10.1103/PhysRevE.99.023304} {\bibfield  {journal} {\bibinfo
  {journal} {Physical Review E}\ }\textbf {\bibinfo {volume} {99}},\ \bibinfo
  {pages} {023304} (\bibinfo {year} {2019})}\BibitemShut {NoStop}%
\bibitem [{\citenamefont {Huembeli}\ \emph {et~al.}(2018)\citenamefont
  {Huembeli}, \citenamefont {Dauphin},\ and\ \citenamefont
  {Wittek}}]{huembeli_identifying_2018}%
  \BibitemOpen
  \bibfield  {author} {\bibinfo {author} {\bibfnamefont {P.}~\bibnamefont
  {Huembeli}}, \bibinfo {author} {\bibfnamefont {A.}~\bibnamefont {Dauphin}},\
  and\ \bibinfo {author} {\bibfnamefont {P.}~\bibnamefont {Wittek}},\
  }\bibfield  {title} {\bibinfo {title} {Identifying quantum phase transitions
  with adversarial neural networks},\ }\href
  {https://doi.org/10.1103/PhysRevB.97.134109} {\bibfield  {journal} {\bibinfo
  {journal} {Physical Review B}\ }\textbf {\bibinfo {volume} {97}},\ \bibinfo
  {pages} {134109} (\bibinfo {year} {2018})}\BibitemShut {NoStop}%
\bibitem [{\citenamefont {Rem}\ \emph {et~al.}(2019{\natexlab{a}})\citenamefont
  {Rem}, \citenamefont {KÃ€ming}, \citenamefont {Tarnowski}, \citenamefont
  {Asteria}, \citenamefont {FlÃ€schner}, \citenamefont {Becker},
  \citenamefont {Sengstock},\ and\ \citenamefont
  {Weitenberg}}]{rem_identifying_2019}%
  \BibitemOpen
  \bibfield  {author} {\bibinfo {author} {\bibfnamefont {B.~S.}\ \bibnamefont
  {Rem}}, \bibinfo {author} {\bibfnamefont {N.}~\bibnamefont {KÃ€ming}},
  \bibinfo {author} {\bibfnamefont {M.}~\bibnamefont {Tarnowski}}, \bibinfo
  {author} {\bibfnamefont {L.}~\bibnamefont {Asteria}}, \bibinfo {author}
  {\bibfnamefont {N.}~\bibnamefont {FlÃ€schner}}, \bibinfo {author}
  {\bibfnamefont {C.}~\bibnamefont {Becker}}, \bibinfo {author} {\bibfnamefont
  {K.}~\bibnamefont {Sengstock}},\ and\ \bibinfo {author} {\bibfnamefont
  {C.}~\bibnamefont {Weitenberg}},\ }\bibfield  {title} {{\bibinfo {title} {Identifying quantum phase transitions using artificial
  neural networks on experimental data}},\ }\href@noop {} {\bibfield  {journal}
  {\bibinfo  {journal} {Nature Physics}\ }\textbf {\bibinfo {volume} {15}},\
  \bibinfo {pages} {917} (\bibinfo {year} {2019}{\natexlab{a}})}\BibitemShut
  {NoStop}%
\bibitem [{Note1()}]{Note1}%
  \BibitemOpen
  \bibinfo {note} {We choose to work with a larger ${\protect \boldsymbol
  {k}}$-space than the minimal $13 \times 13$ space so that we capture Fourier
  amplitudes at wavevectors $\pi $ and $\pi /2$.}\BibitemShut {Stop}%
\bibitem [{\citenamefont {Bishop}(2006)}]{Bishop2006}%
  \BibitemOpen
  \bibfield  {author} {\bibinfo {author} {\bibfnamefont {C.~M.}\ \bibnamefont
  {Bishop}},\ }\href@noop {} {\emph {\bibinfo {title} {Pattern Recognition and
  Machine Learning}}},\ Information Science and Statistics\ (\bibinfo
  {publisher} {{Springer}},\ \bibinfo {address} {{New York}},\ \bibinfo {year}
  {2006})\BibitemShut {NoStop}%
\bibitem [{Note2()}]{Note2}%
  \BibitemOpen
  \bibinfo {note} {This is somewhat similar in concept to the ``learning by
  confusion'' scheme \cite {vanNieuwenburg2017NaturePhys,
  Liu2018Phys.Rev.Lett.}, though---for this complex multiphase system---we
  carry out these adjustments manually.}\BibitemShut {Stop}%
\bibitem [{Note3()}]{Note3}%
  \BibitemOpen
  \bibinfo {note} {This is distinct from the common choice of using one neural
  network with multineuron output for multiphase detection\cite
  {Venderley2018Phys.Rev.Lett., Rem2019Nat.Phys.,
  Bohrdt2019Nat.Phys.}.}\BibitemShut {Stop}%
\bibitem [{Note4()}]{Note4}%
  \BibitemOpen
  \bibinfo {note} {We restrict $w({\protect \boldsymbol {x}})$ to be symmetric
  under reflections and rotations of the spatial coordinates for simplicity of
  parametrization.}\BibitemShut {Stop}%
\bibitem [{\citenamefont {Meyes}\ \emph {et~al.}(2019)\citenamefont {Meyes},
  \citenamefont {Lu}, \citenamefont {{de Puiseau}},\ and\ \citenamefont
  {Meisen}}]{Meyes2019ArXiv}%
  \BibitemOpen
  \bibfield  {author} {\bibinfo {author} {\bibfnamefont {R.}~\bibnamefont
  {Meyes}}, \bibinfo {author} {\bibfnamefont {M.}~\bibnamefont {Lu}}, \bibinfo
  {author} {\bibfnamefont {C.~W.}\ \bibnamefont {{de Puiseau}}},\ and\ \bibinfo
  {author} {\bibfnamefont {T.}~\bibnamefont {Meisen}},\ }\bibfield  {title}
  {\bibinfo {title} {Ablation {{Studies}} in {{Artificial Neural Networks}}},\
  }\href@noop {} {\bibfield  {journal} {\bibinfo  {journal} {arXiv:1901.08644
  [cs, q-bio]}\ } (\bibinfo {year} {2019})}\BibitemShut {NoStop}%
\bibitem [{\citenamefont {Metlitski}\ \emph {et~al.}(2009)\citenamefont
  {Metlitski}, \citenamefont {Fuertes},\ and\ \citenamefont
  {Sachdev}}]{metlitski2009entanglement}%
  \BibitemOpen
  \bibfield  {author} {\bibinfo {author} {\bibfnamefont {M.~A.}\ \bibnamefont
  {Metlitski}}, \bibinfo {author} {\bibfnamefont {C.~A.}\ \bibnamefont
  {Fuertes}},\ and\ \bibinfo {author} {\bibfnamefont {S.}~\bibnamefont
  {Sachdev}},\ }\bibfield  {title} {\bibinfo {title} {{Entanglement entropy in
  the $O(N)$ model}},\ }\href {https://doi.org/10.1103/PhysRevB.80.115122}
  {\bibfield  {journal} {\bibinfo  {journal} {Physical Review B}\ }\textbf
  {\bibinfo {volume} {80}},\ \bibinfo {pages} {115122} (\bibinfo {year}
  {2009})}\BibitemShut {NoStop}%
\bibitem [{\citenamefont {Zurek}\ \emph {et~al.}(2005)\citenamefont {Zurek},
  \citenamefont {Dorner},\ and\ \citenamefont {Zoller}}]{Zurek2005PRL}%
  \BibitemOpen
  \bibfield  {author} {\bibinfo {author} {\bibfnamefont {W.~H.}\ \bibnamefont
  {Zurek}}, \bibinfo {author} {\bibfnamefont {U.}~\bibnamefont {Dorner}},\ and\
  \bibinfo {author} {\bibfnamefont {P.}~\bibnamefont {Zoller}},\ }\bibfield
  {title} {\bibinfo {title} {Dynamics of a quantum phase transition},\ }\href
  {https://doi.org/10.1103/PhysRevLett.95.105701} {\bibfield  {journal}
  {\bibinfo  {journal} {Physical Review Letters}\ }\textbf {\bibinfo {volume}
  {95}},\ \bibinfo {pages} {105701} (\bibinfo {year} {2005})}\BibitemShut
  {NoStop}%
\bibitem [{\citenamefont {Metlitski}(2020)}]{metlitski2020boundary}%
  \BibitemOpen
  \bibfield  {author} {\bibinfo {author} {\bibfnamefont {M.~A.}\ \bibnamefont
  {Metlitski}},\ }\bibfield  {title} {\bibinfo {title} {{Boundary criticality
  of the O($N$) model in $d = 3$ critically revisited}},\ }\href
  {https://arxiv.org/abs/2009.05119} {\bibfield  {journal} {\bibinfo  {journal}
  {arXiv:2009.05119 [cond-mat.str-el]}\ } (\bibinfo {year} {2020})}\BibitemShut
  {NoStop}%
\bibitem [{\citenamefont {Kalinowski}\ \emph {et~al.}(2021)\citenamefont
  {Kalinowski}, \citenamefont {Samajdar}, \citenamefont {Melko}, \citenamefont
  {Lukin}, \citenamefont {Sachdev},\ and\ \citenamefont
  {Choi}}]{Kalinowski2021}%
  \BibitemOpen
  \bibfield  {author} {\bibinfo {author} {\bibfnamefont {M.}~\bibnamefont
  {Kalinowski}}, \bibinfo {author} {\bibfnamefont {R.}~\bibnamefont
  {Samajdar}}, \bibinfo {author} {\bibfnamefont {R.~G.}\ \bibnamefont {Melko}},
  \bibinfo {author} {\bibfnamefont {M.~D.}\ \bibnamefont {Lukin}}, \bibinfo
  {author} {\bibfnamefont {S.}~\bibnamefont {Sachdev}},\ and\ \bibinfo {author}
  {\bibfnamefont {S.}~\bibnamefont {Choi}},\ }\bibfield  {title} {\bibinfo
  {title} {Bulk and {{Boundary Quantum Phase Transitions}} in a {{Square
  Rydberg Atom Array}}},\ }\href@noop {} {\bibfield  {journal} {\bibinfo
  {journal} {in preparation}\ } (\bibinfo {year} {2021})}\BibitemShut {NoStop}%
\bibitem [{\citenamefont {Samajdar}\ \emph {et~al.}(2021)\citenamefont
  {Samajdar}, \citenamefont {Ho}, \citenamefont {Pichler}, \citenamefont
  {Lukin},\ and\ \citenamefont {Sachdev}}]{Samajdar.2021}%
  \BibitemOpen
  \bibfield  {author} {\bibinfo {author} {\bibfnamefont {R.}~\bibnamefont
  {Samajdar}}, \bibinfo {author} {\bibfnamefont {W.~W.}\ \bibnamefont {Ho}},
  \bibinfo {author} {\bibfnamefont {H.}~\bibnamefont {Pichler}}, \bibinfo
  {author} {\bibfnamefont {M.~D.}\ \bibnamefont {Lukin}},\ and\ \bibinfo
  {author} {\bibfnamefont {S.}~\bibnamefont {Sachdev}},\ }\bibfield  {title}
  {\bibinfo {title} {{Quantum phases of Rydberg atoms on a kagome lattice}},\
  }\href {https://doi.org/10.1073/pnas.2015785118} {\bibfield  {journal}
  {\bibinfo  {journal} {Proceedings of the National Academy of Sciences of the
  United States of America}\ }\textbf {\bibinfo {volume} {118}},\ \bibinfo
  {pages} {e2015785118} (\bibinfo {year} {2021})},\ \Eprint
  {https://arxiv.org/abs/2011.12295} {2011.12295} \BibitemShut {NoStop}%
\bibitem [{\citenamefont {Verresen}\ \emph {et~al.}(2021)\citenamefont
  {Verresen}, \citenamefont {Lukin},\ and\ \citenamefont
  {Vishwanath}}]{Verresen.2021}%
  \BibitemOpen
  \bibfield  {author} {\bibinfo {author} {\bibfnamefont {R.}~\bibnamefont
  {Verresen}}, \bibinfo {author} {\bibfnamefont {M.~D.}\ \bibnamefont
  {Lukin}},\ and\ \bibinfo {author} {\bibfnamefont {A.}~\bibnamefont
  {Vishwanath}},\ }\bibfield  {title} {\bibinfo {title} {{Prediction of Toric
  Code Topological Order from Rydberg Blockade}},\ }\href
  {https://doi.org/10.1103/PhysRevX.11.031005} {\bibfield  {journal} {\bibinfo
  {journal} {Physical Review X}\ }\textbf {\bibinfo {volume} {11}},\ \bibinfo
  {pages} {031005} (\bibinfo {year} {2021})}\BibitemShut {NoStop}%
\bibitem [{\citenamefont {Semeghini}\ \emph {et~al.}(2021)\citenamefont
  {Semeghini}, \citenamefont {Levine}, \citenamefont {Keesling}, \citenamefont
  {Ebadi}, \citenamefont {Wang}, \citenamefont {Bluvstein}, \citenamefont
  {Verresen}, \citenamefont {Pichler}, \citenamefont {Kalinowski},
  \citenamefont {Samajdar}, \citenamefont {Omran}, \citenamefont {Sachdev},
  \citenamefont {Vishwanath}, \citenamefont {Greiner}, \citenamefont
  {Vuletic},\ and\ \citenamefont {Lukin}}]{semeghini_probing_2021}%
  \BibitemOpen
  \bibfield  {author} {\bibinfo {author} {\bibfnamefont {G.}~\bibnamefont
  {Semeghini}}, \bibinfo {author} {\bibfnamefont {H.}~\bibnamefont {Levine}},
  \bibinfo {author} {\bibfnamefont {A.}~\bibnamefont {Keesling}}, \bibinfo
  {author} {\bibfnamefont {S.}~\bibnamefont {Ebadi}}, \bibinfo {author}
  {\bibfnamefont {T.~T.}\ \bibnamefont {Wang}}, \bibinfo {author}
  {\bibfnamefont {D.}~\bibnamefont {Bluvstein}}, \bibinfo {author}
  {\bibfnamefont {R.}~\bibnamefont {Verresen}}, \bibinfo {author}
  {\bibfnamefont {H.}~\bibnamefont {Pichler}}, \bibinfo {author} {\bibfnamefont
  {M.}~\bibnamefont {Kalinowski}}, \bibinfo {author} {\bibfnamefont
  {R.}~\bibnamefont {Samajdar}}, \bibinfo {author} {\bibfnamefont
  {A.}~\bibnamefont {Omran}}, \bibinfo {author} {\bibfnamefont
  {S.}~\bibnamefont {Sachdev}}, \bibinfo {author} {\bibfnamefont
  {A.}~\bibnamefont {Vishwanath}}, \bibinfo {author} {\bibfnamefont
  {M.}~\bibnamefont {Greiner}}, \bibinfo {author} {\bibfnamefont
  {V.}~\bibnamefont {Vuletic}},\ and\ \bibinfo {author} {\bibfnamefont {M.~D.}\
  \bibnamefont {Lukin}},\ }\bibfield  {title} {\bibinfo {title} {Probing
  topological spin liquids on a programmable quantum simulator},\ }\href
  {https://doi.org/10.1126/science.abi8794} {\bibfield  {journal} {\bibinfo
  {journal} {Science}\ }\textbf {\bibinfo {volume} {374}},\ \bibinfo {pages}
  {1242} (\bibinfo {year} {2021})}\BibitemShut {NoStop}%
\bibitem [{\citenamefont {Fishman}\ \emph {et~al.}(2020)\citenamefont
  {Fishman}, \citenamefont {White},\ and\ \citenamefont
  {Stoudenmire}}]{itensor}%
  \BibitemOpen
  \bibfield  {author} {\bibinfo {author} {\bibfnamefont {M.}~\bibnamefont
  {Fishman}}, \bibinfo {author} {\bibfnamefont {S.~R.}\ \bibnamefont {White}},\
  and\ \bibinfo {author} {\bibfnamefont {E.~M.}\ \bibnamefont {Stoudenmire}},\
  }\bibfield  {title} {\bibinfo {title} {{The \mbox{ITensor} Software Library
  for Tensor Network Calculations}},\ }\href {https://arxiv.org/abs/2007.14822}
  {\bibfield  {journal} {\bibinfo  {journal} {arXiv:2007.14822 [cs.MS]}\ }
  (\bibinfo {year} {2020})}\BibitemShut {NoStop}%
\bibitem [{\citenamefont {Paszke}\ \emph {et~al.}(2019)\citenamefont {Paszke},
  \citenamefont {Gross}, \citenamefont {Massa}, \citenamefont {Lerer},
  \citenamefont {Bradbury}, \citenamefont {Chanan}, \citenamefont {Killeen},
  \citenamefont {Lin}, \citenamefont {Gimelshein}, \citenamefont {Antiga},
  \citenamefont {Desmaison}, \citenamefont {Kopf}, \citenamefont {Yang},
  \citenamefont {DeVito}, \citenamefont {Raison}, \citenamefont {Tejani},
  \citenamefont {Chilamkurthy}, \citenamefont {Steiner}, \citenamefont {Fang},
  \citenamefont {Bai},\ and\ \citenamefont {Chintala}}]{paszke_pytorch_2019}%
  \BibitemOpen
  \bibfield  {author} {\bibinfo {author} {\bibfnamefont {A.}~\bibnamefont
  {Paszke}}, \bibinfo {author} {\bibfnamefont {S.}~\bibnamefont {Gross}},
  \bibinfo {author} {\bibfnamefont {F.}~\bibnamefont {Massa}}, \bibinfo
  {author} {\bibfnamefont {A.}~\bibnamefont {Lerer}}, \bibinfo {author}
  {\bibfnamefont {J.}~\bibnamefont {Bradbury}}, \bibinfo {author}
  {\bibfnamefont {G.}~\bibnamefont {Chanan}}, \bibinfo {author} {\bibfnamefont
  {T.}~\bibnamefont {Killeen}}, \bibinfo {author} {\bibfnamefont
  {Z.}~\bibnamefont {Lin}}, \bibinfo {author} {\bibfnamefont {N.}~\bibnamefont
  {Gimelshein}}, \bibinfo {author} {\bibfnamefont {L.}~\bibnamefont {Antiga}},
  \bibinfo {author} {\bibfnamefont {A.}~\bibnamefont {Desmaison}}, \bibinfo
  {author} {\bibfnamefont {A.}~\bibnamefont {Kopf}}, \bibinfo {author}
  {\bibfnamefont {E.}~\bibnamefont {Yang}}, \bibinfo {author} {\bibfnamefont
  {Z.}~\bibnamefont {DeVito}}, \bibinfo {author} {\bibfnamefont
  {M.}~\bibnamefont {Raison}}, \bibinfo {author} {\bibfnamefont
  {A.}~\bibnamefont {Tejani}}, \bibinfo {author} {\bibfnamefont
  {S.}~\bibnamefont {Chilamkurthy}}, \bibinfo {author} {\bibfnamefont
  {B.}~\bibnamefont {Steiner}}, \bibinfo {author} {\bibfnamefont
  {L.}~\bibnamefont {Fang}}, \bibinfo {author} {\bibfnamefont {J.}~\bibnamefont
  {Bai}},\ and\ \bibinfo {author} {\bibfnamefont {S.}~\bibnamefont
  {Chintala}},\ }\bibfield  {title} {\bibinfo {title} {{{PyTorch}}: {{An
  Imperative Style}}, {{High}}-{{Performance Deep Learning Library}}},\ }in\
  \href@noop {} {\emph {\bibinfo {booktitle} {Advances in {{Neural Information
  Processing Systems}} 32}}},\ \bibinfo {editor} {edited by\ \bibinfo {editor}
  {\bibfnamefont {H.}~\bibnamefont {Wallach}}, \bibinfo {editor} {\bibfnamefont
  {H.}~\bibnamefont {Larochelle}}, \bibinfo {editor} {\bibfnamefont
  {A.}~\bibnamefont {Beygelzimer}}, \bibinfo {editor} {\bibfnamefont
  {F.}~\bibnamefont {d'{Alch{\'e}-Buc}}}, \bibinfo {editor} {\bibfnamefont
  {E.}~\bibnamefont {Fox}},\ and\ \bibinfo {editor} {\bibfnamefont
  {R.}~\bibnamefont {Garnett}}}\ (\bibinfo  {publisher} {{Curran Associates,
  Inc.}},\ \bibinfo {year} {2019})\ pp.\ \bibinfo {pages}
  {8024--8035}\BibitemShut {NoStop}%
\bibitem [{\citenamefont {Kingma}\ and\ \citenamefont
  {Ba}(2017)}]{kingma_adam_2017}%
  \BibitemOpen
  \bibfield  {author} {\bibinfo {author} {\bibfnamefont {D.~P.}\ \bibnamefont
  {Kingma}}\ and\ \bibinfo {author} {\bibfnamefont {J.}~\bibnamefont {Ba}},\
  }\bibfield  {title} {\bibinfo {title} {Adam: {{A Method}} for {{Stochastic
  Optimization}}},\ }\href@noop {} {\bibfield  {journal} {\bibinfo  {journal}
  {arXiv:1412.6980 [cs]}\ } (\bibinfo {year} {2017})}\BibitemShut {NoStop}%
\bibitem [{\citenamefont {Ioffe}\ and\ \citenamefont
  {Szegedy}(2015)}]{ioffe_batch_2015}%
  \BibitemOpen
  \bibfield  {author} {\bibinfo {author} {\bibfnamefont {S.}~\bibnamefont
  {Ioffe}}\ and\ \bibinfo {author} {\bibfnamefont {C.}~\bibnamefont
  {Szegedy}},\ }\bibfield  {title} {\bibinfo {title} {Batch {{Normalization}}:
  {{Accelerating Deep Network Training}} by {{Reducing Internal Covariate
  Shift}}},\ }\href {https://arxiv.org/abs/1502.03167} {\bibfield  {journal}
  {\bibinfo  {journal} {arXiv:1502.03167v3 [cs.LG]}\ } (\bibinfo {year}
  {2015})}\BibitemShut {NoStop}%
\bibitem [{\citenamefont {Saffman}\ \emph {et~al.}(2010)\citenamefont
  {Saffman}, \citenamefont {Walker},\ and\ \citenamefont
  {M{\o}lmer}}]{Saffman2010Rev.Mod.Phys.}%
  \BibitemOpen
  \bibfield  {author} {\bibinfo {author} {\bibfnamefont {M.}~\bibnamefont
  {Saffman}}, \bibinfo {author} {\bibfnamefont {T.~G.}\ \bibnamefont
  {Walker}},\ and\ \bibinfo {author} {\bibfnamefont {K.}~\bibnamefont
  {M{\o}lmer}},\ }\bibfield  {title} {\bibinfo {title} {Quantum information
  with {{Rydberg}} atoms},\ }\href {https://doi.org/10.1103/RevModPhys.82.2313}
  {\bibfield  {journal} {\bibinfo  {journal} {Reviews of Modern Physics}\
  }\textbf {\bibinfo {volume} {82}},\ \bibinfo {pages} {2313} (\bibinfo {year}
  {2010})}\BibitemShut {NoStop}%
\bibitem [{\citenamefont {Weimer}\ \emph {et~al.}(2010)\citenamefont {Weimer},
  \citenamefont {M{\"u}ller}, \citenamefont {Lesanovsky}, \citenamefont
  {Zoller},\ and\ \citenamefont {B{\"u}chler}}]{Weimer2010NaturePhys}%
  \BibitemOpen
  \bibfield  {author} {\bibinfo {author} {\bibfnamefont {H.}~\bibnamefont
  {Weimer}}, \bibinfo {author} {\bibfnamefont {M.}~\bibnamefont {M{\"u}ller}},
  \bibinfo {author} {\bibfnamefont {I.}~\bibnamefont {Lesanovsky}}, \bibinfo
  {author} {\bibfnamefont {P.}~\bibnamefont {Zoller}},\ and\ \bibinfo {author}
  {\bibfnamefont {H.~P.}\ \bibnamefont {B{\"u}chler}},\ }\bibfield  {title}
  {\bibinfo {title} {A {{Rydberg}} quantum simulator},\ }\href
  {https://doi.org/10.1038/nphys1614} {\bibfield  {journal} {\bibinfo
  {journal} {Nature Physics}\ }\textbf {\bibinfo {volume} {6}},\ \bibinfo
  {pages} {382} (\bibinfo {year} {2010})}\BibitemShut {NoStop}%
\bibitem [{\citenamefont {Wilk}\ \emph {et~al.}(2010)\citenamefont {Wilk},
  \citenamefont {Ga{\"e}tan}, \citenamefont {Evellin}, \citenamefont {Wolters},
  \citenamefont {Miroshnychenko}, \citenamefont {Grangier},\ and\ \citenamefont
  {Browaeys}}]{Wilk2010Phys.Rev.Lett.}%
  \BibitemOpen
  \bibfield  {author} {\bibinfo {author} {\bibfnamefont {T.}~\bibnamefont
  {Wilk}}, \bibinfo {author} {\bibfnamefont {A.}~\bibnamefont {Ga{\"e}tan}},
  \bibinfo {author} {\bibfnamefont {C.}~\bibnamefont {Evellin}}, \bibinfo
  {author} {\bibfnamefont {J.}~\bibnamefont {Wolters}}, \bibinfo {author}
  {\bibfnamefont {Y.}~\bibnamefont {Miroshnychenko}}, \bibinfo {author}
  {\bibfnamefont {P.}~\bibnamefont {Grangier}},\ and\ \bibinfo {author}
  {\bibfnamefont {A.}~\bibnamefont {Browaeys}},\ }\bibfield  {title} {\bibinfo
  {title} {Entanglement of {{Two Individual Neutral Atoms Using Rydberg
  Blockade}}},\ }\href {https://doi.org/10.1103/PhysRevLett.104.010502}
  {\bibfield  {journal} {\bibinfo  {journal} {Physical Review Letters}\
  }\textbf {\bibinfo {volume} {104}},\ \bibinfo {pages} {010502} (\bibinfo
  {year} {2010})}\BibitemShut {NoStop}%
\bibitem [{\citenamefont {Vedral}\ \emph {et~al.}(1997)\citenamefont {Vedral},
  \citenamefont {Plenio}, \citenamefont {Rippin},\ and\ \citenamefont
  {Knight}}]{Vedral1997Phys.Rev.Lett.}%
  \BibitemOpen
  \bibfield  {author} {\bibinfo {author} {\bibfnamefont {V.}~\bibnamefont
  {Vedral}}, \bibinfo {author} {\bibfnamefont {M.~B.}\ \bibnamefont {Plenio}},
  \bibinfo {author} {\bibfnamefont {M.~A.}\ \bibnamefont {Rippin}},\ and\
  \bibinfo {author} {\bibfnamefont {P.~L.}\ \bibnamefont {Knight}},\ }\bibfield
   {title} {\bibinfo {title} {Quantifying {{Entanglement}}},\ }\href
  {https://doi.org/10.1103/PhysRevLett.78.2275} {\bibfield  {journal} {\bibinfo
   {journal} {Physical Review Letters}\ }\textbf {\bibinfo {volume} {78}},\
  \bibinfo {pages} {2275} (\bibinfo {year} {1997})}\BibitemShut {NoStop}%
\bibitem [{\citenamefont {Elben}\ \emph {et~al.}(2020)\citenamefont {Elben},
  \citenamefont {Kueng}, \citenamefont {Huang}, \citenamefont {{van Bijnen}},
  \citenamefont {Kokail}, \citenamefont {Dalmonte}, \citenamefont {Calabrese},
  \citenamefont {Kraus}, \citenamefont {Preskill}, \citenamefont {Zoller},\
  and\ \citenamefont {Vermersch}}]{Elben2020Phys.Rev.Lett.}%
  \BibitemOpen
  \bibfield  {author} {\bibinfo {author} {\bibfnamefont {A.}~\bibnamefont
  {Elben}}, \bibinfo {author} {\bibfnamefont {R.}~\bibnamefont {Kueng}},
  \bibinfo {author} {\bibfnamefont {H.-Y.~R.}\ \bibnamefont {Huang}}, \bibinfo
  {author} {\bibfnamefont {R.}~\bibnamefont {{van Bijnen}}}, \bibinfo {author}
  {\bibfnamefont {C.}~\bibnamefont {Kokail}}, \bibinfo {author} {\bibfnamefont
  {M.}~\bibnamefont {Dalmonte}}, \bibinfo {author} {\bibfnamefont
  {P.}~\bibnamefont {Calabrese}}, \bibinfo {author} {\bibfnamefont
  {B.}~\bibnamefont {Kraus}}, \bibinfo {author} {\bibfnamefont
  {J.}~\bibnamefont {Preskill}}, \bibinfo {author} {\bibfnamefont
  {P.}~\bibnamefont {Zoller}},\ and\ \bibinfo {author} {\bibfnamefont
  {B.}~\bibnamefont {Vermersch}},\ }\bibfield  {title} {\bibinfo {title}
  {Mixed-{{State Entanglement}} from {{Local Randomized Measurements}}},\
  }\href {https://doi.org/10.1103/PhysRevLett.125.200501} {\bibfield  {journal}
  {\bibinfo  {journal} {Physical Review Letters}\ }\textbf {\bibinfo {volume}
  {125}},\ \bibinfo {pages} {200501} (\bibinfo {year} {2020})}\BibitemShut
  {NoStop}%
\bibitem [{\citenamefont {Huang}\ \emph {et~al.}(2020)\citenamefont {Huang},
  \citenamefont {Kueng},\ and\ \citenamefont {Preskill}}]{Huang2020Nat.Phys.}%
  \BibitemOpen
  \bibfield  {author} {\bibinfo {author} {\bibfnamefont {H.-Y.}\ \bibnamefont
  {Huang}}, \bibinfo {author} {\bibfnamefont {R.}~\bibnamefont {Kueng}},\ and\
  \bibinfo {author} {\bibfnamefont {J.}~\bibnamefont {Preskill}},\ }\bibfield
  {title} {\bibinfo {title} {Predicting many properties of a quantum system
  from very few measurements},\ }\href
  {https://doi.org/10.1038/s41567-020-0932-7} {\bibfield  {journal} {\bibinfo
  {journal} {Nature Physics}\ }\textbf {\bibinfo {volume} {16}},\ \bibinfo
  {pages} {1050} (\bibinfo {year} {2020})}\BibitemShut {NoStop}%
\bibitem [{\citenamefont {Ketterer}\ \emph {et~al.}(2020)\citenamefont
  {Ketterer}, \citenamefont {Wyderka},\ and\ \citenamefont
  {G{\"u}hne}}]{Ketterer2020Quantum}%
  \BibitemOpen
  \bibfield  {author} {\bibinfo {author} {\bibfnamefont {A.}~\bibnamefont
  {Ketterer}}, \bibinfo {author} {\bibfnamefont {N.}~\bibnamefont {Wyderka}},\
  and\ \bibinfo {author} {\bibfnamefont {O.}~\bibnamefont {G{\"u}hne}},\
  }\bibfield  {title} {\bibinfo {title} {Entanglement characterization using
  quantum designs},\ }\href {https://doi.org/10.22331/q-2020-09-16-325}
  {\bibfield  {journal} {\bibinfo  {journal} {Quantum}\ }\textbf {\bibinfo
  {volume} {4}},\ \bibinfo {pages} {325} (\bibinfo {year} {2020})}\BibitemShut
  {NoStop}%
\bibitem [{\citenamefont {Knips}\ \emph {et~al.}(2020)\citenamefont {Knips},
  \citenamefont {Dziewior}, \citenamefont {K{\l}obus}, \citenamefont
  {Laskowski}, \citenamefont {Paterek}, \citenamefont {Shadbolt}, \citenamefont
  {Weinfurter},\ and\ \citenamefont {Meinecke}}]{Knips2020npjQuantumInf}%
  \BibitemOpen
  \bibfield  {author} {\bibinfo {author} {\bibfnamefont {L.}~\bibnamefont
  {Knips}}, \bibinfo {author} {\bibfnamefont {J.}~\bibnamefont {Dziewior}},
  \bibinfo {author} {\bibfnamefont {W.}~\bibnamefont {K{\l}obus}}, \bibinfo
  {author} {\bibfnamefont {W.}~\bibnamefont {Laskowski}}, \bibinfo {author}
  {\bibfnamefont {T.}~\bibnamefont {Paterek}}, \bibinfo {author} {\bibfnamefont
  {P.~J.}\ \bibnamefont {Shadbolt}}, \bibinfo {author} {\bibfnamefont
  {H.}~\bibnamefont {Weinfurter}},\ and\ \bibinfo {author} {\bibfnamefont
  {J.~D.~A.}\ \bibnamefont {Meinecke}},\ }\bibfield  {title} {\bibinfo {title}
  {Multipartite entanglement analysis from random correlations},\ }\href
  {https://doi.org/10.1038/s41534-020-0281-5} {\bibfield  {journal} {\bibinfo
  {journal} {npj Quantum Information}\ }\textbf {\bibinfo {volume} {6}},\
  \bibinfo {pages} {51} (\bibinfo {year} {2020})}\BibitemShut {NoStop}%
\bibitem [{\citenamefont {Neven}\ \emph {et~al.}(2021)\citenamefont {Neven},
  \citenamefont {Carrasco}, \citenamefont {Vitale}, \citenamefont {Kokail},
  \citenamefont {Elben}, \citenamefont {Dalmonte}, \citenamefont {Calabrese},
  \citenamefont {Zoller}, \citenamefont {Vermersch}, \citenamefont {Kueng},\
  and\ \citenamefont {Kraus}}]{Neven2021ArXiv}%
  \BibitemOpen
  \bibfield  {author} {\bibinfo {author} {\bibfnamefont {A.}~\bibnamefont
  {Neven}}, \bibinfo {author} {\bibfnamefont {J.}~\bibnamefont {Carrasco}},
  \bibinfo {author} {\bibfnamefont {V.}~\bibnamefont {Vitale}}, \bibinfo
  {author} {\bibfnamefont {C.}~\bibnamefont {Kokail}}, \bibinfo {author}
  {\bibfnamefont {A.}~\bibnamefont {Elben}}, \bibinfo {author} {\bibfnamefont
  {M.}~\bibnamefont {Dalmonte}}, \bibinfo {author} {\bibfnamefont
  {P.}~\bibnamefont {Calabrese}}, \bibinfo {author} {\bibfnamefont
  {P.}~\bibnamefont {Zoller}}, \bibinfo {author} {\bibfnamefont
  {B.}~\bibnamefont {Vermersch}}, \bibinfo {author} {\bibfnamefont
  {R.}~\bibnamefont {Kueng}},\ and\ \bibinfo {author} {\bibfnamefont
  {B.}~\bibnamefont {Kraus}},\ }\bibfield  {title} {\bibinfo {title}
  {Symmetry-resolved entanglement detection using partial transpose moments},\
  }\href@noop {} {\bibfield  {journal} {\bibinfo  {journal} {arXiv:2103.07443
  [cond-mat]}\ } (\bibinfo {year} {2021})}\BibitemShut {NoStop}%
\bibitem [{\citenamefont {Zhou}\ \emph {et~al.}(2020)\citenamefont {Zhou},
  \citenamefont {Zeng},\ and\ \citenamefont {Liu}}]{PhysRevLett.125.200502}%
  \BibitemOpen
  \bibfield  {author} {\bibinfo {author} {\bibfnamefont {Y.}~\bibnamefont
  {Zhou}}, \bibinfo {author} {\bibfnamefont {P.}~\bibnamefont {Zeng}},\ and\
  \bibinfo {author} {\bibfnamefont {Z.}~\bibnamefont {Liu}},\ }\bibfield
  {title} {\bibinfo {title} {{Single-Copies Estimation of Entanglement
  Negativity}},\ }\href {https://doi.org/10.1103/PhysRevLett.125.200502}
  {\bibfield  {journal} {\bibinfo  {journal} {Physical Review Letters}\
  }\textbf {\bibinfo {volume} {125}},\ \bibinfo {pages} {200502} (\bibinfo
  {year} {2020})}\BibitemShut {NoStop}%
\bibitem [{\citenamefont {Yu}\ \emph {et~al.}(2021)\citenamefont {Yu},
  \citenamefont {Imai},\ and\ \citenamefont
  {G{\"u}hne}}]{Yu2021Phys.Rev.Lett.}%
  \BibitemOpen
  \bibfield  {author} {\bibinfo {author} {\bibfnamefont {X.-D.}\ \bibnamefont
  {Yu}}, \bibinfo {author} {\bibfnamefont {S.}~\bibnamefont {Imai}},\ and\
  \bibinfo {author} {\bibfnamefont {O.}~\bibnamefont {G{\"u}hne}},\ }\bibfield
  {title} {\bibinfo {title} {Optimal {{Entanglement Certification}} from
  {{Moments}} of the {{Partial Transpose}}},\ }\href
  {https://doi.org/10.1103/PhysRevLett.127.060504} {\bibfield  {journal}
  {\bibinfo  {journal} {Physical Review Letters}\ }\textbf {\bibinfo {volume}
  {127}},\ \bibinfo {pages} {060504} (\bibinfo {year} {2021})}\BibitemShut
  {NoStop}%
\bibitem [{\citenamefont {{van Nieuwenburg}}\ \emph {et~al.}(2017)\citenamefont
  {{van Nieuwenburg}}, \citenamefont {Liu},\ and\ \citenamefont
  {Huber}}]{vanNieuwenburg2017NaturePhys}%
  \BibitemOpen
  \bibfield  {author} {\bibinfo {author} {\bibfnamefont {E.~P.~L.}\
  \bibnamefont {{van Nieuwenburg}}}, \bibinfo {author} {\bibfnamefont {Y.-H.}\
  \bibnamefont {Liu}},\ and\ \bibinfo {author} {\bibfnamefont {S.~D.}\
  \bibnamefont {Huber}},\ }\bibfield  {title} {\bibinfo {title} {Learning phase
  transitions by confusion},\ }\href {https://doi.org/10.1038/nphys4037}
  {\bibfield  {journal} {\bibinfo  {journal} {Nature Physics}\ }\textbf
  {\bibinfo {volume} {13}},\ \bibinfo {pages} {435} (\bibinfo {year}
  {2017})}\BibitemShut {NoStop}%
\bibitem [{\citenamefont {Venderley}\ \emph {et~al.}(2018)\citenamefont
  {Venderley}, \citenamefont {Khemani},\ and\ \citenamefont
  {Kim}}]{Venderley2018Phys.Rev.Lett.}%
  \BibitemOpen
  \bibfield  {author} {\bibinfo {author} {\bibfnamefont {J.}~\bibnamefont
  {Venderley}}, \bibinfo {author} {\bibfnamefont {V.}~\bibnamefont {Khemani}},\
  and\ \bibinfo {author} {\bibfnamefont {E.-A.}\ \bibnamefont {Kim}},\
  }\bibfield  {title} {\bibinfo {title} {Machine {{Learning
  Out}}-of-{{Equilibrium Phases}} of {{Matter}}},\ }\href
  {https://doi.org/10.1103/PhysRevLett.120.257204} {\bibfield  {journal}
  {\bibinfo  {journal} {Physical Review Letters}\ }\textbf {\bibinfo {volume}
  {120}},\ \bibinfo {pages} {257204} (\bibinfo {year} {2018})}\BibitemShut
  {NoStop}%
\bibitem [{\citenamefont {Rem}\ \emph {et~al.}(2019{\natexlab{b}})\citenamefont
  {Rem}, \citenamefont {K\"aming}, \citenamefont {Tarnowski}, \citenamefont
  {Asteria}, \citenamefont {Fl\"aschner}, \citenamefont {Becker}, \citenamefont
  {Sengstock},\ and\ \citenamefont {Weitenberg}}]{Rem2019Nat.Phys.}%
  \BibitemOpen
  \bibfield  {author} {\bibinfo {author} {\bibfnamefont {B.~S.}\ \bibnamefont
  {Rem}}, \bibinfo {author} {\bibfnamefont {N.}~\bibnamefont {K\"aming}},
  \bibinfo {author} {\bibfnamefont {M.}~\bibnamefont {Tarnowski}}, \bibinfo
  {author} {\bibfnamefont {L.}~\bibnamefont {Asteria}}, \bibinfo {author}
  {\bibfnamefont {N.}~\bibnamefont {Fl\"aschner}}, \bibinfo {author}
  {\bibfnamefont {C.}~\bibnamefont {Becker}}, \bibinfo {author} {\bibfnamefont
  {K.}~\bibnamefont {Sengstock}},\ and\ \bibinfo {author} {\bibfnamefont
  {C.}~\bibnamefont {Weitenberg}},\ }\bibfield  {title} {\bibinfo {title}
  {{Identifying Quantum Phase Transitions Using Artificial Neural Networks on
  Experimental Data}},\ }\href {https://doi.org/10.1038/s41567-019-0554-0}
  {\bibfield  {journal} {\bibinfo  {journal} {Nature Physics}\ }\textbf
  {\bibinfo {volume} {15}},\ \bibinfo {pages} {917} (\bibinfo {year}
  {2019}{\natexlab{b}})}\BibitemShut {NoStop}%
\bibitem [{\citenamefont {Bohrdt}\ \emph {et~al.}(2019)\citenamefont {Bohrdt},
  \citenamefont {Chiu}, \citenamefont {Ji}, \citenamefont {Xu}, \citenamefont
  {Greif}, \citenamefont {Greiner}, \citenamefont {Demler}, \citenamefont
  {Grusdt},\ and\ \citenamefont {Knap}}]{Bohrdt2019Nat.Phys.}%
  \BibitemOpen
  \bibfield  {author} {\bibinfo {author} {\bibfnamefont {A.}~\bibnamefont
  {Bohrdt}}, \bibinfo {author} {\bibfnamefont {C.~S.}\ \bibnamefont {Chiu}},
  \bibinfo {author} {\bibfnamefont {G.}~\bibnamefont {Ji}}, \bibinfo {author}
  {\bibfnamefont {M.}~\bibnamefont {Xu}}, \bibinfo {author} {\bibfnamefont
  {D.}~\bibnamefont {Greif}}, \bibinfo {author} {\bibfnamefont
  {M.}~\bibnamefont {Greiner}}, \bibinfo {author} {\bibfnamefont
  {E.}~\bibnamefont {Demler}}, \bibinfo {author} {\bibfnamefont
  {F.}~\bibnamefont {Grusdt}},\ and\ \bibinfo {author} {\bibfnamefont
  {M.}~\bibnamefont {Knap}},\ }\bibfield  {title} {\bibinfo {title}
  {Classifying snapshots of the doped {{Hubbard}} model with machine
  learning},\ }\href {https://doi.org/10.1038/s41567-019-0565-x} {\bibfield
  {journal} {\bibinfo  {journal} {Nature Physics}\ }\textbf {\bibinfo {volume}
  {15}},\ \bibinfo {pages} {921} (\bibinfo {year} {2019})}\BibitemShut
  {NoStop}%
\end{thebibliography}%


%
